\documentclass[12pt, a4paper, oneside]{Thesis} 
\graphicspath{{Pictures/}} 
\usepackage[square, numbers, comma, sort&compress]{natbib} 
\usepackage{amsmath,amssymb}
\usepackage[displaymath,textmath,sections,graphics,floats]{preview}
\usepackage{verbatim}
\usepackage[utf8]{inputenc} 
\usepackage{multimedia}
\usepackage{breqn}
\usepackage{url}
\usepackage{braket}

\usepackage[nodayofweek,level]{datetime}
\newcommand{\mydate}{\formatdate{29}{11}{2013}}

\newcommand{\bea}{\begin{eqnarray}}
\newcommand{\eea}{\end{eqnarray}}

\hypersetup{urlcolor=blue, colorlinks=true} 
\title{\ttitle} 

\begin{document}
\frontmatter 
\setstretch{1.3} 
\fancyhead{} 
\rhead{\thepage} 
\lhead{} 
\pagestyle{fancy} 
\newcommand{\HRule}{\rule{\linewidth}{0.5mm}} 
\hypersetup{pdftitle={\ttitle}}
\hypersetup{pdfsubject=\subjectname}
\hypersetup{pdfauthor=\authornames}
\hypersetup{pdfkeywords=\keywordnames}
\begin{titlepage}
\begin{center}
\textsc{\LARGE \univname}\\[1.5cm] 
\textsc{\Large Master's Thesis}\\[0.5cm] 
\HRule \\[0.4cm] 
{\huge \bfseries \ttitle}\\[0.4cm] 
\HRule \\[1.5cm] 
 
\begin{minipage}{0.4\textwidth}
\begin{flushleft} \large
\emph{Author:}\\
\href{http://www.AlexanderGallego.com}{\authornames} 
\end{flushleft}
\end{minipage}
\begin{minipage}{0.4\textwidth}
\begin{flushright} \large
\emph{Supervisor:} \\
\href{http://www.AntonioEnea.com}{\supname} 
\end{flushright}
\end{minipage}\\[3cm]
 
\large \textit{A thesis submitted in fulfilment of the requirements\\ for the degree of \degreename}\\[0.3cm] 
\textit{in the}\\[0.4cm]
\groupname\\\deptname\\[2cm] 
 
{\large \mydate}\\[4cm] 
\vfill
\end{center}
\end{titlepage}
\Declaration{

\addtocontents{toc}{\vspace{1em}} 

I, \authornames, declare that this thesis titled, `\ttitle' and the work presented in it are my own. I confirm that:

\begin{itemize} 
\item[\tiny{$\blacksquare$}] This work was done wholly or mainly while in candidature for a research degree at this University.
\item[\tiny{$\blacksquare$}] Where any part of this thesis has previously been submitted for a degree or any other qualification at this University or any other institution, this has been clearly stated.
\item[\tiny{$\blacksquare$}] Where I have consulted the published work of others, this is always clearly attributed.
\item[\tiny{$\blacksquare$}] Where I have quoted from the work of others, the source is always given. With the exception of such quotations, this thesis is entirely my own work.
\item[\tiny{$\blacksquare$}] I have acknowledged all main sources of help.
\item[\tiny{$\blacksquare$}] Where the thesis is based on work done by myself jointly with others, I have made clear exactly what was done by others and what I have contributed myself.\\
\end{itemize}
 
 
Date: \mydate\\
\rule[1em]{25em}{0.5pt} 
}

\clearpage 
\pagestyle{empty} 
\null\vfill 
\textit{``Happiness is only real when shared''}
\begin{flushright}
Chris McCandless
\end{flushright}
\vfill\vfill\vfill\vfill\vfill\vfill\null 
\clearpage 
\addtotoc{Abstract} 
\abstract{\addtocontents{toc}{\vspace{1em}} 
We study the effects of a general type of features of the inflaton potential on the spectrum and bispectrum of primordial curvature perturbations. These features correspond to a discontinuity in the $n$-th order derivative of the potential which are dumped exponentially away from the value of the field where the feature happens.
Interestingly we find that different values of the amplitude and of the order of the feature can lead to the same effects on the power spectrum on both large and short scales, and on the bispectrum at small scales. Only taking into account the bispectrum at large scales it is possible to resolve this degeneracy.
We provide fully numerical calculations and analytical approximations for the spectrum and the bispectrum, which are in good agreement with each other. The analytical approximation allows to  to determine the class of features which give the same spectrum and can only be distinguished with the bispectrum.
\clearpage 
\setstretch{1.3} 
\acknowledgements{\addtocontents{toc}{\vspace{1em}} 
Agradezco con mucho cariño a mi esposa Karen, mis padres, mis hermanos y a todos mi amigos (los que por fortuna serían muchos para nombrarlos en este espacio). También agradezco a mi asesor, el profesor Antonio Romano, al profesor Diego Restrepo y a mis amigos y compañeros del Instituto de Física.
}

\clearpage 
\pagestyle{fancy} 
\lhead{\emph{Contents}} 
\tableofcontents 
\lhead{\emph{List of Figures}} 
\listoffigures 
\clearpage 
\lhead{\emph{Physical Constants}} 
\listofconstants{lrcl} 
{
Speed of Light & $c$ & $=$ & $1$\\
Boltzmann constant & $k_B$ & $=$ & $1$\\
Reduced Planck constant & $\hbar$ & $=$ & $1$\\
Reduced Planck mass & $M_p$ & $=$ & $1$\\ 
}
\setstretch{1.3} 
\pagestyle{empty} 
\dedicatory{A mis padres: Hern\'an y Libia.\\
A mis mejores amigos: mis hermanos.\\
Y a mi amor y mejor amiga: Karen.
}
\addtocontents{toc}{\vspace{2em}} 
\mainmatter 
\pagestyle{fancy} 
\chapter{Introduction} 

\label{Chapter1} 

\lhead{Chapter 1. \emph{Introduction}} 
In the last few decades the outstanding advances in observational cosmology have set up the Standard Model of Cosmology \citep{et, wmapcpr, pxvi}. There is a significant set of observational data from experiments like the Cosmic Background Explorer (COBE) Satellite, the Sloan Digital Sky Survey (SDSS), the Wilkinson Microwave Anisotropy Probe (WMAP), and the Planck mission \citep{wmapcpr,pxvi}. There are also many other ground-based and sub-orbital experiments \citep{gbe1,gbe2}. Recent experiments reveal a universe that is $13.81$Gyrs old, and made of $4.9\%$ Baryonic Matter. The rest is made of Dark Matter, $26.8\%$, and $68.3\%$ of a unknown type of energy, called Dark Enery \citep{wmapcpr, pxvi,et,pncmb}.

According to the Cosmological principle, the universe is isotropic and homogeneous\citep{mc, eu}. However, results from the CfA survey, and the SDSS, reveled an inhomogeneous distribution of the galaxies in the observable universe. This showed that the universe is only homogeneous at large scales (at about $100$Mpc). Moreover, results from COBE, WMAP, and Planck showed that our universe is not perfectly isotropic \citep{wmapfmr, pxvi,pxxii}.

The Cosmic Microwave Background (CMB) radiation is the radiation coming from the last scattering surface emitted when the universe was only about $378,000$ years old. At this time, protons and electrons combined to form neutral light atoms (recombination) and then photons started to travel freely through space (decoupling) \citep{wmapcpr, pxxii}\footnote{Recombination and decoupling occurred at different times \cite{eu,mc}.}. Although this radiation is extremely isotropic, it is still possible to observe small fluctuations in the temperature map, determined by $\Delta T/T \sim 10^{-5}$ \citep{wmapcpr,wmapfmr, pxxii, xc}. The universe was very young at the recombination epoch. Thus, a detailed measurement and subsequent study of this radiation can gives us a very valuable information about the physics of the early universe \citep{xc}.

In a perfectly homogeneous and isotropic universe, the process of formation of stars, and development of more complex structures (due to gravitational instability) would never have begun \citep{mc,inf, hael}. Thus, any model trying to explain the evolution of the universe must include some level of inhomogeneity to account for the formation of Large-Scale Structures and anisotropies in the CMB. For instance WMAP and Planck data are very well fitted by the $\Lambda$CDM model \citep{wmapcpr, pxvi}. And many different inflationary models can account for the formation of anisotropies in the universe \citep{xc,aersi}.

Cosmological inflation is a period of the universe's evolution during which the scale factor is accelerating \citep{ag, linde, inf}. But instead of being a replacement for the Big Bang, inflation is actually more like an odd-on: after the primordial inflationary period, the universe returns to a non-accelerating phase in which all the success of the Big Bang model is recovered \citep{inf}.

During inflation, space-time itself fluctuates quantum mechanically about a background in an accelerated expanding phase. These microscopic 
fluctuations were spread out to macroscopic scales, where they eventually become small classical fluctuations in the space-time. From then on,  
slightly over-dense regions started the process of gravitational collapse, forming the stars and galaxies \citep{inf, pncmb, mc, eu}.
 
For a long time, observations of the fluctuations in the CMB and the Large-Scale Structures have been focused mainly on the \emph{Gaussian} contribution 
as measured by the two-point correlations of the density fluctuations (or by its Fourier transform, known as the \emph{power spectrum}) \citep{et, m, inf}.
It has been precisely the study of the power spectrum which has settled the numerical value of many of the cosmological parameters in the 
$\Lambda$CDM model \citep{et}.

The main justification to go beyond the study of the Gaussian primordial fluctuations is discriminate between models of inflation which can generate the same power spectrum \citep{xc, rb, modifiedg}. Thus, in order to distinguish these models we study the generation of fluctuations in slow-roll inflationary models, and then compare these results with the non-Gaussianity on the correlation functions obtained experimentally\citep{et, xc, hael, wmapcpr,wmapfmr,pxvi,pxxii}.

\chapter{The Standard $\Lambda$CDM Model of Cosmology} 

\label{Chapter2} 

\lhead{Chapter 2. \emph{The Standard $\Lambda$CDM Model of Cosmology}} 
When we look at the sky we see a nearly isotropic universe. This means that the universe looks almost the same in all directions. Observations also indicate that at large-scales the universe is homogeneous \citep{cosmichomogeneity,sdss}, which means that if we move from one random point to another in the universe then everything looks the same. The \emph{cosmological principle} is based on these empirical observations. It states that at large-scales the visible parts of our universe are homogeneous and isotropic \citep{peeblessm, primack}. 

In 1916, Einstein published his theory of General Relativity (GR), which related the geometry of space-time with the energy-matter content of the universe. A few years later, Alexander Friedmann (1922) and Georges Lemaître (1927) independently applied the cosmological principle to GR. They solved the Einstein equations by assuming a homogeneous and isotropic spacetime. The resulting equations revealed the possibility of an expanding universe. In a 1927 article Lemaître even suggested an estimated value of the rate of expansion. It was only two years later in 1929 that Edwin Hubble confirmed the existence of this expansion and determined its rate, now called the Hubble rate parameter $H$. It was the birth of modern cosmology: for the first time in history we had a compelling and testable theory of the universe.

At the beginning of this chapter we assume that the universe is described by the Big Bang model of cosmology \citep{itoc,bbn}. In Section \ref{c2ot} we review a series of classical observational tests which confirm the predictions of the model. Then in Section \ref{c2pbbm} we point out some shortcomings of the Big Bang that will lead us beyond the standard Big Bang model. The cosmological inflationary theory is introduced in Section \ref{c2pbbm} in order to solve the problems that the Big Bang model is insufficient for. The current standard model of cosmology called the $\Lambda$CDM model \citep{wmapcpr,pxvi} is described in Section \ref{c2lcdm}. 
\section{The Big Bang Model}\label{c2bbm}
The Big Bang model which rests on two theoretical principles: the cosmological principle and General Relativity as the correct theory of gravity on cosmological scales. In this model the energy-matter content of the universe is divided into four primary components \citep{primack, peeblessm}: i) radiation, which is composed of relativistic or nearly relativistic particles such as photons and neutrinos, ii) ``ordinary matter'' composed of protons, neutrons, and electrons. It is generally called baryonic matter, iii) cold dark matter (CDM) which refers to exotic non-baryonic matter that interacts only weakly with ordinary matter, and iv) dark energy (DE), characterized by a large, negative pressure. This is the only form of matter that can accelerate the expansion of the universe. The first three components are commonly united in just one form simply called the matter component \citep{pxvi}.

The Einstein field equations describing the dynamics of the universe with metric tensor $g_{\mu \nu}$ are given by
\begin{equation}\label{efe}
 R_{\mu \nu} - \frac{1}{2} \mathcal{R} g_{\mu \nu} + \Lambda g_{\mu \nu} = 8\pi G T_{\mu \nu},
\end{equation}
where we have taken the speed of light $c=1$. $R_{\mu \nu}$ and $T_{\mu \nu}$ are the Riemann and energy-momentum tensors, respectively. $\mathcal{R}$ is the Ricci scalar, and $\Lambda$ and $G$ are the cosmological and gravitational constants, respectively. The metric signature that we adopt is $(-1,1,1,1)$. When the cosmological principle is assumed the spacetime takes the form of the Friedmann-Lemaître-Robertson-Walker (FLRW) spacetime, with metric
\begin{equation}
  ds^2 = g_{\mu \nu} dx^\mu dx^\nu = - dt^2 + a^2(t) \left(  \frac{dr^2}{1-K r^2}+ r^2d\theta + r^2 \sin^2\theta d\phi^2  \right),
\end{equation}
where $a(t)$ is the scale factor\citep{w, xc}. The constant of curvature is denoted by $K$ with three possible set of values $K<0,K=0, \mbox{ and } K>0$, corresponding to open, flat, and closed universes \citep{w}, respectively.
The Einstein field equations give
\begin{equation}\label{fe}
 \mbox{Friedmann equation \hspace{17 mm}} \left( \frac{\dot {a} }{a} \right)^2 = \frac{8 \pi G }{3} \rho - \frac{K}{a^2},
\end{equation}
\begin{equation}\label{ae}
  \mbox{Acceleration equation \hspace{16 mm}}  \frac{\ddot{a} }{a} = -\frac{4 \pi G }{3} \left( \rho + 3P \right),
\end{equation}
where dots indicate derivatives with respect to time. The Hubble rate parameter which measures how rapidly the scale factor changes is defined as $H\equiv \dot a/a$. From the energy conservation condition $\nabla^\mu T_{\mu \nu}=0$ we can derive a third equation which will be useful later
 \begin{equation}\label{fluide}
  \mbox{Fluid equation \hspace{27 mm}} \frac{d \rho}{dt} + 3 \frac{\dot {a} }{a} \left( \rho + P \right)= 0,
\end{equation}
where $\rho$ and $P$ are the energy density and pressure in the universe, respectively. Here $\rho$ is the sum of several different components of species: radiation, baryonic matter, CDM, and DE. The fluid equation expresses the energy conservation for the fluid as the universe expands.

From the equations of motion it is clear that the rate of expansion $H$ is related to the energy-matter content of the universe $\rho$ and to the geometry of spacetime $K$. 

The fluid equation Eq. \eqref{fluide} can be written as
\begin{equation}\label{fluide2}
 \frac{d}{dt}(\rho a^3) + P \frac{d}{dt}a^3 =0.
\end{equation}
The comoving volume $V=a^3$ is the volume of a region expanding together with the cosmic fluid, thus the energy in a comoving volume is given by $U=\rho a^3$. Then Eq. \eqref{fluide} takes the simple form
\begin{equation}\label{fluide3}
 dU+ P dV=0.
\end{equation}
For a fluid in equilibrium the first law of thermodynamics states 
\begin{equation}
 dU +P dV =T dS,
\end{equation}
where $T$ is the temperature and $S$ the entropy \citep{eu}. An adiabatic process is one in which $dS=0$. This is precisely the case of Eq. \eqref{fluide3} which 
shows that a homogeneous and isotropic universe with a perfect fluid expands adiabatically \citep{gron}. In case that the perfect fluid obeys the barotropic relation
\begin{equation}\label{barotropic}
 P=w \rho,
\end{equation}
Eq. \eqref{fluide2} is expressed as
\begin{equation}
 \frac{d}{dt}(\rho a^3) + w \rho \frac{d}{dt}a^3 =0 \, .
\end{equation}
In this chapter we adopt the usual convention of writing the values at the present time with a subscript $0$. The solution is given in terms of the present value of the density $\rho_0$
\begin{equation}\label{density}
 \rho= \rho_0 a^{-3(w+1)}\, .
\end{equation}
For radiation we have $w=1/3$, for matter, which have effectively zero pressure, $w=0$, and for a cosmological constant $w=-1$. Thus
\begin{align}
    \rho_{\gamma} &= \rho_{0,\gamma} a^{-4} \, , \\
    \rho_{m} &= \rho_{0,m} a^{-3},
\end{align}
where $\rho_{\Lambda}$ is constant. As the scale factor increases, the density of radiation will decrease faster than for matter. The density of a cosmological constant does not change as the universe expands. Recent observations \citep{wmapcpr,pxvi} show that today matter and a cosmological constant dominate the energy content in the universe, but since $\rho_{\gamma} \propto a^{-4}$, as $a$ decreases as time goes back, $\rho$ increases, thus radiation was the dominant content in the early universe. There is a special value of $\rho$ for which $K=0$, it is known as the critical density
\begin{equation}
  \rho_c(t) \equiv \frac{3 H^2(t)}{8 \pi G}.
\end{equation}
An useful dimensionless quantity called the density parameter is defined as
\begin{equation}
  \Omega(t) \equiv \frac{\rho(t)}{\rho_c(t)}.
\end{equation}
With this definition the Friedmann equation Eq. \eqref{fe} can be written as
\begin{equation}\label{fe2}
 \Omega_{tot}+\Omega_K = 1,
\end{equation}
where $\Omega_{tot}= \Omega_m+\Omega_{\Lambda}$ is the sum of all the energy-matter components and $\Omega_k= -K/(aH)^2$ is the density parameter associated with the curvature term. Planck reported the values for $\Omega_m$ and $\Omega_{\Lambda}$ parameters by fitting the model to the data \citep{pxvi,bbn}. The best fit parameter values for the density parameters are $\Omega_m=0.314 \pm 0.020$ and $\Omega_{\Lambda}=0.686 \pm 0.020$. From this results and from Eq. \eqref{fe2} it is safe to conclude that the universe is nearly flat \citep{pxvi}. Thus, from now on, we consider a flat universe $K=0$ with spacetime metric
\begin{equation}
  ds^2 = - dt^2 + a(t)^2 d \vec{x}^2,
\end{equation}
where $\vec x$ are the comoving spatial coordinates and the fundamental distance measure is the distance on this comoving grid. For example, an important comoving distance is the distance light could have traveled since a time $t_i$. In a time $dt$, light travels a physical distance $dt=a dx$ so the total comoving distance light could have traveled, which is called \textit{comoving horizon} is
\begin{equation}\label{conformaltime}
 \tau = \int_{t_i}^t \frac{dt'}{a(t')}.
\end{equation}
It is called the comoving horizon because no information could have propagate further than $\tau$ in the comoving grid since $t_i$. If we now consider $t_i=0$ as the beginning of time then regions in the universe separated by a distance greater than $\tau$ are not causally connected. In the next chapters we will use $\tau$, also called the \textit{conformal time}, to study the evolution of perturbations.

In the case of a flat universe the evolution of the scale factor in terms of time is easy to find. Substituting the density \eqref{density} into Eq. \eqref{fe}, the Friedmann equation can be written 
\begin{equation}
 \left( \frac{\dot {a} }{a} \right)^2 = \frac{8 \pi G }{3} \rho_0 a^{-3(w+1)},
\end{equation}
which can be further written as
\begin{equation}
 \frac{d a}{dt}= \sqrt{\frac{8\pi G \rho_0}{3}} a^{-\frac{3}{2}(w+\frac{1}{3})}.
\end{equation}
The term in the square root is constant in time. If we choose the scale factor at the present time $t_0$ as $a_0=1$, the solutions for the different components are
\begin{align}
  a_{\gamma}(t) &= \left( \frac{t}{t_0} \right)^{1/3} \label{aradiation} \, , \\
  a_{m}(t) &= \left( \frac{t}{t_0} \right)^{2/3}  \label{amatter} \, , \\
  a_{\Lambda}(t) &= \exp\left[{\sqrt{\frac{8\pi \rho_{\Lambda}}{3}} (t-t_0)} \right]. \label{alambda}
\end{align}
As can be seen from the last equation a constant energy density gives a very rapid expansion. And the Hubble constant for each component is
\begin{align}
  H_{\gamma}(t) &= \frac{1}{2t}\, , \\
  H_{m}(t) &= \frac{2}{3t} \, ,\\
  H_{\Lambda}(t) &= \sqrt{\frac{8\pi \rho_{\Lambda}}{3}}= \mbox{constant}.
\end{align}
If we suppose that the universe have been matter-dominated by most of its time, then by measuring the Hubble rate today $H_0=(2/3)t_0^{-1}$ and the age of the universe $t_0$ separately we could have a powerful test for the Big Bang model. In the following section we will confront the theory of an expanding universe with the astronomical observations. After this tests we will see the need of extending the Big Bang theory by adding cosmological inflation, the new model will be called the $\Lambda$CDM model \citep{pxvi}.
\section{Observational Tests}\label{c2ot}
Nowadays we have good evidence that the universe is indeed expanding. This means that early in its history the distance between galaxies were smaller than they are today. It was then an extremely hot and dense state which began to expand rapidly. This early development of the universe was known as the Hot Big-Bang. The success of the Big Bang theory rests on three observational pillars: i) expansion of the universe, ii) nucleosynthesis of light elements, and iii) origin of the cosmic microwave background (CMB) radiation. We review now this measurable signatures which strongly support the notion that the universe evolved from a hot dense, homogeneous, and isotropic gas as proposed by the Big Bang model.
\paragraph{Expansion of the universe}
According to the Big Bang the universe is expanding in all directions, as it can be seen in the equations of motion from GR [Eqs.  \eqref{fe}-\eqref{ae}]. The Hubble diagram Fig. \ref{hubblediagramplot} is the most direct evidence of this prediction \citep{hubblediagram, pxvi}. The Hubble's law states that the recessional velocity $v$ of a galaxy is proportional to its distance $d$ from us, $v=H_0 d$. Current values of the Hubble constant are parameterized by the \textit{little} $h$ as \citep{mc, pxvi}
\begin{equation}
 H_0=100h \mbox{ km s}^{-1}\mbox{Mpc}^{-1},
\end{equation}
where Mpc denotes the astronomical length scale megaparsec equal to $3.086\times10^{22}$ m. From Planck results \citep{pxvi} the present value of $h$ is $h=0.673 \pm 0.012$. As the galaxies are receding from us their emitted wavelength $\lambda_{emit}$ should be stretched out so that the observed wavelength $\lambda_{obs}$ is larger than $\lambda_{emit}$. A measure of this stretching factor is called the redshift $z$ defined by
\begin{equation}
 z \equiv \frac{\lambda_{obs} - \lambda_{emit}}{\lambda_{emit}} = \frac{a_0 - a}{a},
\end{equation}
and since $a_0=1$, the farthest objects are at larger redshift. Moreover, current experiments show that the expansion rate of the universe is accelerating \citep{acceleration1, acceleration2}. The discovery of cosmic acceleration has indicated the presence of DE in the universe \citep{wmapcpr, pxvi}, since, as we saw above in Eq. \eqref{alambda}, a cosmological constant can accelerate the expansion rate of the universe. It also has stimulated new physics beyond the standard model from modifications to GR as possible explanations of its presence \citep{modifiedg}.  
\begin{figure}
 \centering
 \includegraphics[scale=0.6]{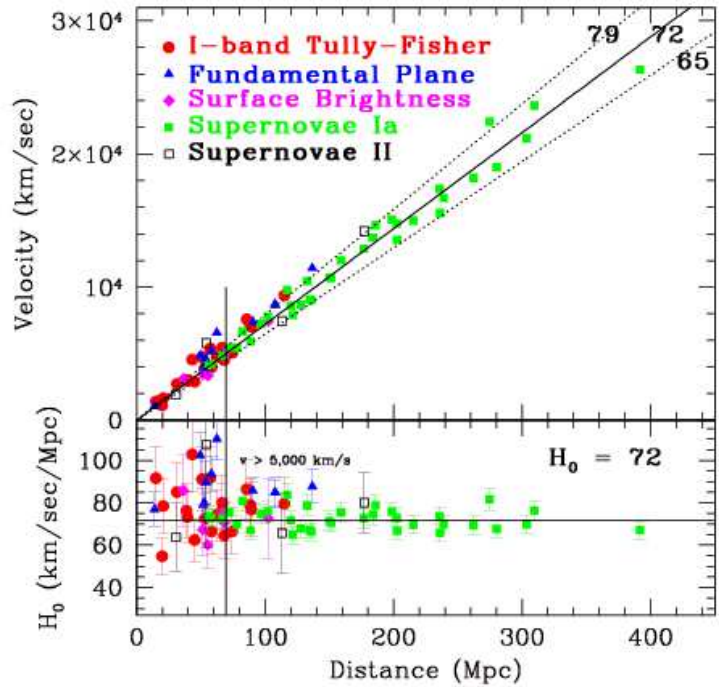}
 \caption[Hubble diagram]{Hubble diagram from the Hubble Space Telescope Key Project. \textit{Top}: Hubble diagram of distance vs. velocity for secondary
distance indicators calibrated by Cepheids using five different measures of distance. \textit{Bottom}: Value of $H_0$ as a function of distance. Image credit: Freedman et al. \citep{hubblediagram}}
 \label{hubblediagramplot}
\end{figure}
\paragraph{Big-bang nucleosynthesis}
The Big-bang nucleosynthesis (BBN) is a nonequilibrium process that took place just a few minutes after the Big-Bang when nucleons were synthesised into the light elements such as D, $^3$H, $^3$He, and $^4$He. To study the BBN we need some basic assumptions from GR and the standard model of particle physics. Less than one second after the Big-Bang the universe was very hot with a temperature larger than $10^{10}$K. It was a rapidly expanding plasma, with most of its energy in the form of radiation and high energy relativistic particles, such as a photons and neutrinos. This high radiation made impossible the formation of any atom or nucleus for if it were produced, it would be soon destroyed by these relativistic particles. As the universe expanded the temperature dropped well below the binding energies of typical nuclei MeV$/k_B$ giving rise to the light elements \citep{nucleosynthesis, bbn}. These predictions of the BBN are in good agreement with current observations \citep{bbn}. For instance in the case of primordial abundance of deuterium these observations correspond to probes of clouds at high redshift on the line of sight of distant quasars. While $^4$He is determined from observations in ionized hydrogen regions of compact blue galaxies \citep{bbn}. 

In the Big Bang model one of the most important parameters is the baryon to photon ratio, $\eta = n_b /n_\gamma$ \citep{nucleosynthesis, bbn} where the photon number density $n_\gamma$ is determined from the CMB temperature and $n_b$ is related to the baryonic density $\Omega_b$ \citep{bbn}. Fig. \ref{bbnplot} shows the abundances of $^4$He, D, $^3$He and $^7$Li from the stellar and extragalactic observational experiments (as the ones mentioned above) and from the WMAP and Planck missions \citep{bbn}. From these and other observations the successful predictions of BBN make it a cornerstone of the Big Bang cosmology \citep{pxvi, bbn, nucleosynthesis}.
\begin{figure}
 \centering
 \includegraphics[scale=0.8]{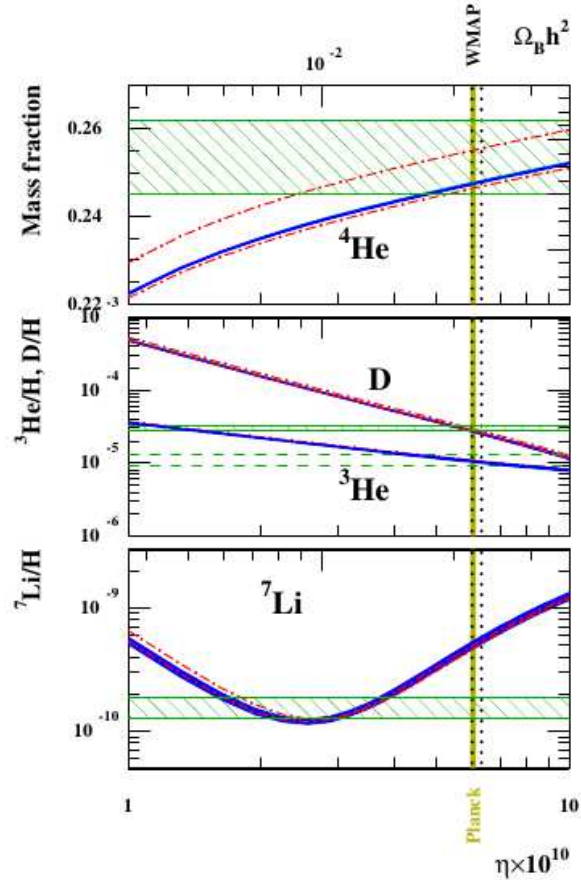}
 \caption[Abundances of light elements]{Abundances of $^4$He, D, $^3$He and $^7$Li (blue lines) as a function of $\Omega_b h^2$ (top) or $\eta$ (bottom) \citep{bbn}. The observational abundances are represented by the green horizontal areas. The WMAP  and Planck baryonic densities results are represented by the dotted-black and solid-yellow lines, respectively. Image credit: Coc et al. \citep{bbn}. }
 \label{bbnplot}
\end{figure}
\paragraph{Cosmic microwave background Radiation} In addition to the assumption of a homogeneous and isotropic universe, we also consider a universe whose rate of expansion $H$ is less than the interaction rate per particle $\Gamma\equiv n \sigma v$ \citep{eu}, where $n$ is the number density of the target interacting particles, $\sigma$ is the cross section, and $v$ is the average relative velocity. This last assumption tells us that after the Big-Bang particles were in thermal equilibrium \citep{eu,peebles}. Which is not surprising since homogeneity and isotropy imply no temperature gradients and hence no heat flow. Photons then  remain tightly coupled to electrons via Compton scattering. Any hydrogen atom produced was immediately ionized due to the low baryon to photon ratio $\eta$. As the temperature dropped below 1eV, when the universe was roughly 378,000 years old, protons and electrons combined to form neutral light atoms. This period is known as the recombination epoch. Decoupling occurred shortly afterwards, when photons started to travel freely through space since their energy was not enough to continue been scattered by electrons. These photons constitute the relic radiation coming from the last scattering surface that we observe today. And, since before decoupling photons were at thermal equilibrium, this radiation should have a blackbody spectrum today. Thus when we observe them, we see a uniform radiation coming from all directions in the sky. The best-stimate CMB photon temperature today is $T_0=2.72548 \pm 0.00057$K \citep{temperature, pxvi}. Fig. \ref{cmbintensity} shows the spectrum of the CMB radiation for various experiments (See Ref. \citep{cmbspectrum} and references therein). The dashed line corresponds to the theoretical curve of a blackbody spectrum. The agreement between theory and experiments is remarkable and is one of the most powerful predictions of the Big Bang.
\begin{figure}
 \centering
 \includegraphics[scale=0.5]{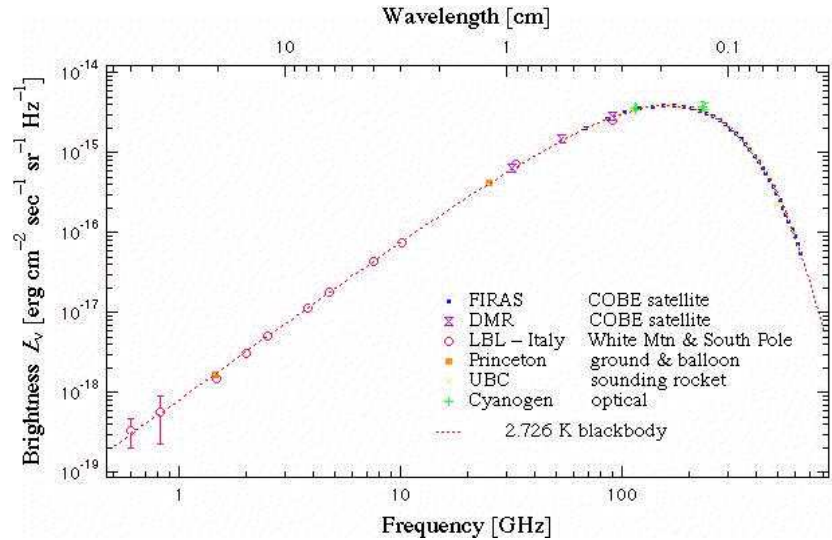}
 \caption[CMB spectrum from COBE]{Brightness of the cosmic microwave background radiation as a function of wavelength (top) or frequency (bottom) from Differential Microwave Radiometers (DMR), Far Infrared Absolute Spectrophotometer (FIRAS) and other ground-based and sub-orbital experiments. The theoretical curve for a blackbody spectrum is represented here by the dashed line. Image credit: G. Smoot \citep{cmbspectrum}. }
 \label{cmbintensity}
\end{figure}
\section{Problems with the Big Bang Model}\label{c2pbbm}
The Big Bang theory has encountered remarkable successes, however it is not complete, and a few unsatisfactory facts remain unexplained. It seems that the model necessitates a larger theoretical framework. We review now the main problems: i) flatness problem, ii) horizon problem, and iii) primordial perturbations.
\paragraph{The flatness problem}
In the Big Bang model, the scale factor behaves like $a \propto t^{q}$ with $q<1$, see Eqs. \eqref{aradiation} and \eqref{amatter}. Rewriting Eq. \eqref{fe2} in the form
\begin{equation}\label{flatness}
 |\Omega_{tot}-1|= \frac{|K|}{a^2 H^2}\, , 
\end{equation}
it can be seen that $|\Omega_{tot}-1|$ is an increasing function of time in either case. That means that the flat geometry is a unstable situation for the universe: any small departure from this stringent initial condition would quickly turn the universe into open or closed. And considering the current value of $\Omega_{tot} \approx 1$ from Planck \citep{wmapcpr, pxvi} it requires an extreme fine-tuning in the early universe, something that the Big Bang paradigm cannot explain.
\paragraph{The horizon problem} 
To understand the horizon problem consider the maximal physical distance that can be covered by a light ray $d_{hor}= a \tau$ since the beginning of time $t_i=0$. From Eq. \eqref{conformaltime} this distance is given by (remember that  $a \propto t^{q}$)
\begin{equation}
 d_{hor}(t)= a(t) \int_{0}^t \frac{dt'}{a(t')}= \frac{t}{1-q}= \frac{q}{1-q} H^{-1},
\end{equation}
which is called the (particle) \textit{horizon} \citep{linf} and it is of the order of the so-called \textit{Hubble radius} $H^{-1}$. We thus have that for a finite time $d_{hor}$ is finite, meaning that the scales at which particles or perturbations were connected at early times had to be small. How could then different regions in the universe, which are separated by large distances, have nearly the same temperature and other physical properties if they could not have been causally connected?
\paragraph{Primordial perturbations} 
The Big Bang assumes a uniform, isotropic universe from its beginning. But this uniformity would not allow the formation of galaxies, or clusters of galaxies in the universe. We know that gravitational instability is the responsible for the structure in the universe. After the Big Bang, as the temperature continued to descend due to the expansion, the energy density of the universe began to be dominated by matter instead of radiation. The primordial perturbations in the matter density accumulated matter from the surrounding areas. In this way initially overdense regions began to grow, independently of how small this overdensity was. Correspondingly, the slightly underdense regions grew more underdense. Hence we need an explanation of how this \textit{primordial fluctuations} were created in order to give rise to underdense/overdense regions in the universe. A detailed search for these primordial perturbations started with the COBE satellite, then with the WMAP, and culminated in the recent Planck mission as can be seen in Fig. \ref{resolution}. Hotter (colder) regions of the sky are represented by red (blue). The temperature fluctuations correspond to regions of slightly different densities at early times, representing the seeds of all future structure: clusters of galaxies, galaxies and stars \citep{wmapfmr,pxxii}. In the COBE mission the resolution to measure the fluctuations were of about $7^\circ$. While WMAP managed to go down to about $0.5^\circ$. The amazing results from Planck come from a resolution of $0.1^\circ$ which allows it to measure temperature fluctuations down to a $10^{-6}$K \citep{pi}. The Planck map of the entire sky is shown in Fig. \ref{planckmap}.
\begin{figure}
 \centering
 \includegraphics[scale=0.5]{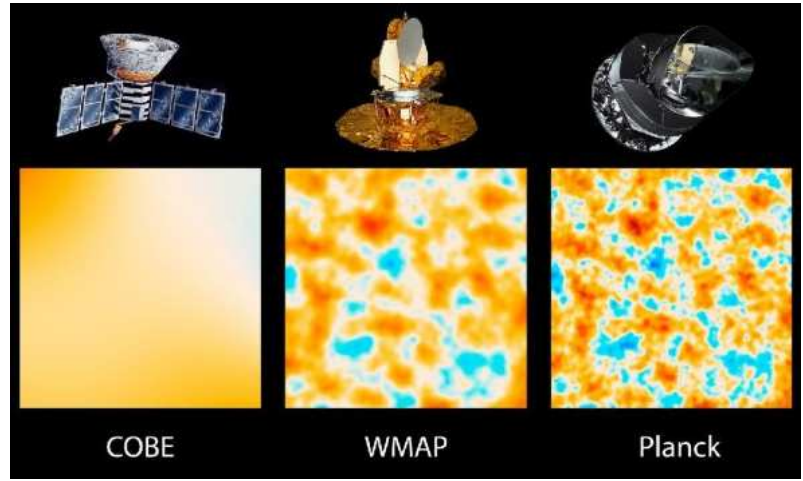}
 \caption[Evolution of satellites]{Comparison of the three satellite's measures of the CMB temperature. The image shows tiny small variations (anisotropies) in the temperature readings, specially from WMAP and Planck missions. Hotter (colder) regions of the sky are represented by red (blue). Image credit: NASA/JPL-Caltech/ESA.}
 \label{resolution}
\end{figure}
\begin{figure}
 \centering
 \includegraphics[scale=0.5]{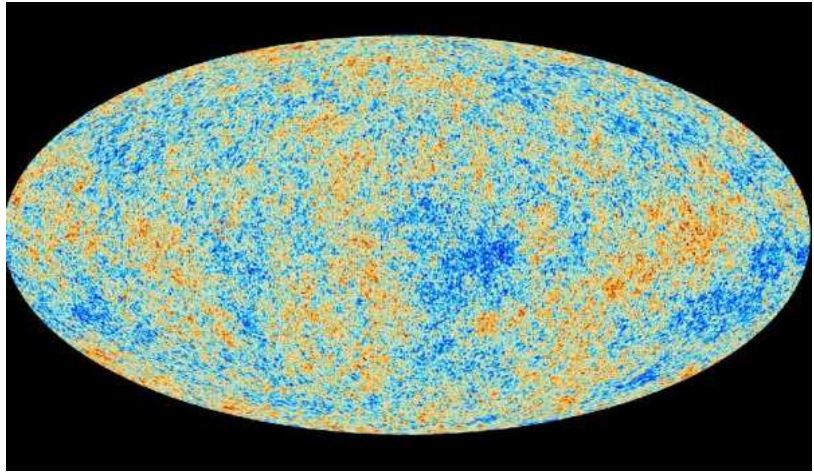}
 \caption[Full sky map by Planck]{All-sky image of the Cosmic microwave background as seen by Planck. Hotter (colder) regions of the sky are represented by red (blue). Image credit: ESA and the Planck Collaboration.}
 \label{planckmap}
\end{figure}

In the next section we will see how the inflationary cosmology theory, when added to the Big Bang, can solved the shortcomings mentioned above.
\section{Solutions to the Big Bang problems from Inflation}\label{sbbi}
In order to solve the problems of the Big Bang it is necessary to introduce cosmological inflation \citep{ag, pxvi, pxxii}. In the next chapter we will propose a precise and quantitative theory of inflation characterized by a scalar field as the energy-matter component of the universe before the Big Bang. For the moment we will only consider its qualitative definition and implications. 

Inflation is any period of the universe’s evolution during which the scale factor is accelerating
\begin{equation}\label{inflationdefinition}
 \mbox{INFLATION} \Leftrightarrow \ddot a >0.
\end{equation}
We will see that from this simple definition the unsolved problems of the Big Bang theory can be relieved.
\paragraph{The flatness problem}
This is the simplest problem that can be solved. From the inflationary condition Eq. \eqref{inflationdefinition} we have that
\begin{equation}
 \ddot a >0  \Leftrightarrow \frac{d}{dt}\dot a>0 \Leftrightarrow  \frac{d}{dt}(a H) >0,
\end{equation}
which indicates that during inflation $aH$ is an increasing function of time. Thus from Eq. \eqref{flatness} we conclude that as $aH$ increases $|\Omega_{tot}-1|$  has to decrease, i.e., $\Omega_{tot} \approx 1$ in concordance with Planck results \citep{pxvi}.
\paragraph{The horizon problem}
The Hubble radius is the distance over which particles can travel in the course of one expansion time \citep{mc}. Another useful quantity is the \textit{comoving Hubble radius} $H^{-1}/a=(aH)^{-1}$. If particles are separated by distances greater than $(aH)^{-1}$ they cannot communicate with each other today, but it is possible that they were in causal contact early on \citep{linf}. As opposed to the comoving horizon $\tau$ in which particles that are separated by a distance larger than $\tau$ could have never communicated, since it is the maximal comoving distance travel by light. A solution to the horizon problem comes from a shrinkage of the comoving Hubble radius so that particles that were initially in causal contact with one another early on can now no longer communicate. We thus would need $(aH)^{-1}$ to decrease, then $aH$ must increase. But this is simply the inflationary condition met above since
\begin{equation}
   \frac{d}{dt}(a H) >0 \Leftrightarrow \frac{d}{dt}\dot a>0 \Leftrightarrow \ddot a >0.
\end{equation}
\paragraph{Primordial perturbations}
The quantum fluctuations of the inflaton field generates the primordial perturbations responsible for the formation of structures. According to this, the universe should exhibit anisotropies from its early times.  Fig. \ref{cmbplot} shows the anisotropy spectrum (power spectrum) of the CMB temperature from Planck \citep{pi}. Planck has now measured the first seven acoustic peaks to high precision that are well fit by inflation and the standard Big Bang model which we will called $\Lambda$CDM \citep{wmapcpr,pxvi}.
\begin{figure}
 \centering
 \includegraphics[scale=0.5]{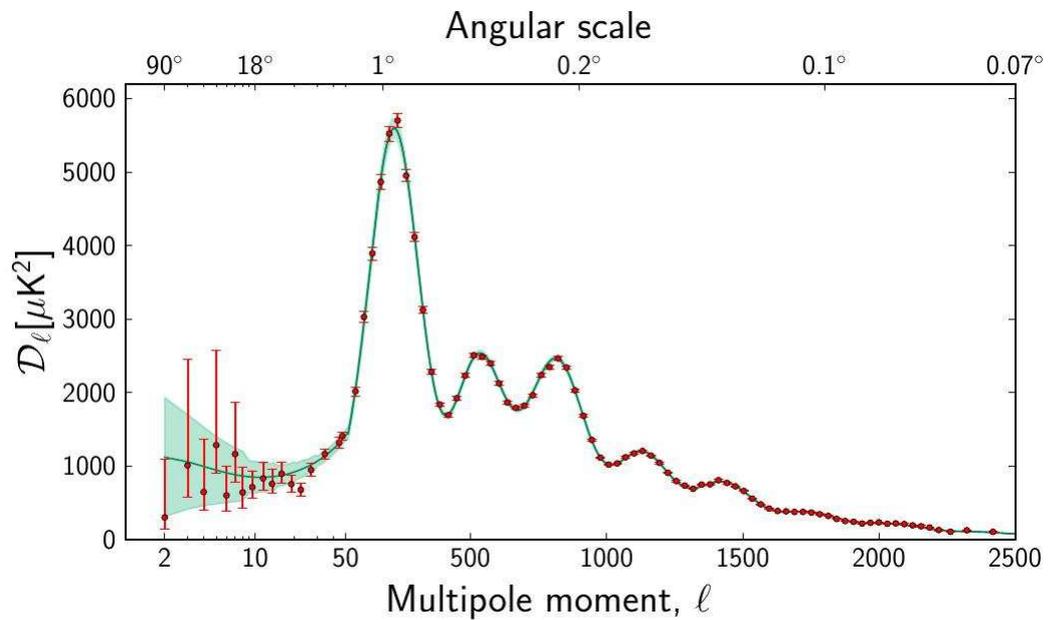}
 \caption[CMB anisotropies by Planck]{The temperature angular power spectrum of the primary CMB from Planck \citep{pi}. The vertical scale is $\mathcal{D}_l=l(l +1)C_l/2\pi$ vs the angular scale (top) or the multipole-moment (bottom). Cosmic variance is represented by the shaded area and is also included in the error bars on individual points. The horizontal axis is logarithmic up to $l= 50$, and linear beyond. The green line is the theoretical prediction from the $\Lambda$CDM model. Image credit: Ade et al. Planck 2013 results \citep{pi}.}
 \label{cmbplot}
\end{figure}
\section{The Standard $\Lambda$CDM Model of Cosmology}\label{c2lcdm}
The remarkable agreement between experiments and the Big Bang plus the inflationary theory have set up the standard model of cosmology. From now on we will simply call it the $\Lambda$CDM model \citep{wmapcpr, pxvi}. The model is based upon an expanding universe whose dynamics are governed by General Relativity with a spatially-flat metric and whose constituents are dominated by CDM and, recently, by a cosmological constant $\Lambda$. The primordial seeds of structure formation generated during inflation are adiabatic and have a gaussian distribution with a nearly scale-invariant spectrum \citep{pxvi}. This model is described by only six parameters.


\chapter{Cosmological Perturbations} 
\label{Chapter3} 
\lhead{Chapter 3. \emph{Cosmological Perturbations}} 

The cosmological inflationary scenario provides a possible mechanism for generating the spatial and matter variations in the very early universe. These are the 
seed fluctuations for the observed structures and CMB anisotropies, hence it is the finest and leading theory explaining the early universe \citep{ag, linde, inf, et, aersi}.
\section{Inflation}
From the acceleration equation, Eq. \eqref{ae}, we can see that in order to fulfill the inflation condition $\ddot a >0$ we must need that
\begin{equation}\label{pd}
 P < -\frac{1}{3}\rho,
\end{equation}
and since the energy density is always positive, the pressure must be negative. A negative pressure can be modeled by a cosmological constant $\Lambda$ with $w=-1$ as we saw before, Eq. \eqref{barotropic}. Also from Eq. \eqref{alambda} we see that $\Lambda$ induces an accelerated expansion. However, inflation must come to an end after some amount of time \citep{pxxii}, thus $\Lambda$ would have to decay into the ``conventional'' particles \citep{inf}. And postulating that a cosmological constant can decay away, converting its energy into another forms is not a viable way to bring inflation to an end \citep{inf}. Besides, we do not want to spoil the success of the Big Bang model. So in typical models, inflation occurs when the universe is extremely young, before nucleosynthesis, perhaps around $10^{-34}$s. Hence, at this large energy scale of around $10^{16}$GeV we must consider the fundamental interactions of particles. 

\subsection{The inflaton}
\label{inflaton}
As the universe expands it cools down. It is known that when a physical system is heated or cooled the properties of the system can change dramatically, leading to phase transitions \citep{inf}. In fact, an example of a phase transition could have been when quarks first condensed to form hadrons \citep{eu}. In fundamental particle physics a phase transition can be controlled by a form of matter known as a scalar field. Thus, in the simplest inflationary universe, a scalar field would decay away after the phase transition, ending inflation and bringing a large amount of expansion in order to solved the flatness and horizon problems. And since we are considering a field at very high energies, we know from quantum mechanics that the field have quantum fluctuations in spacetime. Moreover, spacetime itself fluctuates quantum mechanically about a \textit{background} that is expanding at an accelerating rate. These fluctuations are then responsible for the generation of matter and spatial variations in the very early universe \citep{inf, hael}.

We consider the simplest inflationary model, composed of a single scalar field $\phi$ called the \textit{inflaton}\footnote{The case of multi-field inflation can be studied in Refs. \citep{linf, xc}.}. We first write down the energy-momentum tensor for $\phi$ and then investigate if this kind of field could have a negative pressure $P$ and then if it is actually responsible for an accelerated expansion. The energy-momentum tensor can be written \citep{inf} as
\begin{equation}
 T_{\mu \nu}= \partial_{\mu}\phi \partial_{\nu}\phi - g_{\mu \nu}\left[ \frac{1}{2} g^{\alpha \beta} \partial_{\alpha}\phi \partial_{\beta}\phi + V(\phi) \right],
\end{equation}
where $V(\phi)$ is the potential energy for the field.
The energy density $\rho$ is the time-time component $T^{0 0}$
\begin{equation}\label{t00}
 \rho = \frac{1}{2}\dot \phi^2 + V(\phi).
\end{equation}
The first term is the kinetic energy density. The pressure $P$ is $T^i{}_i$ (no sum over $i$), thus we have
\begin{equation}\label{tii}
  P= \frac{1}{2}\dot \phi^2 - V(\phi).
\end{equation}
Thus, for a scalar field to have negative pressure it is necessary that its kinetic energy be less than the potential. The analogy of a scalar field trapped in a false vacuum to achieved a negative pressure was used in the initial formulation of inflation by A. Guth \citep{ag}. But it was soon realized that such scenario was unfeasible \citep{agw}. To avoid this and several other problems as a consequence of a trapped field the inflaton was considered slowly rolling toward its true vacuum \citep{mc}. The small change in its kinetic energy implies $\phi$ is almost constant and then its total energy is all potential $V$, which remains constant with time. A spacetime with a constant positive vacuum energy density $V_0$ is called a \textit{de Sitter space}. We have already seen the effect of constant energy density on the evolution of the universe, Eq. \eqref{alambda}. In the case of a universe dominated by a scalar field density $\rho$ we have that
\begin{equation}
 \frac{\dot a}{a} = H = \sqrt{\frac{8\pi G \rho}{3}} \approx \mbox{constant},
\end{equation}
hence
\begin{equation}
 a(t) \approx a(t_i) e^{H (t-t_i)},
\end{equation}
where $t_i$ is a beginning time for inflation and $a(t_i)$ the initial value of the scale factor. Thus the inflaton $\phi$ can be responsible for an accelerated expansion.

We first study the zero-mode background evolution of spacetime and the inflaton. Then we move onto perturbations in spacetime and the scalar field $\phi$.
\subsection{The background evolution}\label{background}
We could in principle determine the dynamics of $\phi$ in the background from the Einstein equations by simply replacing the energy density and pressure of the inflaton [Eqs. \eqref{t00} and \eqref{tii}] into Eqs. \eqref{fe}-\eqref{ae}. In this chapter we follow a different approach though, since we should also consider later the quantum dynamics of the inflaton and the spacetime itself under perturbations. And this approach is more conveniently achieved by the ADM formalism in which the starting point for the formulation is the Lagrangian \cite{adm,xc}.

The dynamics  of the spacetime and the scalar field $\phi$ driving inflation are governed by the action \citep{linf, inf}
\begin{equation}\label{action1}
  S = \frac{M_{Pl}}{2} \int d^4x\sqrt{-g} R + \int d^4x \sqrt{-g} \left[ -\frac{1}{2} g^{\mu \nu} \partial_\mu \phi \partial_\nu \phi -V(\phi)
\right],
\end{equation}
where $ M_{Pl} = (8\pi G)^{-1/2}$ is the reduced Planck mass. The first term is the Hilbert-Einstein action. The second term represents the minimal coupling of the inflaton to gravity through the background metric tensor \citep{linf, xc, inf}
\begin{equation}
  ds^2 = g_{\mu \nu} dx^\mu dx^\nu = - dt^2 + a(t)^2 d \vec{x}^2,
\end{equation}
where $\vec{x}$ are the comoving spatial coordinates. The classical equations of motion are given by
\begin{equation}\label{bea}
 \left(\frac{\dot a}{a}\right)^2= \frac{1}{3 M^2_{Pl}}\left( \frac{1}{2} \dot \phi^2 + V(\phi) \right)
\end{equation}
\begin{equation}\label{bephi}
  \ddot \phi + 3H\dot \phi + \partial_{\phi}V = 0,
\end{equation}
\begin{equation}\label{bce}
  \dot H = - \frac{\dot \phi^2}{2 M^2_P},
\end{equation}
where dots and $\partial_{\phi}$ indicate derivatives with respect to time and the scalar field, respectively. The last equation is redundant since it can be derived from the first two; it is called the the continuity equation.

The amount of inflation is quantified by the ratio of the scale factor at the final time to its value at some initial time. Normally, this ratio is a large number, thus it is customary to take the logarithm to give the number of $e$-foldings
\begin{equation}
  N(t) = \ln \left[\frac{a(t)}{a(t_i)}\right].
\end{equation}
It is also called the number of Hubble times, since $N=\int_{t_i}^{t} H(t') dt'$. In order to obtain at least $N \sim 60$ \citep{inf, xc}, we need to impose that the Hubble parameter $H$ does not change much within a Hubble time $H^{-1}$. This requisite gives rise to the first \emph{slow-roll} condition \citep{hael,linf,xc}
\begin{equation}\label{sreps}
  \epsilon \equiv -\frac{\dot H}{H^2} \ll \mbox{\textit{O}}(1).
\end{equation}
The second slow-roll condition is given by the requirement that $\epsilon$ does not change much within a Hubble time \citep{xc}
\begin{equation}\label{sreta}
  \eta \equiv \frac{\dot \epsilon}{\epsilon H} \ll \mbox{\textit{O}}(1).
\end{equation}
From the definition equations it can be seen that the slow-roll parameters are dimensionless. By applying the background solution Eq. \eqref{bce} the slow-roll parameter $\epsilon$ can also be written as
\begin{equation}\label{sreps2}
 \epsilon = \frac{\dot \phi^2}{2 M_P^2 H^2} \ll \mbox{\textit{O}}(1),
\end{equation}
where we can see that the kinetic energy of the scalar field is neglected, thus the energy driving inflation is dominated almost entirely by the potential $V$ \citep{linf, xc}. From the discussion of section \ref{inflaton} this result was expected. The Hubble parameter can be approximated then by
\begin{equation}
 H^2 \approx \frac{1}{3 M^2_P} V.
\end{equation}
The second condition can also be rewritten as
\begin{equation}
 \eta = 2\left( \frac{\ddot \phi}{\dot \phi H} + \epsilon \right) \ll \mbox{\textit{O}}(1),
\end{equation}
so $\ddot \phi$ is negligible and Eq. \eqref{bephi} be expressed by the attractor equation \citep{xc}
\begin{equation}\label{attractors}
 3H \dot \phi + \partial_{\phi}V \approx 0.
\end{equation}
The slow-roll parameter thus described impose restrictions on the rolling veloctity of the scalar field $\phi$. They are also called the Hubble flow function (HFF) slow-roll parameters and can be interpreted as measure of the deviation from a purely de Sitter background \citep{pxxii}. For instance if the scalar field is exactly constant then $\epsilon=0$, and the spacetime is a purely de Sitter space given an exact exponential expansion. The hierarchy of these parameters is defined as $\epsilon_1 =-\dot H/ H^2, \epsilon_{i+1} =\dot \epsilon_i/ (H \epsilon_i)$, for $i \le 1$. An equivalent definition of the slow-roll parameter is given by \citep{pxxii}
\begin{align}
 \epsilon_V & \equiv \frac{M^2_P}{2} \left( \frac{\partial_{\phi}V}{V} \right)^2 \ll \mbox{\textit{O}}(1), \label{srpeps} \\
 \eta &\equiv M^2_P  \frac{\partial_{\phi}\partial_{\phi}V}{V} \ll \mbox{\textit{O}}(1). \label{srpeta}
\end{align}
By using \eqref{attractors} they are related to the HFF slow-roll parameters by \citep{pxxii}
\begin{equation}
 \epsilon = \epsilon_V \, , \hspace{17 mm}  \eta_V = -2\eta_V + 4\epsilon_V.
\end{equation}
Although the various definitions of the slow-roll parameters are equivalent, the definitions Eqs. \eqref{sreps} and \eqref{sreta} are more general \citep{xc}.
\section{non-Gaussianity}\label{NG}
An important prediction of inflationary cosmology is that there should be small departures from the large-scale homogeneity observed in the universe. Inflation also predicts that these perturbations have a characteristic spectrum and bispectrum. During inflation, space-time itself fluctuates quantum mechanically about a background in an accelerated expanding phase. These are the seed fluctuations for the observed structures and CMB anisotropies. The microscopic fluctuations were spread out to macroscopic scales, where they eventually become small classical fluctuations in space-time. From then on, slightly over-dense regions started the process of gravitational collapse, forming the stars and galaxies \citep{inf, pncmb, mc, eu}.

To understand the primordial fluctuations of a quantum field $\zeta$, we need to define a quantity which brings into mutual relation its fluctuations in different places. That quantity is given by the $n$\emph{-point correlation functions} \citep{pe,hael}
\begin{equation}
\Braket{\Omega(t) | \zeta(\vec{x}_1, t) \zeta(\vec{x}_2, t) \ldots \zeta(\vec{x}_n, t) | \Omega(t) },
\end{equation}
where these functions represent the extent to which fluctuations in different places are correlated with each other in the ground state of the interacting theory $\ket{\Omega(t)}$. In case of the early universe, the fluctuations are small compared with the background, then the correlation functions can be evaluated by perturbation theory \citep{pe, hael}. The dynamics governing the field $\zeta(\vec{x}, t)$ is determined by the action $S$ with the associated Lagrange density $\mathcal{L}\left[\zeta\right]$. Expanding $S$ as a series in powers of the perturbation field we have
\begin{equation}
  S=\int d^4x  \left\{ \mathcal{L}^{(0)} \left[ \zeta \right]+ \mathcal{L}^{(1)} \left[ \zeta \right] + \mathcal{L}^{(2)} \left[ \zeta \right] + \mathcal{L}^{(3)} \left[ \zeta \right]+ \ldots \right \}.
\end{equation}
A theory is called \emph{Gaussian} if all the higher order interaction Lagrange beyond the quadratic are zero \citep{hael, a2, inf}, i.e. if $\mathcal{L}^{(i)}\left[ \zeta \right]=0 \mbox{ for } i>2$. We have that higher oder interactions beyond the cubic one can be written in terms of powers of the lower order correlators \citep{hael}, thus in the case of a Gaussian distribution we have that
\begin{equation}\label{sguassian}
  S = \int d^4x    \left \{ \mathcal{L}^{(0)} \left[ \zeta \right] + \mathcal{L}^{(2)} \left[ \zeta \right] \right \},
\end{equation}
is an exact expression. Any departure from Gaussianity is called \emph{non-Gaussianity} \citep{et, inf, hael}, i.e. if $\mathcal{L}^{(i)}\left[ \zeta \right] \ne 0 \mbox{ for } i>2$. In Eq. \eqref{sguassian} the term $\mathcal{L}^{(1)} \left[ \zeta \right]$ vanishes since the background evolution is always a solution to the equations of motion \citep{m,hael}.

In many single-field slow-roll models the inflaton is weakly coupled to gravity, hence the predicted non-Gaussianity is very small and it would be very hard to detect \citep{linf, et, pncmb}. There is, however, a mechanism for producing a large, and detectable amount of non-Gaussianity, which corresponds 
to the violation of any of the following conditions \cite{et}:
\begin{enumerate}
  \item \textbf{Single field.} \, In the generation of the primordial seed fluctuations, there was only the action of \emph{one} quantum field, $\phi$. 
This field was responsible for driving inflation (there is no consideration to multi-fields).

\item \textbf{Slow-roll.} \, During inflation, the evolution of the scalar field was very slow compared to the Hubble time $H^{-1}$.

\item \textbf{Canonical kinetic energy.} \, The kinetic energy of the quantum field is such that the speed of propagation of fluctuations is equal to 
the speed of light.

\item \textbf{Initial vacuum state.} \, The quantum field was in the ``Bunch-Davies" vacuum state \cite{xc}, right before the generation of the 
quantum fluctuations during the inflationary phase.
\end{enumerate}
In chapter \ref{Chapter4} we will study the introduction of a feature in the potential of the inflaton which leads to an enhancement of the the slow-roll parameter $\eta$. In this case $\epsilon$ is small but $\eta$ is large thus, due to this violation of the slow-roll condition, we expect a large contribution to the three-point correlation function from the curvature perturbations.
\section{Curvature perturbations}
\label{curvatureperturbations}
The spacetime primordial variations in the inflationary model have their origin in the quantum behavior of both the field and the spacetime itself. In section \ref{background} we studied the classical quantities $\phi$ and $g_{\mu \nu}$ in a background spacetime. Now we want to study the quantum perturbations $\delta \phi$ and $\delta g_{\mu \nu}$ on the classical fields
\begin{align}
  \phi(t) & \to \phi(t) + \delta \phi(t, \vec x), \\
  g_{\mu \nu}(t) & \to g_{\mu \nu}(t)+ \delta g_{\mu \nu}(t, \vec x).
\end{align}
We only consider scalar perturbations since the tensor perturbation do not contribute to the bispectrum \citep{aer}. To begin with, we need to find the different scalar functions responsible for the general perturbations $\delta \phi$ and $\delta g_{\mu \nu}$ to the classical background. The first one is the scalar fluctuation to the scalar field $\delta \phi$. To determine the other scalars from the metric perturbation we divide $\delta g_{\mu \nu}$ into blocks
\begin{equation}
 \delta g_{\mu \nu}=
 \begin{pmatrix}
 \delta g_{00}(t, \vec x) & \delta g_{0i}(t, \vec x)\\
 \delta g_{i0}(t, \vec x) & \delta g_{ii}(t, \vec x)
\end{pmatrix}.
\end{equation}
We can now determine each of these scalar field. The easiest to determine is the one given by $g_{00}$ which provides one scalar field. The term $\delta_{0i}$ can be viewed as a spatial vector generated by taken the gradient of a scalar field $B$. The term $\delta g_{ij}$ contains two scalar fields, one is its trace and the other one is given by $\xi$ which contributes to $\delta g_{ij}$ by taking two spatial derivatives $\partial_i \partial_j \xi$. We  thus have five separate scalar field, and four of them correspond to the metric perturbations. However, in single field inflation only one of the five scalar perturbations have a physical meaning \citep{xc}.

In order to study the fluctuations about the simple, spatially invariant inflationary background we follow the procedure described by Maldacena \citep{m}. It is convenient to use the ADM formalism \citep{adm,m} in which the metric \eqref{action1} takes the form
\begin{align}
  ds^2 &= -N^2 dt^2 + h_{ij}(dx^i+ N^i dt)(dx^j+ N^j dt) \, ,  \\ \notag
  &= -(N^2-N_iN^i)dt^2 + 2N_i dt dx^i+ h_{ij} dx^i dx^j ,
\end{align}
where $h_{ij}$ the spatial metric of the thre-dimensional hypersurfaces embedded in the full spacetime. $N(t,\vec x)$ is the lapse function and $N^i(t,\vec x)$ the shift vector \citep{w}. One advantage of this approach is that the fields $N$ and $N_i$ are both Lagrange multipliers with no underlying dynamics and whose equations of motion are easy to solve.  Also these equations produce two constraints that reduce the number of independent scalar degrees of freedom by two \citep{hael,xc}. Notice that the spatial indices are contracted using the metric $h_{ij}$, for instance $N^i \equiv h^{ij}N_j$. The components of the inverse metric $g^{\mu \nu}$ are 
\begin{equation}
 g^{00}= -\frac{1}{N^2} \, , \hspace{10 mm} g^{0i}= \frac{1}{N^2}N^i , \hspace{10 mm} g^{ij}= \frac{1}{N^2}(N^2 h^{ij}-N^iN^j).
\end{equation}
Inserting the metric $g_{\mu \nu}$ into the full action of the theory Eq. \eqref{action1} we have \citep{hael}
\begin{align}
  S &= \frac{1}{2}\int dt dx^3 \sqrt{h} N \left[ R^{(3)}+ N^{-2}(E_{ij}E^{ij}-E^2)  \right. \\ \notag
    & \left. + N^{-2}(\dot \phi- N^i\partial_i \phi)^2 - h^{ij} \partial_i \phi \partial_j\phi -2V(\phi) \right],
\end{align}
where $R^{(3)}$ is the $3$-D Ricci scalar curvature constructed form $h_{ij}$ and we have set $M_P=1$ to simplify the calculations. We have introduced the spatial tensor $E_{ij}$ which is closely related to the extrinsic curvature $K_{ij}$ \citep{w}. The latter determines the rate of change of the spatial metric $h_{ij}$, as we move along the timelike geodesics orthogonal to the hypersurfaces embedded in the spacetime \citep{w}. The former is defined by
\begin{align}
 E_{ij} &=  \frac{1}{2}\left(\dot{h}_{ij}-\nabla^{(3)}_i N_j -\nabla^{(3)}_j N_i \right) \, , \\ 
 & = \frac{1}{2}\left( \dot h_{ij}- h_{ik}\partial_j N^k - h_{jk} \partial_i N^k - N^k \partial_k h_{ij} \right), \notag
\end{align}
where $\nabla^{(3)}$ is calculated with the curvature $3$-D metric. The definition of $E$ is
\begin{equation}
 E= h_{ij} E^{ij}.
\end{equation}
To proceed with the ADM formalism we now vary the action with respect to the fields $N$ and $N^i$. Varying with respect to $N$ we have
\begin{equation}
  R^{(3)}- N^{-2}(E_{ij}E^{ij}-E^2) - N^{-2}(\dot \phi- N^i\partial_i \phi)^2 - h^{ij} \partial_i \phi \partial_j\phi -2V = 0.
\end{equation}
From Eq. \eqref{bephi} we can replace the pontential for which we have not made any particular consideration, yielding
\begin{equation}\label{constrainte1}
 R^{(3)}+ N^{-2}(E_{ij}E^{ij}-E^2) - N^{-2}(\dot \phi- N^i\partial_i \phi)^2 - h^{ij} \partial_i \phi \partial_j\phi -6H^2+\dot \phi^2 = 0.
\end{equation}
And with respect to $N^i$
\begin{equation}\label{constrainte2}
 \nabla^{(3)}_j \left[ N^{-1} (E_i{}^j- \delta_i{}^j E) \right]- N^{-1} (\dot \phi -N^j \partial_j \phi ) \partial_i \phi = 0.
\end{equation}
A general parameterization of the field's fluctuation is given by
\begin{equation}
 \phi = \phi(t) + \delta\phi(t,\vec x).
\end{equation}
And of the scalar perturbations to the metric can be written down as \cite{mc,hael}
\begin{align}
 N &= 1 + \Phi(t,\vec x),\\
 N^i &= \delta^{ij} \partial_j B(t,\vec x) , \\
 h_{ij} &= a^2 \left[ \left(1+2\zeta(t,\vec x) \right) \delta_{ij} + \partial_i\partial_j \xi \right].
\end{align}
We can remove two of the five scalar perturbations by a time redefinition and a shifting of the spatial coordinates $x^i \to x^i+\delta^{ij}\partial_j f(t,\vec x)$. We use this freedom to choose the coordinates such that $\delta \phi=0$ and $\xi=0$. Hence the parameterizations on the fields are
\begin{align}
 N &= 1 + \Phi(t,\vec x),\\
 N^i &= \delta^{ij} \partial_j B(t,\vec x),\\
 h_{ij} &= a^2 e^{2\zeta(t,\vec x)}\delta_{ij}, \\
 \phi &= \phi(t),
\end{align}
where now $\phi$ has no spatial dependence and $\zeta$ has been slightly redefined for useful purposes when deriving the second and third order action of the theory. In this gauge the constraint equations Eqs. \eqref{constrainte1} and \eqref{constrainte2} are written
\begin{align}
 R^{(3)}- N^{-2}(E_{ij}E^{ij}-E^2) - (1-N^{-2}) \dot \phi^2 -6 H^2 &= 0 \label{constraint12}, \\
 \nabla^{(3)}_j \left[ N^{-1} (E_i{}^j- \delta_i{}^j E) \right] &= 0. \label{constraint22}
\end{align}
We now solve the constraint equations in terms of the scalar field fluctuations. In these coordinates the scalar curvature is written 
\begin{equation}
 R^{(3)}=-4a^{-2}e^{-2\zeta}\left[\partial_i \partial^i \zeta - \frac{1}{2}\partial_i \zeta \partial^i \zeta \right]= -4a^{-2}\partial_i \partial^i \zeta + \mathcal{O}(\zeta^2),
\end{equation}
where we have expanded $\zeta$ up to first order. Similarly
\begin{align}
 E_{ij} &= a^2 e^{2\zeta} \Big\{ \left[ H +\dot \zeta -\partial_k \zeta \partial^k B \right]- \partial_i\partial_j B \Big\} \\
 &= a^2\left[H(1+2\zeta)\delta_{ij}+ \dot \zeta \delta_{ij}- \partial_i\partial_j B \right] +\dots
\end{align}
then to first order in $\zeta$
\begin{equation}
 E_{ij} E^{ij} - E^2 = -6H^2-12H \dot \zeta +4H \partial_i\partial_j B+ \dots 
\end{equation}
In this way the constraint equation Eq.\eqref{constraint12} can be written to first order as
\begin{equation}\label{constraint13}
 -3H \left[2H \Phi - \dot \zeta \right] - \partial_i \partial^i \left[H B a^{-2} \zeta \right] + \dot \phi^2 \Phi =0.
\end{equation}
And Eq. \eqref{constraint22} takes the form
\begin{equation}
 2\partial_i \left[ 2H \Phi - \dot \zeta \right] = 0.
\end{equation}
Thus we have that
\begin{equation}
 \Phi=\frac{1}{2} \frac{\dot \zeta}{H} + \mbox{constant}= \frac{1}{2} \frac{\dot \zeta}{H} ,
\end{equation}
for which the constant is taken as zero in order to recover the original background metric if the perturbations are removed \citep{hael}. Inserting $\Phi$ into Eq. \eqref{constraint13} we have
\begin{equation}
 \partial_i \partial^i \left[H B + a^{-2} \zeta \right] + \frac{\dot \phi^2}{2H} \dot \zeta=0.
\end{equation}
And solving for $B$
\begin{equation}
 B= -\frac{1}{a^2 H} \zeta + \frac{1}{a^2}\chi,
\end{equation}
with
\begin{equation}
 \chi=a^2\frac{\dot \phi^2}{2H^2} \partial^{-2} \dot \zeta= a^2\epsilon \partial^{-2} \dot \zeta,
\end{equation}
where $\partial^{-2}$ is the inverse Laplacian and we have used the expression for $\epsilon$, Eq. \eqref{sreps2}. The definition of $\chi$ given here differs from that of \citep{hael}, but agrees with the one in Refs. \citep{xc, aer}. We have now solved the constraints $N$ and $N^i$ and have obtained expressions for them in terms of only one scalar field perturbation $\zeta$
\begin{align}
 N &= 1 + \frac{1}{2} \frac{\dot \zeta}{H},\\
 N^i &= \partial^i \left[-\frac{1}{a^2 H} \zeta + \frac{1}{a^2}\chi \right], \\
 h_{ij} &= a^2 e^{2\zeta} \delta_{ij}.
\end{align}
To obtain the two and three-point correlation functions we write the full action in the chosen gauge where $\delta \phi=0$
\begin{equation}
 S = \frac{1}{2}\int dt dx^3 \sqrt{h} N \left[ R^{(3)}+ N^{-2}(E_{ij}E^{ij}-E^2) -6H^2 + (N+N^{-1}) \dot \phi^2 \right].
\end{equation}
Then we express all terms $h,R^{(3)}, E_{ij}, \dots$ in terms of the physical scalar field $\zeta$ and expand in terms of $\zeta$ up to the desired order. 

Expanding to second order in the small fluctuation $\zeta$ we have
\begin{equation}\label{s2}
 S_2= M^2_P\int dt d^3x\left[a^3 \epsilon \dot\zeta^2-a\epsilon(\partial \zeta)^2 \right],
\end{equation}
where we have restore the Planck mass $M_P$. Expanding to third order
\begin{dmath}\label{s3}
 S_3=  M^2_P \int dt d^3x\left[a^3 \epsilon^2 \zeta \dot\zeta^2+ a\epsilon^2 \zeta(\partial \zeta)^2- 2a\epsilon \dot \zeta (\partial\zeta)(\partial \chi)+
 \frac{a^3\epsilon}{2}\dot \eta \zeta^2\dot \zeta+ \frac{\epsilon}{2a}(\partial \zeta)(\partial \chi)\partial^2\chi+ \frac{\epsilon}{4a}(\partial^2 \zeta)
 (\partial \chi)^2+ f(\zeta) \frac{\delta L}{\delta \zeta}\bigg|_1 \right]
\end{dmath}
where
\begin{eqnarray}
  \frac{\delta L}{\delta \zeta}\bigg|_1&=& 2a\left(\frac{d\partial^2\chi}{dt}+H \partial^2\chi-\epsilon\partial^2\zeta \right),\\
 f(\zeta) &=& \frac{\eta}{4}\zeta + \mbox{ terms with derivatives on } \zeta
\end{eqnarray}
Here $\delta L/\delta \zeta|_1$ is the variation of the quadratic action with respect to $\zeta$. In the next 
sections we will use the second and third order action expressions to calculate the power spectrum and bispectrum of perturbations, respectively.
\section{Power spectrum and Bispectrum}\label{psb}
\paragraph{Power spectrum}
For a long time, observations of fluctuations in the CMB and LSS have been focused mainly on the Gaussian contribution as measured by the two-point correlations of the density fluctuations or by its Fourier transform \citep{xc, pxxii}
\begin{equation}
 \Braket{ \zeta(\vec{k}_1, t) \zeta(\vec{k}_2, t)  }= (2\pi)^3 \frac{2\pi}{k^3} P_{\zeta}(k)\delta^{(3)}(\vec{k}_1+\vec{k}_2),
\end{equation}
where $P_{\zeta}(k)$ is the power spectrum of curvature perturbations defined by
\begin{equation}\label{ps}
  P_{\zeta}(k) \equiv \frac{2k^3}{(2\pi)^2}|\zeta_k(\tau_e)|^2,
\end{equation}
where $\tau_e$ is the time at the end of inflation \citep{a2,a1}. It has been precisely the study of the power spectrum which has settled the numerical value of many of the 
cosmological parameters in the $\Lambda$CDM model \citep{et, wmapcpr,pxvi}.
\paragraph{Bispectrum}
The three-point correlation function correlates density or temperature fluctuations at three points in space. Its Fourier transform correlates fluctuations with three wave vectors $\vec{k}_i$ $(i=1,2,3)$ and is called the bispectrum $ B_{\zeta}$ given by the relation \citep{pxxii, pxxiv}
\begin{equation}
 \Braket{ \zeta(\vec{k}_1, t) \zeta(\vec{k}_2, t)  \zeta(\vec{k}_3, t) }= (2\pi)^3 B_{\zeta}(k_1,k_2,k_3) \delta^{(3)}(\vec{k}_1+\vec{k}_2+\vec{k}_3),
\end{equation}
and it vanishes if the curvature perturbation is Gaussian \citep{xc,a2}. Therefore, deviations from Gaussianity can be determined by measuring $B_{\zeta}$. A convenient way to do this is to parametrize the level of non-Gaussianity by using a Non-Linearity parameter $f_{NL}$ \citep{xc,pxxiv}. For instance the local model \citep{salopek,figueroa} is defined in terms of a local quadratic correction to the curvature perturbations given by \citep{komatsufnl}
\begin{equation}
  \Phi(\vec x)= \Phi(\vec x)_L + f_{NL} \left[ \Phi(\vec x)_L^2 - \Braket{\Phi(\vec x)_L }^2 \right] \, , 
\end{equation}
where $\Phi(\vec x)_L$ is the linear gaussian part of the perturbation and $f_{NL}$ characterizes the amount of non-Gaussianities. The non-linear coupling constant $f_{NL}$ is also known as the bispectrum amplitude \citep{pxxiv}. 

Note that the bispectrum is a function of the scales $\vec{k}_1,\vec{k}_2, \mbox{ and }\vec{k}_3$, but since one is fixed by the $\delta^{(3)}$ function it only depends on two of them. These vectors form a triangle in Fourier space \citep{et, hael, xc}. The dependence of $B_{\zeta}$ on the scales is divided in two parts. The first one is called the \textit{shape} of the bispectrum and corresponds to the dependence of $B_{\zeta}$ on the momenta ratio $k_2/k_1$ and $k_3/k_1$ having the total momentum $k_{tot} = k_1+k_2 +k_3$ fixed. The second one is called the \textit{running} of the bispectrum corresponding to the dependence of $B_{\zeta}$ on the overall momentum $k_{tot}$ with fixed momenta ratio $k_2/k_1$ and $k_3/k_1$. The most studied shapes are the local form or squeezed form \citep{pxxiv, aer,xc}, corresponding to triangles such that $k_1 \ll k_2=k_3$ and the equilateral form corresponding to $k_1 = k_2=k_3$. In this thesis we will study both shapes thoroughly. 

The Planck constraints on primordial non-Gaussianity on the bispectrum amplitude for the local shape is \citep{pxxiv}
\begin{equation}
 f_{NL}^{local} = 2.7 \pm 5.8,
\end{equation}
and on the equilateral shape
\begin{equation}
 f_{NL}^{equilateral}= -42 \pm 75.
\end{equation}

Following the procedure in Ref. \citep{aer}, where there is a field redefinition and some boundary terms are kept, the third order action Eq. \eqref{s3} can be simplified as
\begin{equation}
 S_3 \supset  M^2_P\int dt d^3x\left[-a^3 \epsilon \eta \zeta \dot\zeta^2-\frac{1}{2}a \epsilon\eta \zeta \partial^2 \zeta \right].
\end{equation}
In the  in-in formalism the above action can be interpreted in the interaction picture. The Hamiltonian after changing to conformal time gives 
\begin{equation}\label{Hamiltonian}
H_{int}(\tau)= M^2_P \int d^3x \, \epsilon \eta a \left[ \zeta \zeta'{}^2 + \frac{1}{2} \zeta^2 \partial^2 \zeta \right].
\end{equation}
The 3-point correlation function at a time $\tau_e$ after horizon crossing is then given by \citep{m,xc}
\begin{equation}
\Braket{\Omega| \zeta(\tau_e,\vec{k}_1) \zeta(\tau_e,\vec{k}_2)  \zeta(\tau_e,\vec{k}_3)|\Omega }= -i \int_{-\infty}^{\tau_e} \Braket{0|\left[ \zeta(\tau_e,\vec{k}_1) \zeta(\tau_e,\vec{k}_2)  \zeta(\tau_e,\vec{k}_3), H_{int}\right]|0 },
\end{equation}
where according to the standard procedure $H_{int}$ evolves the free theory vacuum state $\Ket{0}$ to the interaction vacuum $\Ket{\Omega}$ up to the time where the correlation function is to be evaluated. After substitution of Eq. \ref{Hamiltonian} the expression for the bispectrum $B_{\zeta}$ \citep{inin,xc} is
\begin{dmath}\label{b}
  B_{\zeta}(k_1,k_2,k_3)= 2 (2\pi)^3 M_{P}^2\Im\left[ \zeta(\tau_e,k_1) \zeta(\tau_e,k_2)\zeta(\tau_e,k_3)
	\int^{\tau_e}_{-\infty} d\tau \eta \epsilon a^2 \zeta^*(\tau,k_1)\times
	\left(2\zeta'{}^*(\tau,k_2)\zeta'{}^*(\tau,k_3)-k^2_1 \zeta^*(\tau,k_2)\zeta^*(\tau,k_3)
		\right)  \right] + \mbox{the other two permutations of } k_1, k_2,\mbox{ and } k_3,
\end{dmath}
where $\Im$ is the imaginary part and $\tau_e$ is some time sufficiently after Hubble crossing horizon \citep{a1,a2,a3} when the modes are frozen. To study the squeezed and equilateral limits of the bispectrum it is useful to use the definition function $F_{NL}$ \citep{hann} given by
\begin{equation}\label{FNL}
  F_{NL}(k_1,k_2,k_3;k_*)\equiv \frac{10}{3(2\pi)^7}\frac{(k_1 k_2 k_3)^3}{k_1^3+k_2^3+k_3^3}\frac{1}{P_{\zeta}^2(k_*)} B_{\zeta}(k_1,k_2,k_3),
\end{equation}
where $k_*$ denotes a pivot scale for the power spectrum $P_{\zeta}$ \citep{aer, pxxii}.
\chapter{Effect of features in the potential of the inflaton} 

\label{Chapter4} 

\lhead{Chapter 4. \emph{Features in the potential}} 
In this chapter we study the effects on the power spectrum and bispectrum of primordial curvature perturbations produced by a discontinuity in the $n-th$ order derivative in the inflaton potential and the suppression of the value of the inflaton field during inflation.
\section{Scale Dependent Features}\label{sdf}
In the last decades, the information from the observations of fluctuations in the CMB and the Large-Scale Structures have been extracted from the Gaussian contribution \citep{et, m, inf}. But any information contained in the departure from a perfect Gaussian field is not encoded in the power spectrum, thus it has to be extracted from measurements of higher-order correlation functions. From this measurements we can differentiate inflationary models that can lead to very similar predictions for the power spectrum of primordial perturbations. Thus non-Gaussianity is a sensitive probe of the interaction of the fields driving inflation and therefore contains important information about the fundamental physics during inflation. 

At the end of section \ref{NG} we enumerate various mechanism to produce a large amount of non-Gaussianities. In this thesis we study a model where the slow-roll conditions are violated by the introduction of a ``feature'' in the potential or more generally in the Lagrangian of the inflaton field. In this way we may study how the non-Gaussianities can discriminate between different inflationary models which can predict the same power spectrum.

There are several other reasons that motivate the introduction of features, which have a long history \citep{starobinsky} in the study of primordial fluctuations. For instance, it has been shown  that these models can provide better fits to the apparent low multipole glitch at $l \sim 20-40$ in the study of the angular spectrum of the CMB radiation \citep{constraints1,constraints2}. Moreover, in the Planck 2013 results a new feature was found in the CMB spectrum at $l \sim 1800$ \citep{pxxii}. According to the Planck Collaboration, the features in the temperature power spectrum, particularly the broad dip at $l \sim 1800$, cannot be explained by the standard $\Lambda$CDM model. These departures from the best fit $\Lambda$CDM model spectrum could also be due to unknown systematic effects into the final power spectra \citep{pxxii}. Only when the whole data from the mission can be analyzed this possibility will be investigated. To study the deviations of the primordial power spectrum from a smooth function various models have been proposed (See Ref. \citep{encyclopaedia} for a thorough list and explanation of many different inflationary models). Some models can add a global oscillation, a localized oscillation, or a cutoff to the large scale spectrum, namely the Wiggles, the step-inflation, and the Cutoff models, respectively \citep{pxxii}. For instance, in the case of step-inflation it was shown in Ref. \citep{Adams2} that a large number of ``sudden downward'' steps in the inflaton potential can be obtained from a class of models derived from supergravity theories. In these models the symmetry breaking phase transition of a field coupled to the inflaton ``naturally'' gives rise to the steps, since the mass changes suddenly when each transition occurs.


In this thesis we are interested in the effects that an $n$-th order discontinuity can produce (on the power spectrum and bispectrum) \citep{Adams, a1,a2,starobinsky, pxxii} and on the consequences of having a term which suppresses the value of the scalar field during inflation. This corresponds to a non-localized feature in which the effect of the suppression is higher at the end of inflation.

\section{The Model}\label{model}
We consider a single scalar field $\phi$ coupled to a potential given by
\begin{equation}\label{pot}
V(\phi)= \left\{
\begin{array}{lr}
V_{0b} + \frac{1}{2}m^2 \phi^2 & \phi \ge \phi_0, \\
V_{0a} + \frac{1}{2}m^2 \phi^2 + \lambda \Delta \phi & \phi < \phi_0,\\
\Delta \phi=\phi^n e^{-\left(\frac{\phi_0-\phi}{\phi_0}\right)^2},
\end{array}
\right.
\end{equation}
where $m$ is the mass inflaton. $V_{0b}$ and $V_{0a}$ are the vacuum energies before and 
after the transition, respectively. The transition time is $\tau_0$ and we define $\phi(\tau_0) $ as $ \phi_0$. In Eq. \eqref{pot} $n$ is a positive real number and $\lambda$ is a model parameter whose units in terms of Planck mass are $M_p^{4-n}$. We will see below that this feature in the potential induces an oscillatory ringing in the power spectrum of curvature fluctuations \cite{Adams}. The case for $n=2$ corresponds to a sharp feature in the inflaton mass \citep{aer} as we will see below.

The equations of motion \eqref{bea} and \eqref{bephi} in terms of conformal time $\tau$ \eqref{conformaltime} take the form
\begin{equation}\label{sfe}
  H^2  \equiv \left(\frac{a'}{a^2}\right)^2 = \frac{1}{3 M^2_{Pl}} \left( \frac{1}{2} \frac{{\phi'}^{2}}{a^2} + V(\phi)\right) \, , 
\end{equation}
\begin{equation}\label{be}
  \phi'' + 2\frac{a'}{a} \phi' + a^2 \partial_{\phi}V = 0,
\end{equation}
where primes indicate derivatives with respect to conformal time. Imposing continuity conditions on the potential at $\phi_0$ implies that 
$V_{0a}= V_{0b}-\lambda \phi_0^n$. We assume that this change in the potential energy is small \cite{Adams,a2}. In the next section we will find an 
analytic solution for $a$ and $\phi$ in terms of $\tau$.
\subsection{Analytic solutions to the equations of motion}\label{asem}
In order to obtain analytic solutions for the equations of motion we assume that $V(\phi)$ is dominated by the vacuum 
energy $V_{0}$. In this case $H$ is constant and the scale factor may be approximated by that of a pure de Sitter space
\begin{equation}
 a(\tau)= \frac{-1}{H \tau}.
\end{equation}
We now proceed to obtain the analytic solution of the field. Before the transition Eq. \eqref{be} is written as
\begin{equation}\label{beb}
  \phi'' + 2\frac{a'}{a} \phi' + a^2 m^2 \phi = 0,
\end{equation}
which has the solution
\begin{equation}\label{bsb}
  \phi_b(\tau)= \phi_b^+ a(\tau)^{\lambda^+} + \phi_b^- a(\tau)^{\lambda^-},
\end{equation}
where the slow-roll regime corresponds to $\phi_b^-=0$ and $\lambda^{\pm}$ is defined as
\begin{equation}\label{lambdaplus}
  \lambda^{\pm}=\frac{3}{2}\left( -1\pm \sqrt{1- \left(\frac{2 m}{3 H}\right)^2}\right) \, .
\end{equation}

After the transition Eq. \eqref{be} is written as
\begin{equation}\label{bephia}
\phi'' + 2\frac{a'}{a} \phi' + a^2 \left( m^2 \phi + \lambda \partial_{\phi}\Delta \phi \right) = 0,
\end{equation}
where
\begin{equation}
  \partial_{\phi} \Delta \phi = n \phi ^{n-1} e^{-\left(\frac{\phi _0-\phi}{\phi _0}\right)^2} 
   +  2\phi^n \left(\frac{\phi _0-\phi }{\phi _0{}^2} \right)  e^{-\left(\frac{\phi _0-\phi }{\phi _0}\right)^2}.
\end{equation}
Eq. \eqref{bephia} does not have an exact analytic solution. To obtain an analytic approximation we expand $\phi$ to second order around the 
transition
\begin{equation}
  \phi(\tau)= \phi_0 + \phi'(\tau_0)(\tau-\tau_0)+\frac{1}{2!} \phi''(\tau_0) (\tau-\tau_0)^2+ \mathcal{O}(3),
\end{equation}
and then insert this expansion into $\partial_{\phi}\Delta \phi$ to obtain
\begin{dmath}
  \phi'' + 2\frac{a'}{a} \phi' + a^2 \left[ m^2 \phi + \lambda  \phi _0{}^{n-2} \left( n \phi _0+ \\
  (n+1)(n-2) \phi'(\tau_0) (\tau -\tau_0) + \frac{1}{2}(n+1) (n-2) \phi''(\tau_0) (\tau -\tau_0)^2 \right) \right]= 0,
\end{dmath}
where we have neglected the term $n(n-4)\phi'(\tau_0)^2$ with respect to $(n-2)\phi(\tau_0)\phi''(\tau_0)$ since
\begin{equation}
 \frac{n(n-4)\phi'(\tau_0)^2}{(n-2)\phi(\tau_0)\phi''(\tau_0)} \approx \frac{n(n-4)}{3(n-2)}\frac{m^2}{H^2} \ll 1 , \hspace{10 mm} n \ne 2
\end{equation}
in the last step we use the fact that $m\ll H$ otherwise the scalar field oscillates with a frequency proportional to its mass\footnote{This can be seen in Eqs. \eqref{bsb} and \eqref{lambdaplus}. See also Ref. \citep{mc} exercise 6.7.} then it would not correspond to a slow-roll regime.
The solution is thus given by
\begin{equation}\label{bsa}
  \phi_a(\tau)= \phi_{a}^{{}_{(0)}} +\phi_{a}^{{}_{(1)}} \, (\tau-\tau_0)+\phi_{a}^{{}_{(2)}} \, (\tau-\tau_0)^2 + 
  \phi_a^+ a(\tau)^{\lambda^+} + \phi_a^- a(\tau)^{\lambda^-},
\end{equation}
where
\begin{align}
  \phi_{a}^{{}_{(0)}} &= \begin{aligned}[t]
      &\frac{-\lambda  \phi _{0}^{n-2}}{m^2 \left( m^2-2H^2 \right)} \left[ n (m^2-2 H^2) \phi _{0} \right.\\
      &\left. + 2(n+1)(n-2)H^2\phi _{0}'\tau _0-(n+1)(n-2)H^2 \phi _{0}''\tau _0{}^2 \right] \, , 
       \end{aligned}\\
  \phi_{a}^{{}_{(1)}} &= \begin{aligned}[t]
      &\frac {-\lambda  \phi _{0} {}^{n - 2}} {\left (m^2-2 H^2 \right)} (n + 1) (n - 2) \phi _{0}' \, , 
       \end{aligned}\\
  \phi_{a}^{{}_{(2)}} &= \begin{aligned}[t]
                 & \frac{-\lambda \phi_{0} {}^{n - 2}} {\left (m^2 -2 H^2 \right)} (n + 1) (n - 2)\frac {1} {2} \phi _{0}'' \, .
               \end{aligned}
\end{align}
The quantities $\phi_{a}^{{}_{(i)}} (i=1,2,3.)$ are in terms of $\phi_{0}$, $\phi_{0}'\equiv \phi_b'(\tau_0)$, and 
$\phi_{0}''\equiv \phi_b''(\tau_0)$ for which we already have a solution. The constants of integration $\phi_{a}^{\pm}$ are determined by 
imposing the continuity conditions on $\phi$ and $\phi'$ across $\tau_0$
\begin{dmath}
  \phi_{a}^{\pm} = \frac{\pm a(\tau_0){}^{-\lambda^{\pm} }}{\left(\lambda^--\lambda^+\right) } \left\{ \lambda^\mp \phi _{0}+ \phi_{0}'\tau_0+
	\frac{\lambda  \phi_{0}{}^{n-2}}{m^2} \left[n \lambda^\mp \phi_{0}+\frac{(n+1)(n-2)}{\left(m^2-2 H^2\right)} 
	\left((m^2+2 H^2 \lambda^\mp) \phi_{0}'\tau_0 -\lambda^\mp H^2 \phi_{0}''\tau_0^2\right)\right]\right\}.
\end{dmath}
Thus our complete analytic approximate solution is
\begin{equation}\label{as}
  \phi(\tau){\mbox{\tiny{analytic}}}= \phi_b(\tau) + \phi_a(\tau) \theta(\phi_0-\phi),
\end{equation}
where $\phi_{b}$ and  $\phi_{a}$ are defined in Eqs. \eqref{bsb} and \eqref{bsa}, and $\theta(\phi)$ denotes the Heaviside step function.\\

We choose the model parameters as \cite{aer}
\begin{equation}\label{parameters}
  m=6\times10^{-9}M_{Pl}, \,\,\,\, H=2\times 10^{-7}M_{Pl},\,\,\,\, \phi_b^+=10M_{Pl}.
\end{equation}
From now on we set $M_{Pl}=1$, thus all quantities will be given in Planck units; for instance the transition time is $\tau_0=-M_{Pl}^{-1}=-1$.
\subsection{Background solution and slow-roll parameters}\label{bsandsrp}
In this section we show that our analytic solution is a good approximation for the field inflaton under the de Sitter approximation. To do this we compare the numerical solution for $a$ and $\phi$ [Eqs. \eqref{sfe} and \eqref{be}] with the analytic solution Eq. \eqref{as} in the de Sitter approximation. In the following plots all quantities are in Planck units. It is important to remember that the slow-roll parameters are dimensionless.

In Fig. \ref{phiplot} we plot the numerical and analytical evolution of $\phi$ as a function of conformal time. The evolution corresponds to a slow-roll regime. There is an agreement between the numerical and analytic solutions for the scalar field. Fig. \ref{phidiff33} shows that the percent error between these two solutions is very small: after $60e$-folds this error is only of $0.4\%$. We also found agreement between the numerical and analytic solution for other values of $n$ and $\lambda$. We can conclude then that Eq. \eqref{as} is a good approximation to the background solution, thus in the following sections we use the analytic solution Eq. \eqref{as} to obtain the analytic results for the plots.
\begin{figure}
\centering
  \includegraphics[scale=0.8]{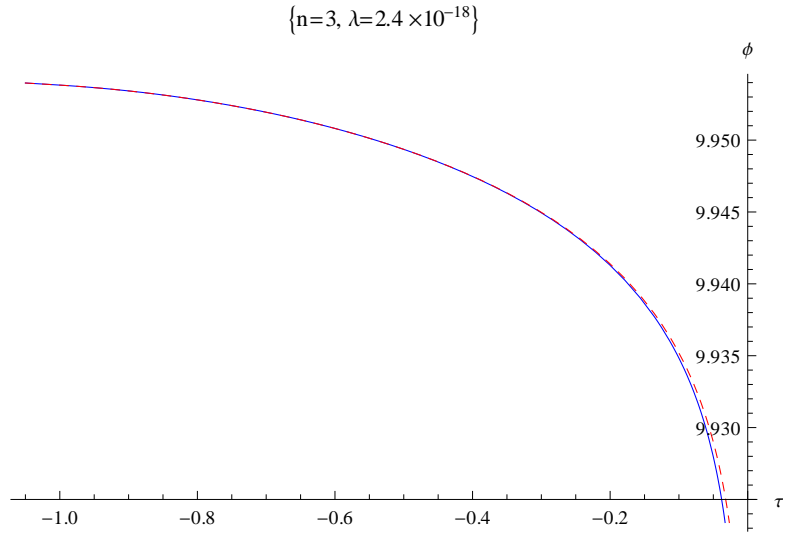}
  \caption{Evolution of $\phi$ for $n=3$ and $\lambda=2.4\times10^{-18}$ in terms of conformal time. The blue and dashed-red lines represent the numerical and analytic results, respectively. The agreement between these two solutions is observed. All quantities are in Planck units.}
  \label{phiplot}
\end{figure}

\begin{figure}
 \centering
 \includegraphics[scale=0.6]{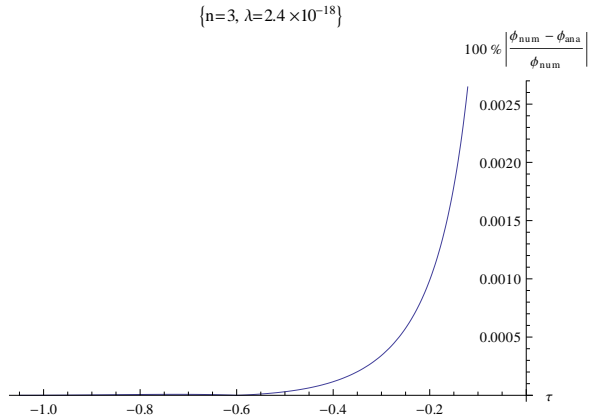}
 \caption{Plots of percent error between the numerical and analytic solutions for the field inflaton $\phi$ for $n=3$ and $\lambda=2.4\times10^{-18}$. Although this percent error is increasing, after $60e$-folds the error is only of $0.4\%$.}
 \label{phidiff33}
\end{figure}
The slow-roll parameters Eqs. \eqref{sreps} and \eqref{sreta} in terms of conformal time take the form
\begin{equation}
  \epsilon= -\frac{{H'}}{a H^2},\,\,\,\,\, \eta= \frac{\epsilon'}{a H\epsilon}.
\end{equation}
We approximate the analytic solutions for the slow-roll parameters after the transition as
\begin{equation}\label{epsilon}
  \epsilon_a(\tau)\approx \frac{1}{2}\left(\lambda^+ \phi_a^+ a(\tau)^{\lambda^+}+ \lambda^- \phi_a^- a(\tau)^{\lambda^-} \right)^2\, ,
\end{equation}
\begin{equation}\label{eta}
  \eta_a(\tau)\approx 2 \lambda^- \frac{ (\frac{\lambda^+}{\lambda^-})^2 \phi_a^+ a(\tau)^{\lambda^+}+ \phi_a^- a(\tau)^{\lambda^-} }{
		(\frac{\lambda^+}{\lambda^-}) \phi_a^+ a(\tau)^{\lambda^+} + \phi_a^- a(\tau)^{\lambda^-} }.
\end{equation}
For comparison, before the transition $\eta= 2 \lambda^+$. Due to the step in the potential the inflaton suffers a strong instant acceleration \citep{Adams}. But still for realistic models, it is required that the step be approximately less than $1\%$ of the overall height of the potential \cite{Adams,a1}. Thus, although $\epsilon$ grows after the transition, we still have that at all times $\dot{\phi}^2 \ll V(\phi)$ and $\epsilon \ll 1$. Furthermore, the step causes a large change in $V''$ and from Eq. \eqref{srpeta} we have that $\eta \propto V''(\phi)$ then $\eta$ will suffer a dramatic change \cite{Adams,inf}. This is the primary source of large non-Gaussianities in the discontinuity of the 2-nd order derivative of the potential as can be seen in Eq. \eqref{b}. Figs. \ref{epsplot33} and \ref{etaplot33} shows the evolution of $\epsilon$ and $\eta$ for $n=3$ and $\lambda=2.4 \times10^{-18}$ over the transition. The analytic Eqs. \eqref{epsilon} and \eqref{eta} are good approximations for the slow-roll parameters. Similar results are obtained for other values of $n$ and $\lambda$ in which there is also an agreement between the numerical and analytic solutions. 
\begin{figure}
 \centering
 \includegraphics[scale=0.6]{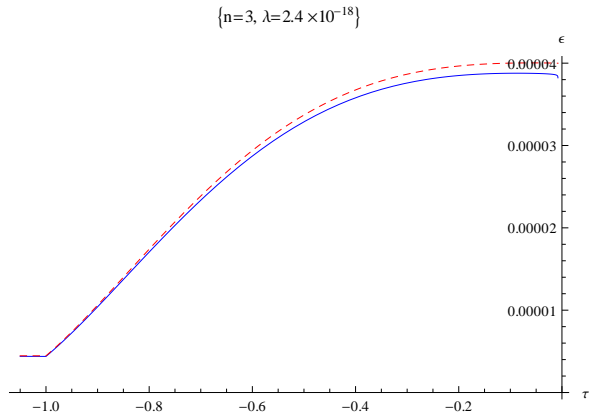}
 \caption{Plots of $\epsilon$ in terms of conformal time showing the evolution over the transition for $n=3$ and $\lambda=2.4 \times10^{-18}$. The blue and dashed-red lines represent the numerical and analytic results, respectively.}
 \label{epsplot33}
\end{figure}

\begin{figure}
 \centering
 \includegraphics[scale=0.6]{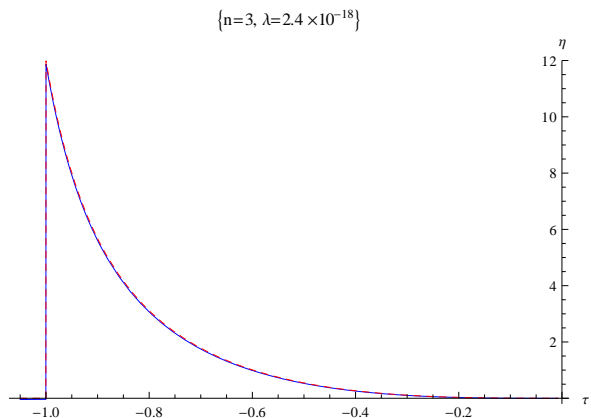}
 \caption{Plots of the slow-roll parameter $\eta$ in terms of conformal time for $n=3$ and $\lambda=2.4 \times10^{-18}$. This violation of $|\eta| \ll 1$ is the primary source of the large non-Gaussianities in the discontinuity of the 2-nd order derivative of the potential, as the leading term in the $F_{NL}$ function Eq. \eqref{FNL} is of order $\eta$. The blue and dashed-red lines represent the numerical and analytic results, respectively.}
 \label{etaplot33}
\end{figure}
\section{Perturbations}
From the action Eq. \eqref{s2} we obtain the linear equation of motion for the curvature perturbation $\zeta$ as
\begin{equation}
 \frac{\partial}{\partial t}\left(a^3\epsilon \frac{\partial \zeta}{\partial t}\right)- a\epsilon\delta^{ij} \frac{\partial^2 \zeta}{\partial x^i\partial x^j}=0 \, . 
\end{equation}
In Fourier space and in terms of conformal time the previous equation is written in the form
\begin{equation}\label{cpe}
  \zeta''_k + 2 \frac{z'}{z} \zeta'_k + k^2 \zeta_k = 0,
\end{equation}
where $z=a\sqrt{2 \epsilon}$ and $k$ is the comoving wavenumber. Towards a better understanding on the evolution of the curvature perturbation it is convenient to define a gauge invariant potential \citep{Adams,aer}
\begin{equation}
 u_k(\tau)= z(\tau) \zeta(\tau,k)
\end{equation}
Then Eq. \eqref{cpe} takes the form\footnote{The definition of $u_k$ in Ref. \citep{Adams} differs from our definition since in our case $z$ is a positive function.}
\begin{equation}
 u_k''+ \left( k^2 - \frac{z''}{z} \right) u_k =0,
\end{equation}
for which the dependence on time has not been written explicitly. Let us now consider the limits $k^2 \gg z''/z$ and $k^2 \ll z''/z$. In the first limit
\begin{equation}
 u_k \propto e^{i k\tau},
\end{equation}
which is the free field solution. In the second limit, $k^2 \ll z''/z$, we have
\begin{equation}
 u_k \propto z,
\end{equation}
which means that the curvature perturbation $\zeta$ is constant
\begin{equation}
  \zeta = \frac{u_k}{z} \propto \mbox{constant}.
\end{equation}
In the slow-roll regime the term $z''/z$ can be written approximately as $z''/z = 2 a^2 H^2$ . The physical wavelength of the perturbation is proportional to $a/k$ and the physical horizon during inflation is given by $H^{-1}$. We thus have that in the first limit the modes are well within the horizon
\begin{equation}
 k \gg a H \hspace{5 mm} \mbox{ modes inside the horizon.}
\end{equation}
While in the second limit
\begin{equation}
 k \ll a H \hspace{5 mm} \mbox{ modes outside the horizon.}
\end{equation}
The transition time $\tau_0$ sets the scale of transition at $k_0=-1/\tau_0=1$.
At the time of transition there is a discontinuity in $\phi''$ which implies that $z''$ contains a Dirac delta function \cite{pp}. 
To evaluate this discontinuity we calculate the contribution of the delta function in $z''$ as 
\begin{equation}\label{d0}
  D_0 \equiv \lim_{\delta \to 0} \int_{\tau_0-\delta}^{\tau_0+\delta}\frac{z''}{z}d\tau = \frac{1}{\phi_0'} \left[ \phi''_{a0}-\phi''_{b0} \right]
  =-n \lambda a(\tau_0)^2 \frac{\phi_0^{n-1}}{\phi_0'}.
\end{equation}

Our choice of vacuum implies that the initial conditions for the mode function and its derivative are given by the Bunch-Davies vacuum at early 
times $\tau \rightarrow \infty$ \cite{m, aer}
\begin{equation}
  \zeta(\tau,k)=\frac{v(\tau,k)}{z(\tau)} \,\,\, \mbox{and} \,\,\, \zeta'(\tau,k)= \partial_{\tau} \left(\frac{v(\tau,k)}{z(\tau)} \right),
	\end{equation}
where
\begin{equation}
  v(\tau,k)=\frac{e^{-i k\tau}}{\sqrt{2k}}\left(1-\frac{i}{k\tau}\right).
\end{equation}

An analytic solution to the mode function is given by \cite{aer}
\begin{equation}\label{r}
  \zeta(\tau,k)=\frac{1}{M_P a(\tau)}\left\{
		\begin{array}{clr}
			\frac{v(\tau,k)}{\sqrt{2\epsilon(\tau)}} & k \le k_0 \mbox{ and } \tau \le \tau_k \\
			\frac{v(\tau,k)}{\sqrt{2\epsilon(\tau_k)}} & k \le k_0 \mbox{ and } \tau > \tau_k \\
			\frac{v(\tau,k)}{\sqrt{2\epsilon(\tau)}} & k > k_0 \mbox{ and } \tau \le \tau_0 \\
			\frac{\alpha(k)v(\tau,k)+\beta(k)v^*(\tau,k)}{\sqrt{2\epsilon(\tau)}} & k > k_0 \mbox{ and } \tau_0 <\tau \le \tau_k \\
			\frac{\alpha(k)v(\tau,k)+\beta(k)v^*(\tau,k)}{\sqrt{2\epsilon(\tau_k)}} & k > k_0 \mbox{ and } \tau > \tau_k ,
\end{array}
\right.
\end{equation}
where we have restored the Planck mass and
\begin{equation}
  \alpha(k)=1+iD_0 |v(\tau_0,k)|^2 \: \mbox{ and } \: \beta(k)= -i D_0 v(\tau_0,k)^2,
	\end{equation}
and $\tau_k=-1/k$ is the horizon crossing time for that mode. We will see in Sec. \ref{rd} that this solution is a good approximation for the power 
spectrum.
\section{Results and Discussions}\label{rd}
\subsection{Background solution and slow-roll parameters}
The numerical results are presented in Figs. \ref{Phin123}-\ref{Etan123}. The time evolution of the inflaton is plotted in Fig.\ref{Phin123} for different types of features of the potential. These effects on the slow-roll parameters can also be seen in the Fig.(\ref{Epsn123}-\ref{Etan123}). When $\lambda$ is kept constant, larger values of $n$  tend to produce larger increments of both $\epsilon$ and $\eta$, while when  $n$ is kept constant, larger values of $\lambda$ tend to produce larger increments of both $\epsilon$ and $\eta$. 
\begin{figure}
 \begin{minipage}{.45\textwidth}
  \includegraphics[scale=0.55]{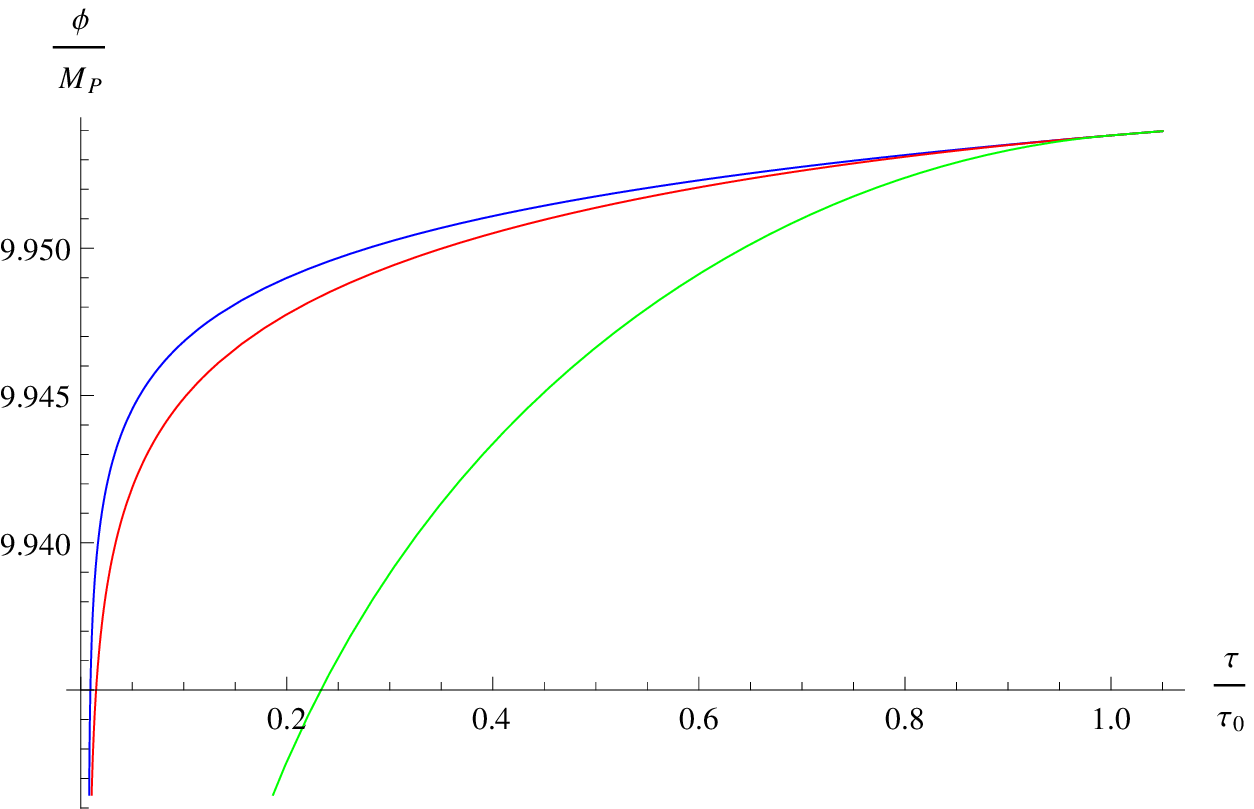}
  \end{minipage}
 \begin{minipage}{.45\textwidth}
  \includegraphics[scale=0.55]{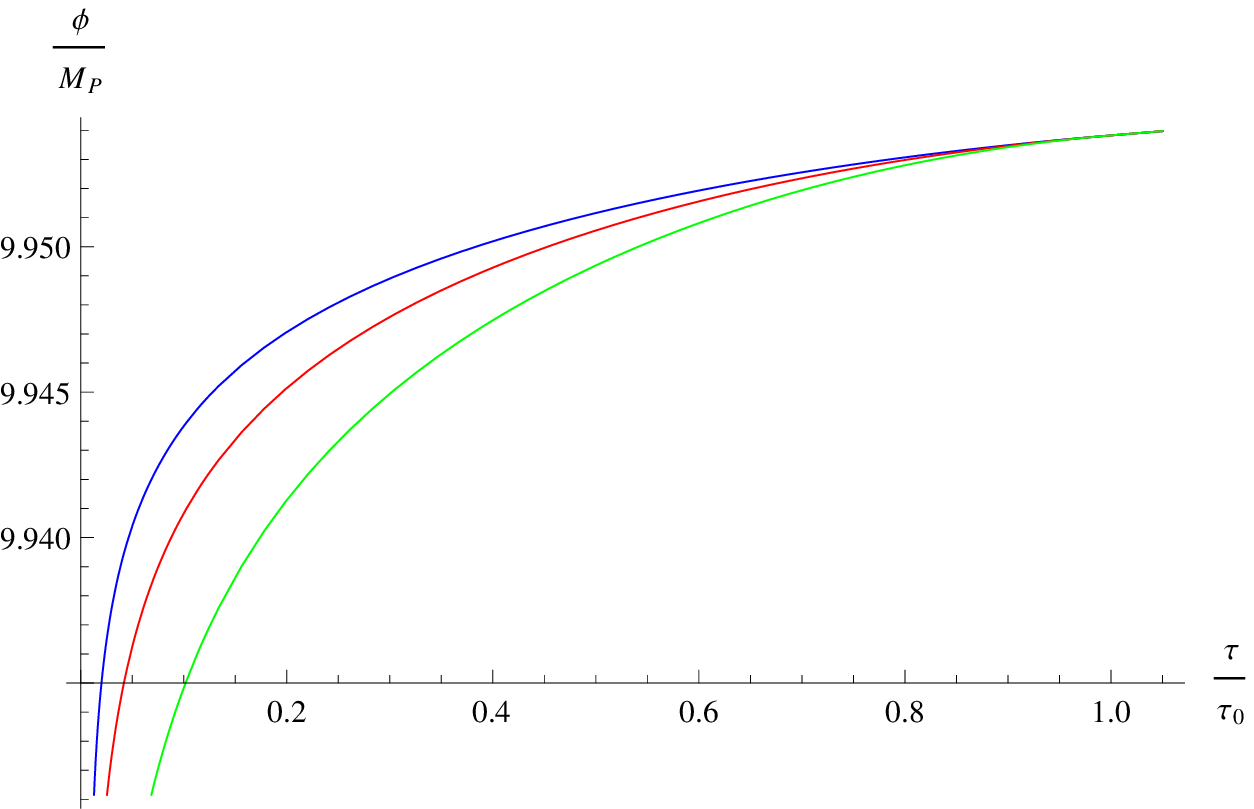}
 \end{minipage}
 \caption{On the left the evolution of $\phi$ is plotted  for  $\lambda=3.8\times10^{-19}$ and $n=2/3$(blue lines), $n=3$ (red lines), and $n=4$ (green lines). On the right the evolution of $\phi$ is plotted for $n=3$ and $\lambda=6.0\times10^{-19}$ (blue lines), $\lambda=1.2\times10^{-18}$ (red lines), and $\lambda=2.4\times10^{-18}$ (green lines).}
\label{Phin123}
\end{figure}

\begin{figure}
 \begin{minipage}{.45\textwidth}
  \includegraphics[scale=0.55]{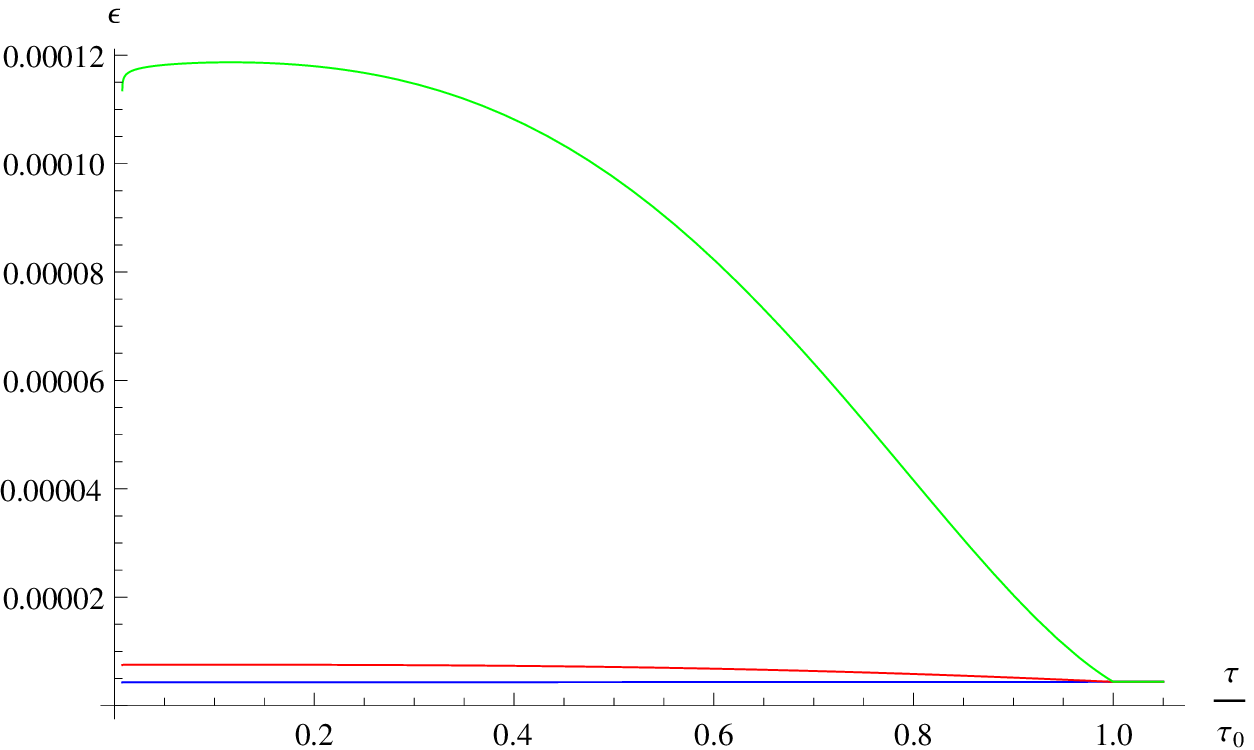}
  \end{minipage}
 \begin{minipage}{.45\textwidth}
  \includegraphics[scale=0.55]{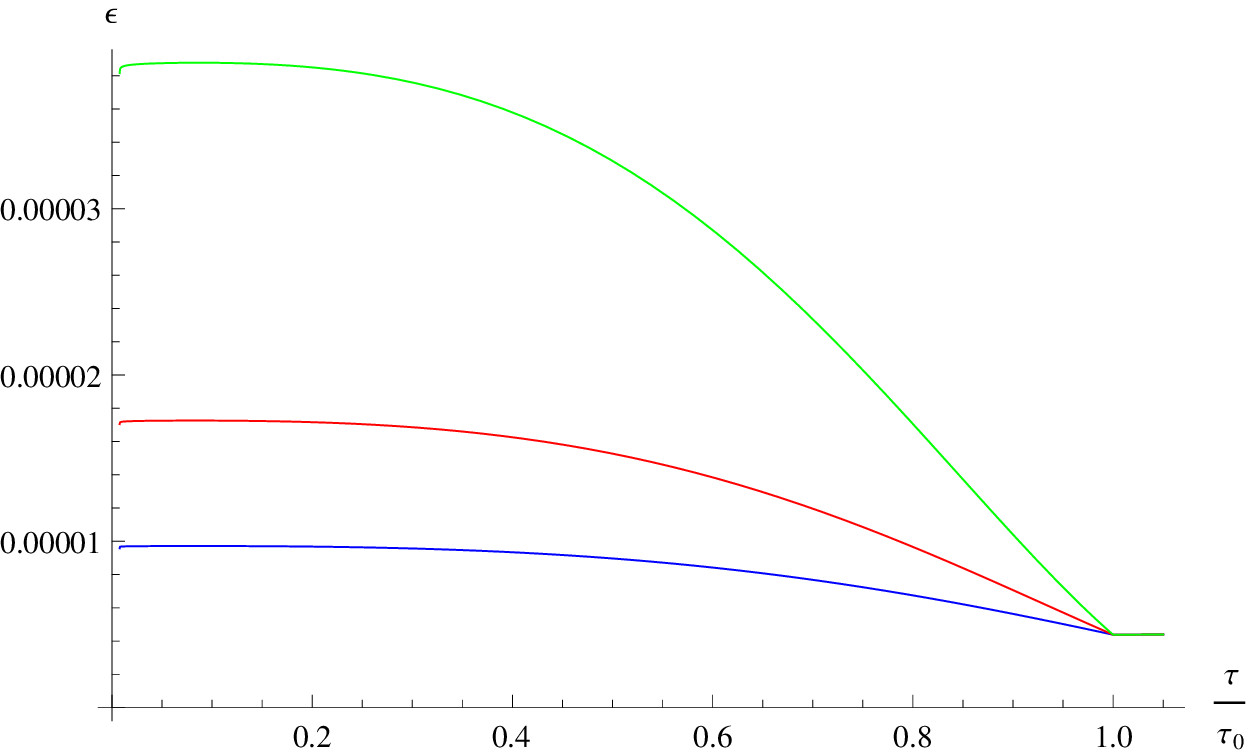}
 \end{minipage}
 \caption{On the left the slow-roll parameters $\epsilon$ is plotted for $\lambda=3.8\times10^{-19}$ and $n=2/3$ (blue lines), $n=3$ (red lines), and $n=4$ (green lines). On the right the slow-roll parameter $\epsilon$ is plotted for $n=3$ and $\lambda=6.0\times10^{-19}$ (blue lines), $\lambda=1.2\times10^{-18}$ (red lines), and $\lambda=2.4\times10^{-18}$ (green lines).}
\label{Epsn123}
\end{figure}

\begin{figure}
 \begin{minipage}{.45\textwidth}
  \includegraphics[scale=0.5]{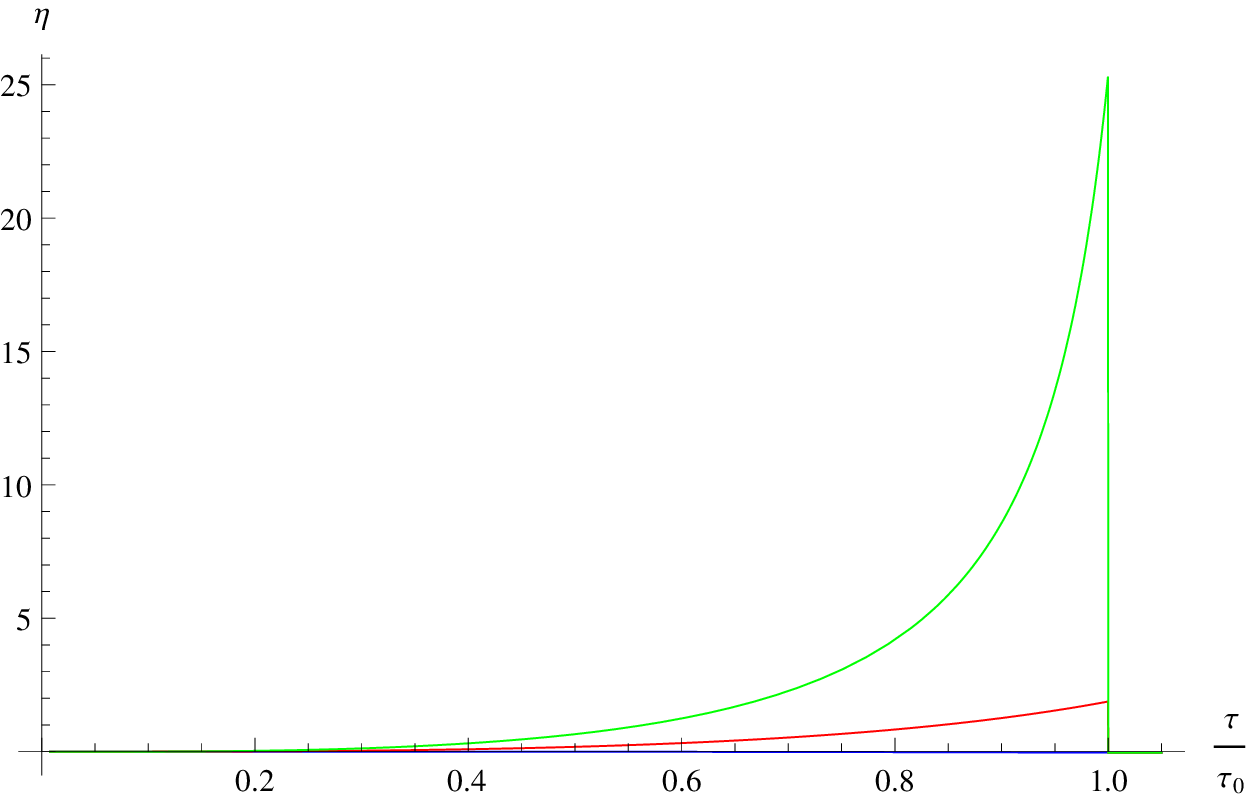}
  \end{minipage}
 \begin{minipage}{.45\textwidth}
  \includegraphics[scale=0.5]{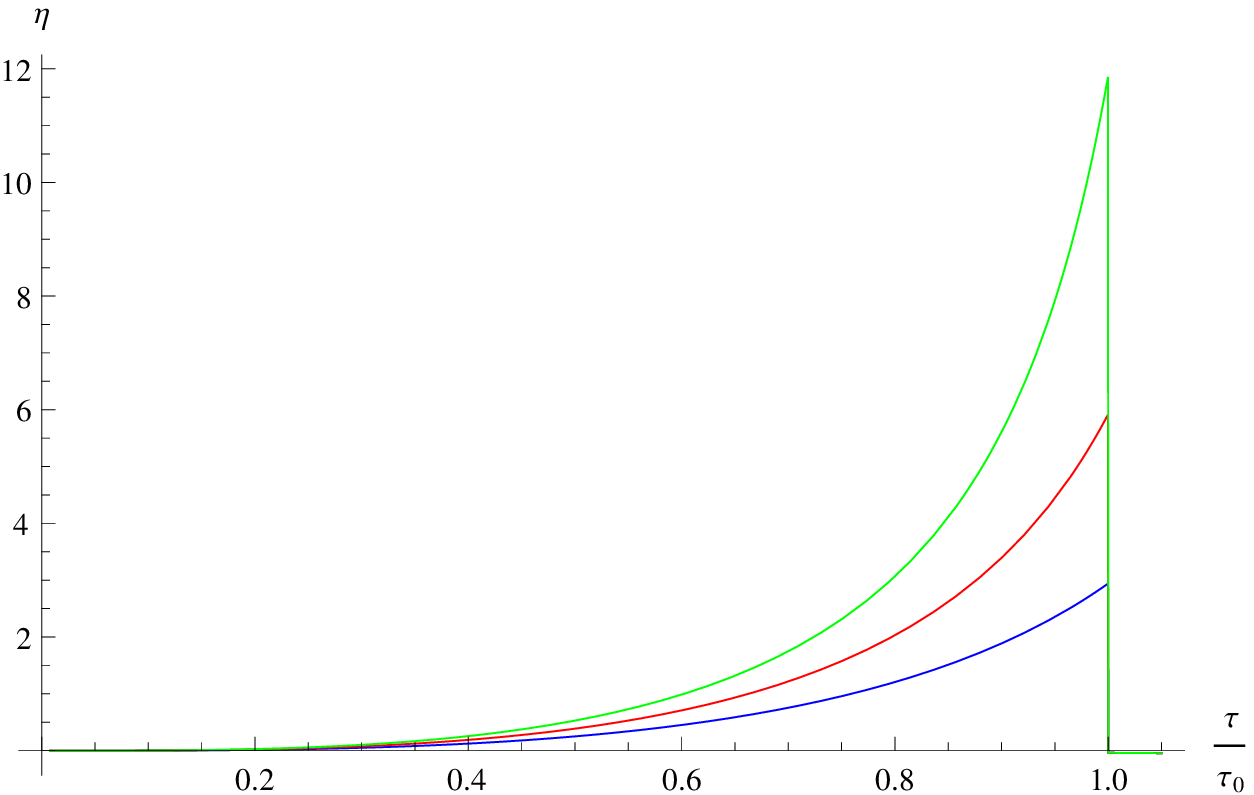}
 \end{minipage}
 \caption{On the left the slow-roll parameter $\eta$ is plotted for $\lambda=3.8\times10^{-19}$ and $n=2/3$ (blue lines), $n=3$ (red lines), and $n=4$ (green lines). On the right the slow-roll parameter $\eta$ is plotted for $n=3$ and $\lambda=6.0\times10^{-19}$ (blue lines), $\lambda=1.2\times10^{-18}$ (red lines), and $\lambda=2.4\times10^{-18}$ (green lines). This violation of the slow-roll condition is the primary source of the large non-Gaussianities in our model since the leading term in the $F_{NL}$ function is of order $\eta$.}
\label{Etan123}
\end{figure}
\subsection{Power spectrum}
In section \ref{psb} we saw that the power spectrum of curvature perturbations was defined as Eq. \eqref{ps}
\begin{equation}\label{ps2}
  P_{\zeta}(k) \equiv \frac{2k^3}{(2\pi)^2}|\zeta_k(\tau_e)|^2.
\end{equation}
For small scales, $k\gg k_0$, there is a simple expression for the power spectrum \cite{pp}
\begin{equation}\label{psa}
  P_{\zeta}(k)^{1/2} = \frac{a(\tau_k) H(\tau_k)^2}{2\pi |\phi'(\tau_k)|} \left[1 + \frac{D_0}{k} \sin(2 k \tau_0) + 
	\frac{D_0^2}{2 k^2}\left[ 1+ \cos(2k\tau_0) \right] \right]^{1/2}.
\end{equation}
In this section we use this expression along with Eq. \eqref{ps2} to compare the power spectrum with the numerical result. 
In Fig. \ref{pthreeplotall33} we compare two different analytic expressions for the power spectrum. In the first one we insert the analytic approximation \eqref{r} into the definition of the power spectrum $P_{\zeta}$ Eq. \eqref{ps}. For the second one we use the analytic expression \eqref{psa}. To plot these functions we use the analytic solution of the scalar field Eq. \eqref{as}. The analytic expressions are compared with the numerical result that we obtained from our code. The parameters used in the plots are $n=3$ and $\lambda=2.4\times10^{-18}$.
\begin{figure}
  \centering
  \includegraphics[scale=0.8]{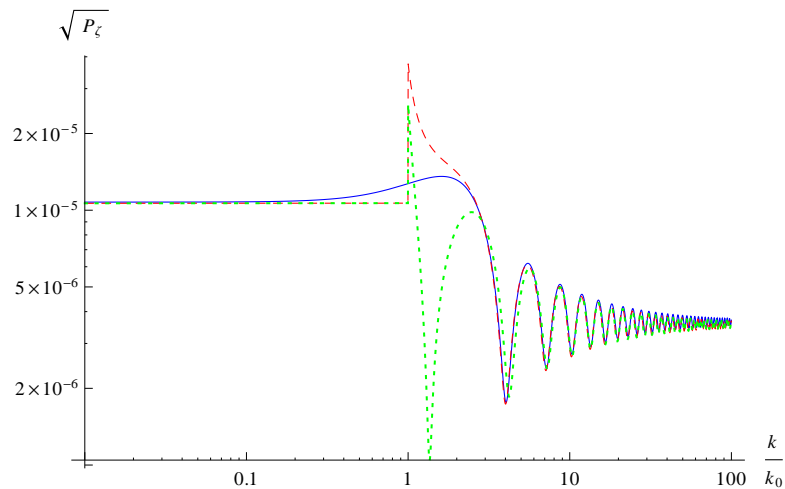}
  \caption{Plots considering two different analytic expressions for the power spectrum: the first one (dashed-red) is given by inserting the analytic approximation \eqref{r} into the definition of $P_{\zeta}$ Eq. \eqref{ps2} and the second one (dotted-green) is given by Eq. \eqref{psa}. These analytic expressions are compared with the numerical result (blue lines). The parameters used in the three plots are $n=3$ and $\lambda=2.4\times10^{-18}$.}
  \label{pthreeplotall33}
\end{figure}
As it can be seen in Fig.(\ref{Refn123ak0}) small scales modes which are sub-horizon at the time of when the feature occurs are affected by the feature, while modes that had already left the horizon are unaffected. In fig.(\ref{LogPn123lines}) the power spectrum of primordial curvature perturbations $P_{\zeta}$ is plotted for different types of features.
\begin{figure}
 \begin{minipage}{.45\textwidth}
  \includegraphics[scale=0.55]{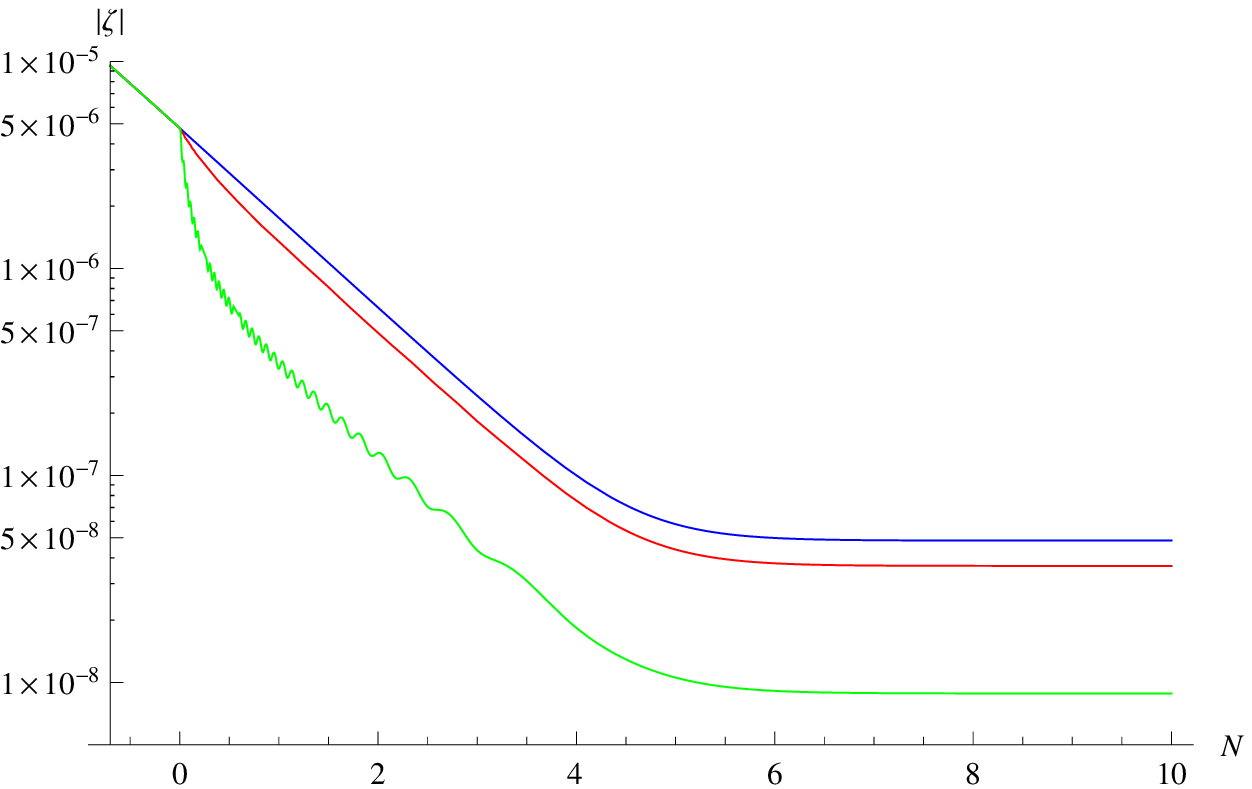}
  \end{minipage}
 \begin{minipage}{.45\textwidth}
  \includegraphics[scale=0.55]{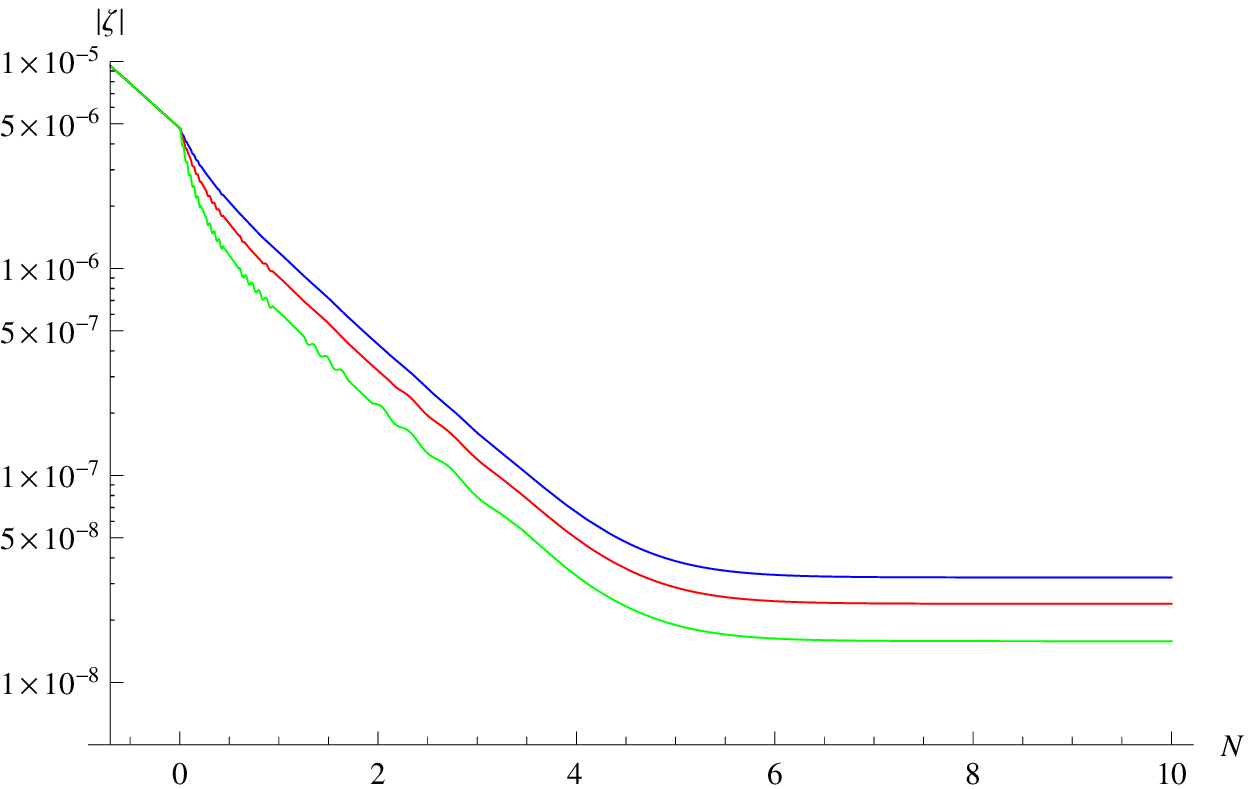}
 \end{minipage}
 \caption{The numerically computed $|\zeta_k|$ is plotted as a function of the number of $e$-folds $N$ after the time of the feature .  The left plots are for $\lambda=3.8\times10^{-19}$ and $n=2/3$ (blue lines), $n=3$ (red lines), and $n=4$ (green lines). The right plots are for $n=3$ and $\lambda=6.0\times10^{-19}$ (blue lines), $\lambda=1.2\times10^{-18}$ (red lines), and $\lambda=2.4\times10^{-18}$ (green lines). All the plots are for the short scale mode with $k=100k_0$ which is sub-horizon when the feature occurs.}
\label{Refn123ak0}
\end{figure}

\begin{figure}
 \begin{minipage}{.45\textwidth}
  \includegraphics[scale=0.55]{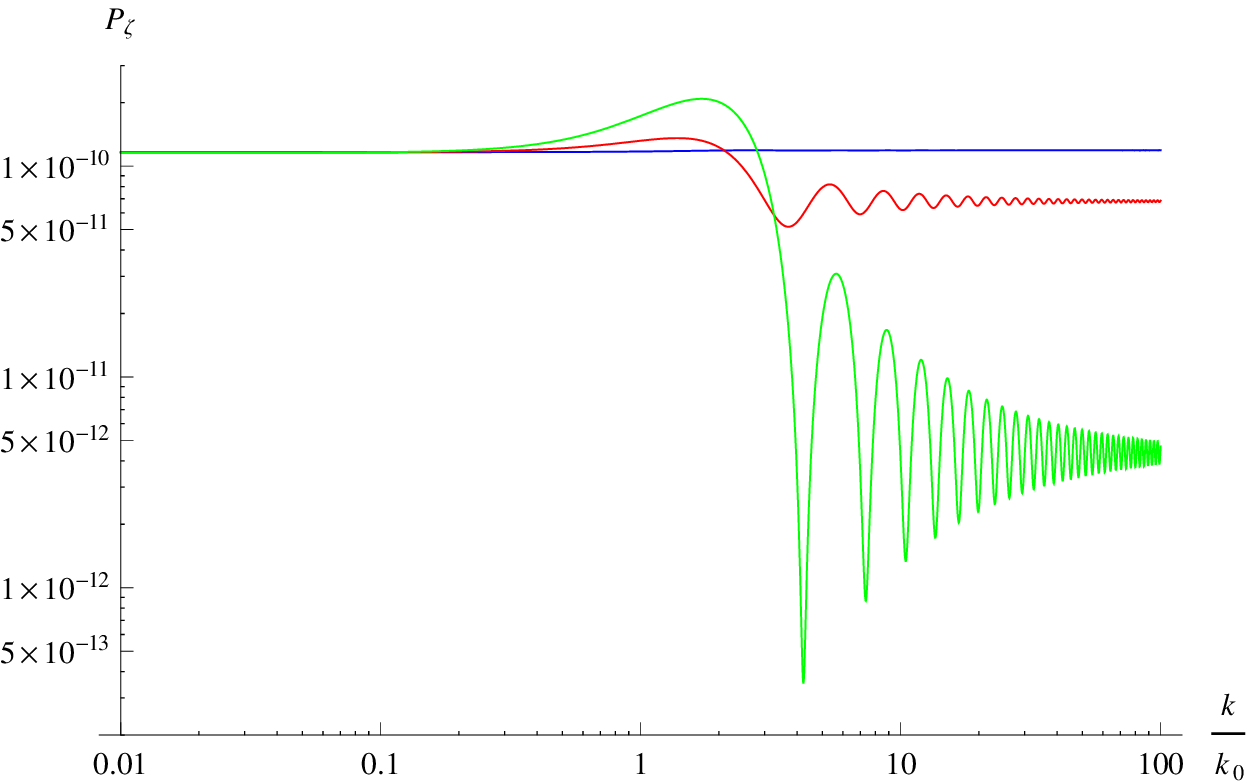}
  \end{minipage}
 \begin{minipage}{.45\textwidth}
  \includegraphics[scale=0.55]{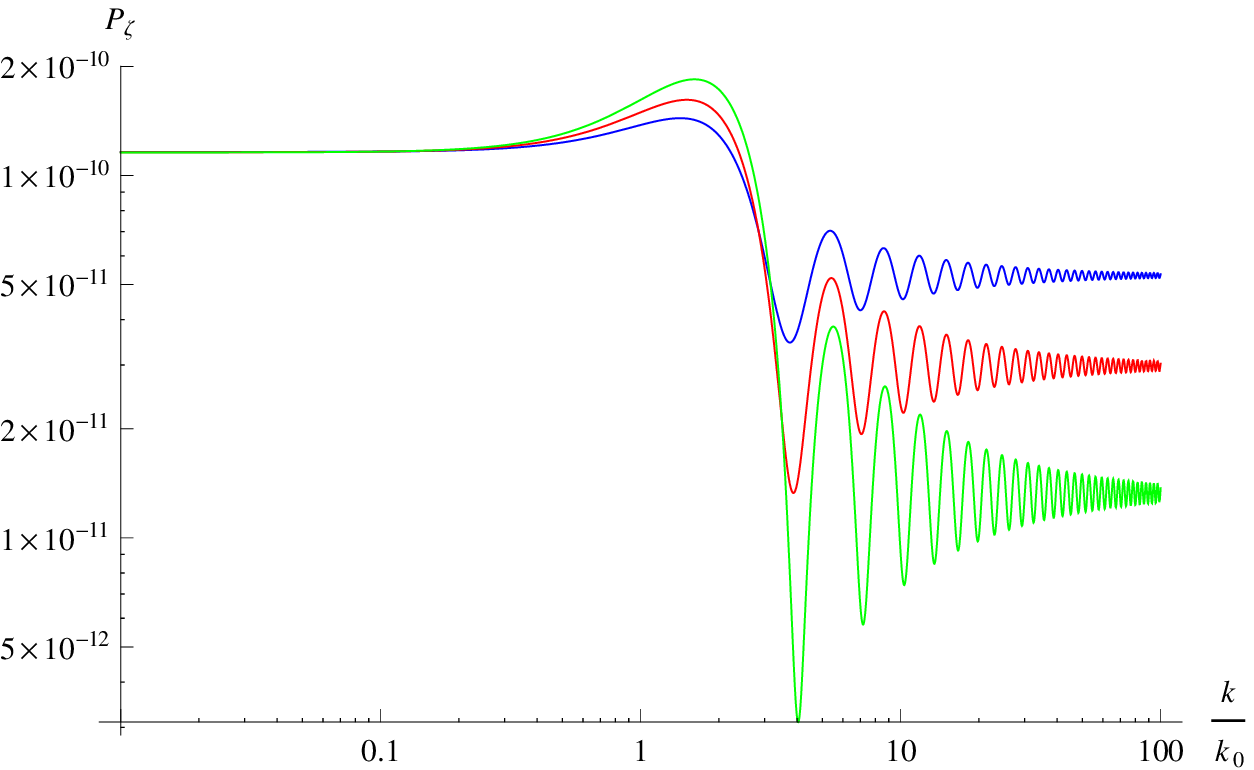}
 \end{minipage}
 \caption{The power spectrum of primordial curvature perturbations $P_{\zeta}$ plotted for different types of features. For the left plots $\lambda$ is constant, $\lambda=3.8\times10^{-19}$, and $n=2/3$ (blue lines), $n=3$ (red lines), and $n=4$ (green lines), For the right plots $n$ is constant, $n=3$, and $\lambda=6.0\times10^{-19}$ (blue lines), $\lambda=1.2\times10^{-18}$ (red lines), and $\lambda=2.4\times10^{-18}$ (green lines).}
\label{LogPn123lines}
\end{figure}
\subsubsection{Models with the same spectrum}
We have found an interesting case in which the the power spectrum for different values of $n$ and $\lambda$ can be the same. For instance for $(n_1,\lambda_1)$ and $(n_2,\lambda_2)$ we can have
\begin{equation}
 P_{\zeta}(k,n_1,\lambda_1)=P_{\zeta}(k,n_2,\lambda_2),
\end{equation}
for all values of $k$. Those values correspond to setting  
\begin{equation}\label{condition}
 n\lambda \phi_0^{n-1} = \mbox{constant}.
\end{equation}
In this case $D_0$ is the same for all models with the same set of parameters Eq. \eqref{parameters}. The interesting fact about this is that the degeneracy can only be broken at the bispectrum level as we will see below. The results are shown in Fig. \ref{P123plotall3} where we use three different values for $n$ and $\lambda$ and obtain the same power spectrum.
\begin{figure}
  \centering
  \includegraphics[scale=0.8]{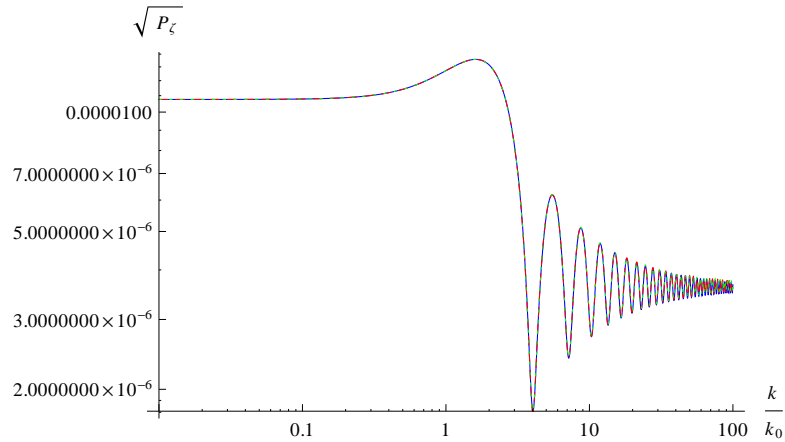}
  \caption{Numerical plots of $P_{\zeta}^{1/2}$ for three different values of $n$ and $\lambda$ for which the power spectrum is exactly the 
  same. The parameters used are (blue) $n=2/3 \mbox{ and } \lambda=2.3\times10^{-15}$, (dashed-red) $n=3 \mbox{ and } \lambda=2.4\times10^{-18}$, 
  and  (green) $n=4 \mbox{ and } \lambda=1.8\times10^{-19}$.}
  \label{P123plotall3}
\end{figure}
\subsection{Bispectrum}
The feature in the potential generates large gaussianities for the cases of large $n$ or large $\lambda$ as is expected from Eq. \eqref{b} given by
\begin{dmath}\label{b2}
  B_{\zeta}(k_1,k_2,k_3)= 2 (2\pi)^3 M_{P}^2\Im\left[ \zeta(\tau_e,k_1) \zeta(\tau_e,k_2)\zeta(\tau_e,k_3)
	\int^{\tau_e}_{\tau_0} d\tau \eta \epsilon a^2 \zeta^*(\tau,k_1)\times
	\left(2\zeta'{}^*(\tau,k_2)\zeta'{}^*(\tau,k_3)-k^2_1 \zeta^*(\tau,k_2)\zeta^*(\tau,k_3)
		\right)  \right] + \mbox{the other two permutations of } k_1, k_2,\mbox{ and } k_3,
\end{dmath}
since as we mentioned above as the inflaton crosses the feature $\eta$ increases dramatically. We evaluate the integral from $\tau_0 \to \tau_e$ since for single field inflationary models $\zeta \to$ constant after Hubble crossing horizon as it freezes out, while for early times $\zeta \to e^{-\imath k\tau}$ which oscillates rapidly, so its contribution to the integral tends to cancel\footnote{See Refs. \citep{a2, a3} for a detailed treatment of the integral for times $\tau<\tau_0$.}\citep{a2, a3}. In this way the integral is dominated by the perturbations leaving the horizon. The numerical results for different values of $n$ and $\lambda$ are plotted in Figs. \ref{FNLn123} - \ref{FNLn123ela} for large and small scales in the squeezed and equilateral limits.
\begin{figure}
 \begin{minipage}{.45\textwidth}
  \includegraphics[scale=0.5]{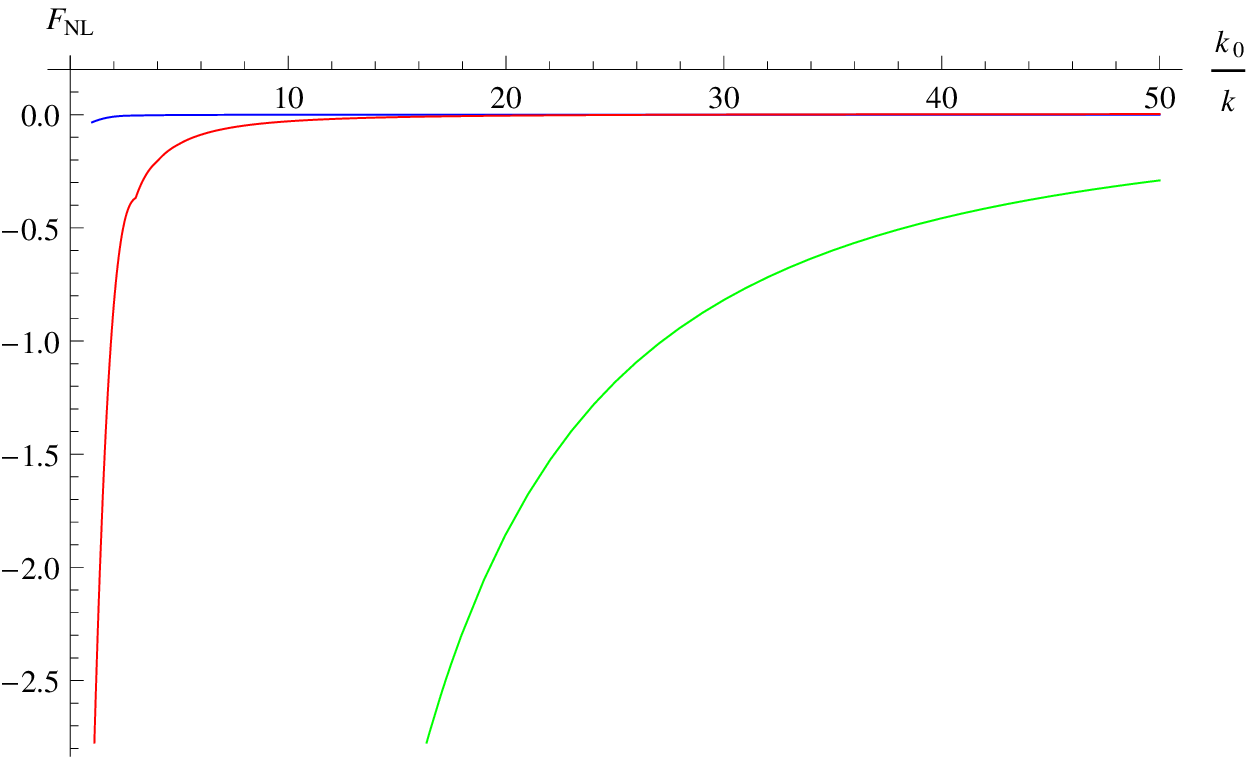}
  \end{minipage}
 \begin{minipage}{.45\textwidth}
  \includegraphics[scale=0.5]{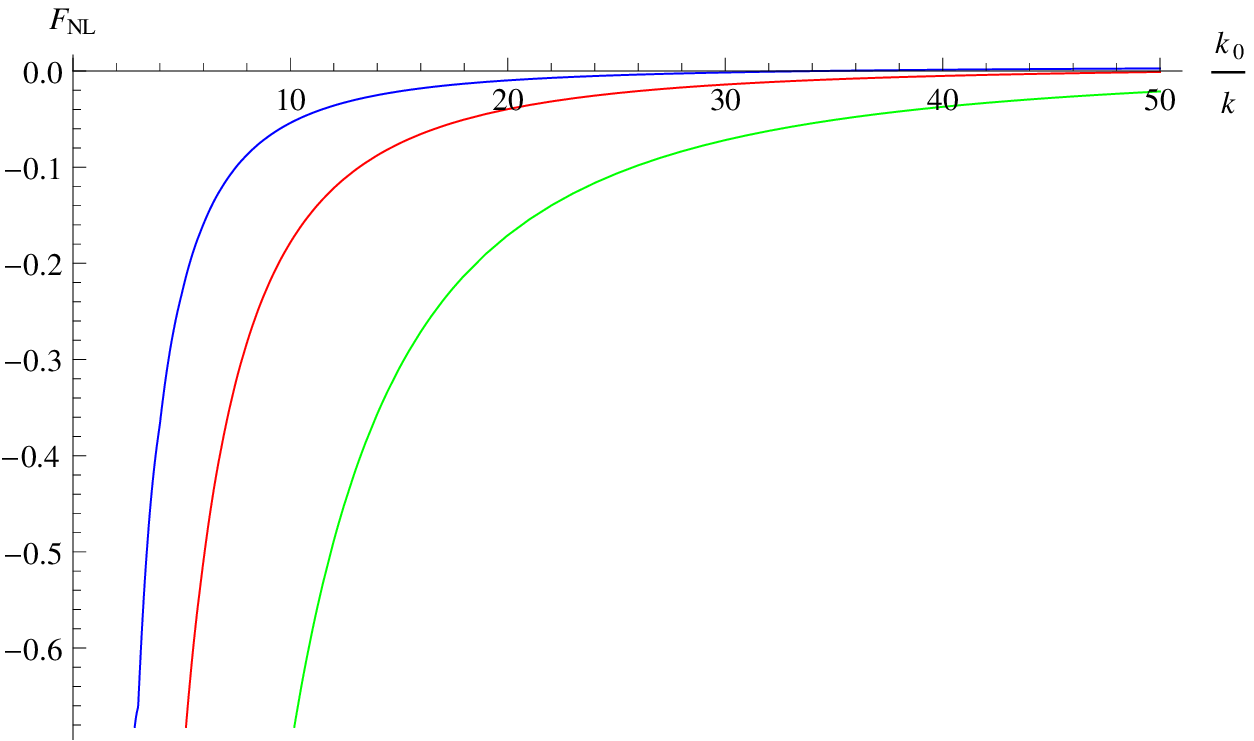}
 \end{minipage}
 \caption{The squeezed limit of the numerically computed bispectrum  $F_{NL}(k_0/500,k,k)$ in plotted for a large scale $k_0/500$. On the right we keep $\lambda$ constant, $\lambda=3.8\times10^{-19}$, while $n=2/3$ (blue lines), $n=3$ (red lines), and $n=4$ (green lines). On the left we keep $n$ constant, $n=3$, while $\lambda=6.0\times10^{-19}$ (blue lines), $\lambda=1.2\times10^{-18}$ (red lines), and $\lambda=2.4\times10^{-18}$ (green lines).}
\label{FNLn123}
\end{figure}

\begin{figure}
 \begin{minipage}{.45\textwidth}
  \includegraphics[scale=0.5]{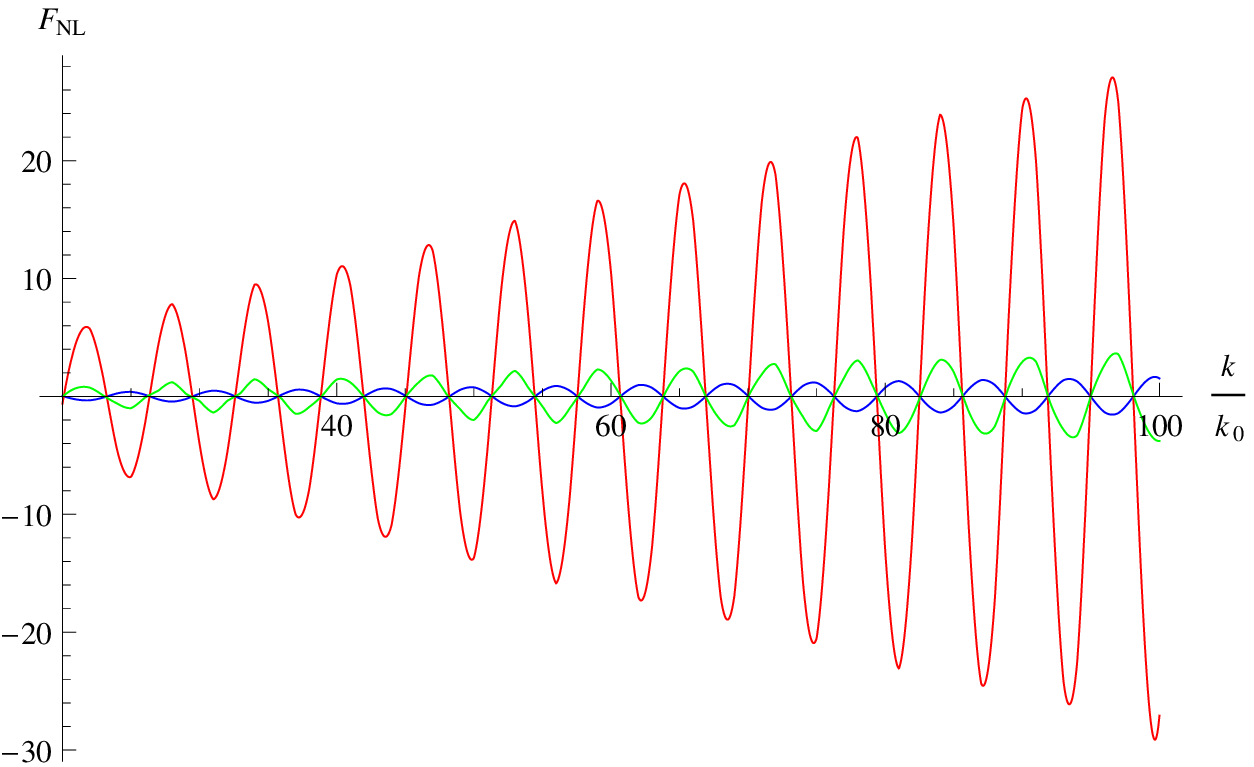}
  \end{minipage}
 \begin{minipage}{.45\textwidth}
  \includegraphics[scale=0.5]{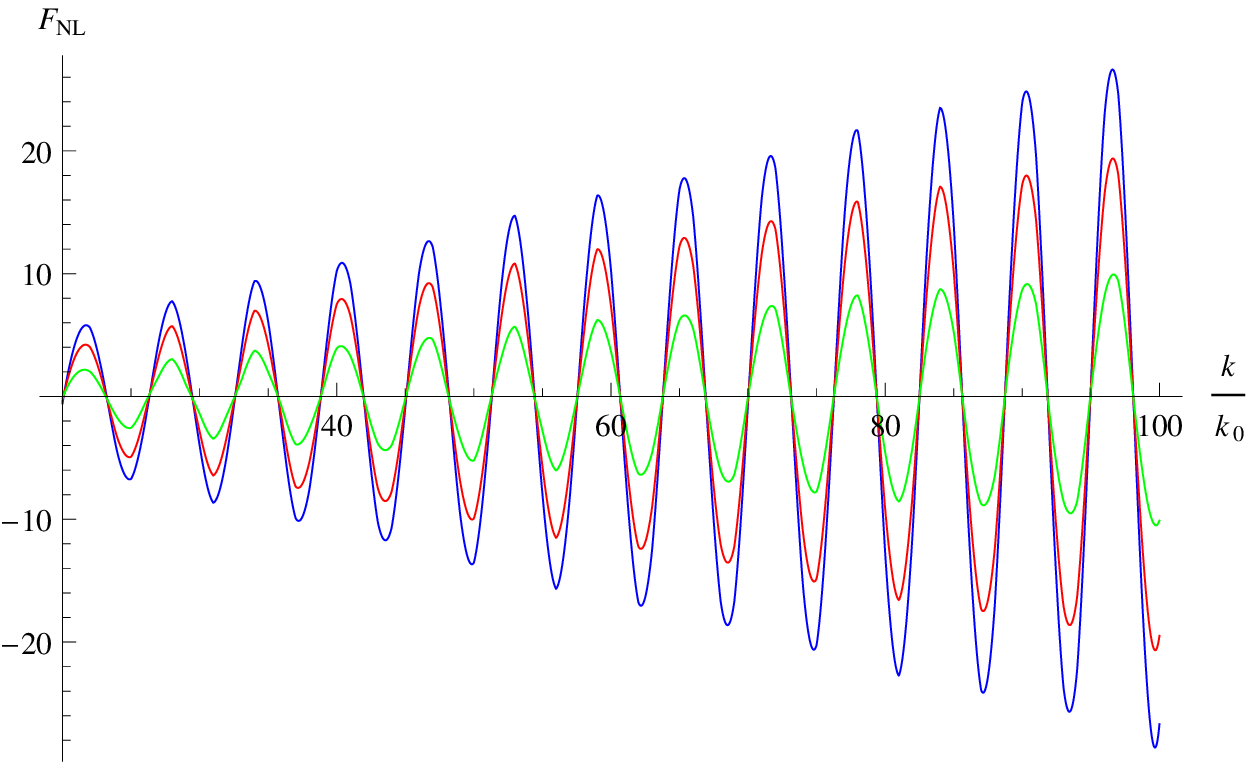}
 \end{minipage}
 \caption{The squeezed limit of the numerically computed bispectrum  $F_{NL}(k,1000k_0,1000k_0)$ is plotted for a small scale $1000 k_0$. On the left $\lambda$ is constant, $\lambda=3.8\times10^{-19}$, while $n=2/3$ (blue lines), $n=3$ (red lines), and $n=4$ (green lines). On the right $n$ is constant, $n=3$, while $\lambda=6.0\times10^{-19}$ (blue lines), $\lambda=1.2\times10^{-18}$ (red lines), and $\lambda=2.4\times10^{-18}$ (green lines).}
\label{FNLn123sla}
\end{figure}

\begin{figure}
 \begin{minipage}{.45\textwidth}
  \includegraphics[scale=0.5]{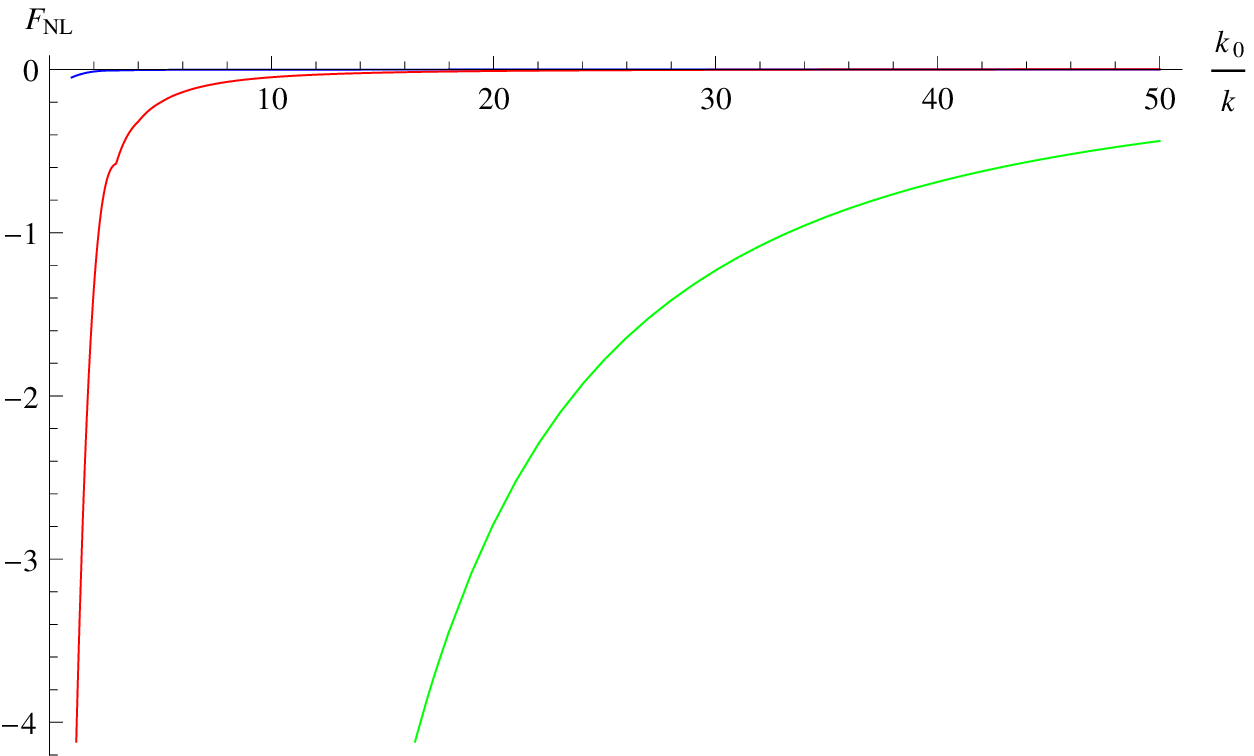}
  \end{minipage}
 \begin{minipage}{.45\textwidth}
  \includegraphics[scale=0.5]{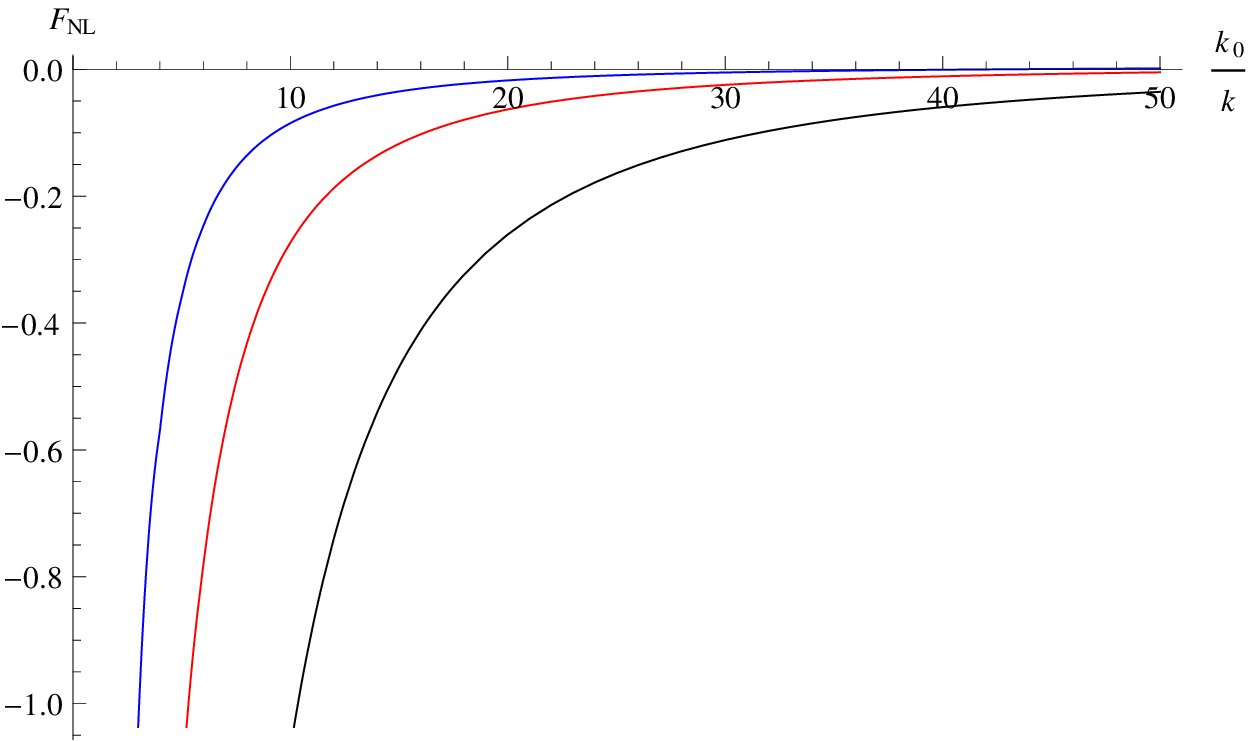}
 \end{minipage}
 \caption{ The equilateral limit of the numerically computed bispectrum  $F_{NL}(k,k,k)$ in plotted for large scales. On the right we keep $\lambda$ constant, $\lambda=3.8\times10^{-19}$, while $n=2/3$ (blue lines), $n=3$ (red lines), and $n=4$ (green lines). On the left we keep $n$ constant, $n=3$, while $\lambda=6.0\times10^{-19}$ (blue lines), $\lambda=1.2\times10^{-18}$ (red lines), and $\lambda=2.4\times10^{-18}$ (green lines).}
\label{FNLn123elb}
\end{figure}

\begin{figure}
 \begin{minipage}{.45\textwidth}
  \includegraphics[scale=0.5]{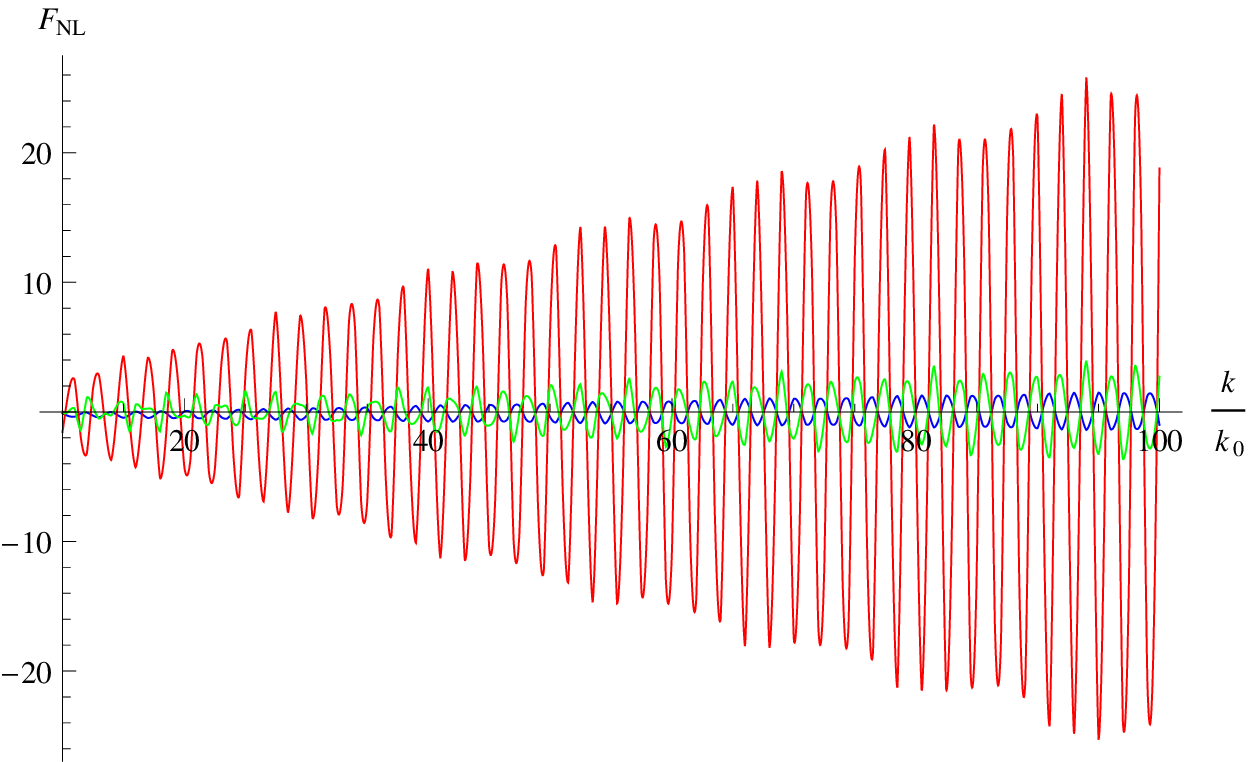}
  \end{minipage}
 \begin{minipage}{.45\textwidth}
  \includegraphics[scale=0.5]{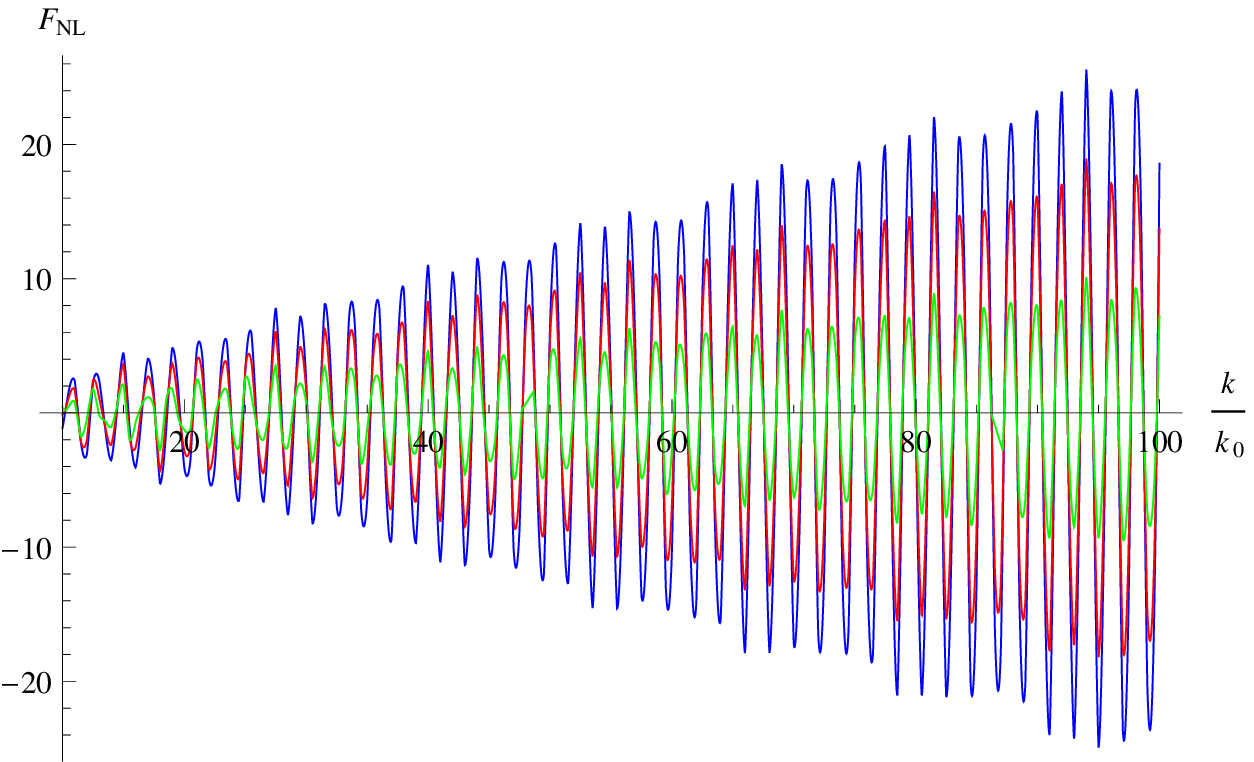}
 \end{minipage}
 \caption{The equilateral limit of the numerically computed bispectrum  $F_{NL}(k,k,k)$ is plotted for small scales. On the left $\lambda$ is constant, $\lambda=3.8\times10^{-19}$, while $n=2/3$ (blue lines), $n=3$ (red lines), and $n=4$ (green lines). On the right $n$ is constant, $n=3$, while $\lambda=6.0\times10^{-19}$ (blue lines), $\lambda=1.2\times10^{-18}$ (red lines), and $\lambda=2.4\times10^{-18}$ (green lines).}
\label{FNLn123ela}
\end{figure}
\subsubsection{Analytical approximation for the bispectrum}
In order to obtain an analytic approximation for the bispectrum we use the analytic expressions for the perturbation Eq. \ref{r} and for the slow-roll parameters Eqs. \ref{epsilon} and \ref{eta}. We also use a series of approximations such as fixing $\epsilon$ and $\eta$ constant after and before horizon crossing for those modes $\epsilon(\tau_k)$ and $\eta(\tau_k)$, except in the time interval around when the feature occurs, when in fact their variation cannot be neglected. We also consider that $\lambda^+ \ll \lambda^-$ throughout the calculation. Another approximation considers the fact that the modes freeze out after horizon crossing then after that the modes are constant and can be taken out of the integral. The analytical approximations for the bispectrum are shown in Figs. \ref{FNLfnlslb1} - \ref{FNLfnlela2} and are in good agreement with the numerical results.
\paragraph{Squeezed limit}
\begin{itemize}
\item In the squeezed limit and for small scales with $k_0 \ll k_1 \ll k_2=k_3\equiv k$ we found an analytic expression for $F_{NL}$ in terms of $D_0$ given by
\bea
F_{NL}^{SL>}(k,k_1)= -\frac{2}{3}\frac{D_0}{k_0} \left( 1 + \frac{D_0}{3k_0} \right)^{-4} \left[ \frac{2k+k_1}{k_0} \frac{k_1}{k_0}\sin \left(\frac{2k + k_1}{k_0}\right) \right. \\ \nonumber
\left. +  \left(2-  \frac{D_0}{k_0}\right) \cos\left(\frac{2k + k_1}{k_0}\right)  -  \frac{D_0}{k_0} \cos\left(\frac{2k-k_1}{k_0}\right)\right],
\eea
where SL stands for squeezed limit and the sign $>$ for small scales.
\item In the squeezed limit and large scales with $k_1 \ll k_2=k_3\equiv k \ll k_0$ the analytic expression for all $n$ is 
\begin{equation}\label{fnlslb}
 F_{NL}^{SL<}(k)= -(n-2)\frac{D_0}{k_0} \left(8+\frac{D_0}{k_0} \right) \left(\frac{k}{k_0}\right)^2,
\end{equation}
where the sign $<$ means for large scales. In this case the sign of $F_{NL}$ is determined by the factor $(n-2)$.
\end{itemize}
\paragraph{Equilateral limit}
\begin{itemize}
 \item In case of the equilateral limit and small scales with $k_0\ll k_1 = k_2=k_3 \equiv k $ the analytic expression is written as
\bea
F_{NL}^{EL>}(k)= -\frac{9}{8}\frac{D_0}{k_0} \left( 1 + \frac{D_0}{3k_0} \right)^{-4} \left[ \frac{k}{k_0}\sin \left(\frac{3k}{k_0}\right) \right. \\ \nonumber
+ \left. \left(3- \frac{7}{3} \frac{D_0}{k_0} \right) \cos\left(\frac{3k}{k_0}\right) \right],
\eea
where EL stands for equilateral limit.
\item In the equilateral limit and for large scales with $k_1 = k_2=k_3 \equiv k \ll k_0 $ the analytic expression is given by
\begin{equation}
 F_{NL}^{EL<}(k)= -(n-2)\frac{4D_0}{3k_0} \left(8+ \frac{D_0}{k_0} \right) \left(\frac{k}{k_0}\right)^2,
\end{equation}
where again the sign of the $F_{NL}$ function is determined by $n$.
\end{itemize}

\begin{figure}
  \begin{minipage}{.5\textwidth}
  \includegraphics[scale=0.55]{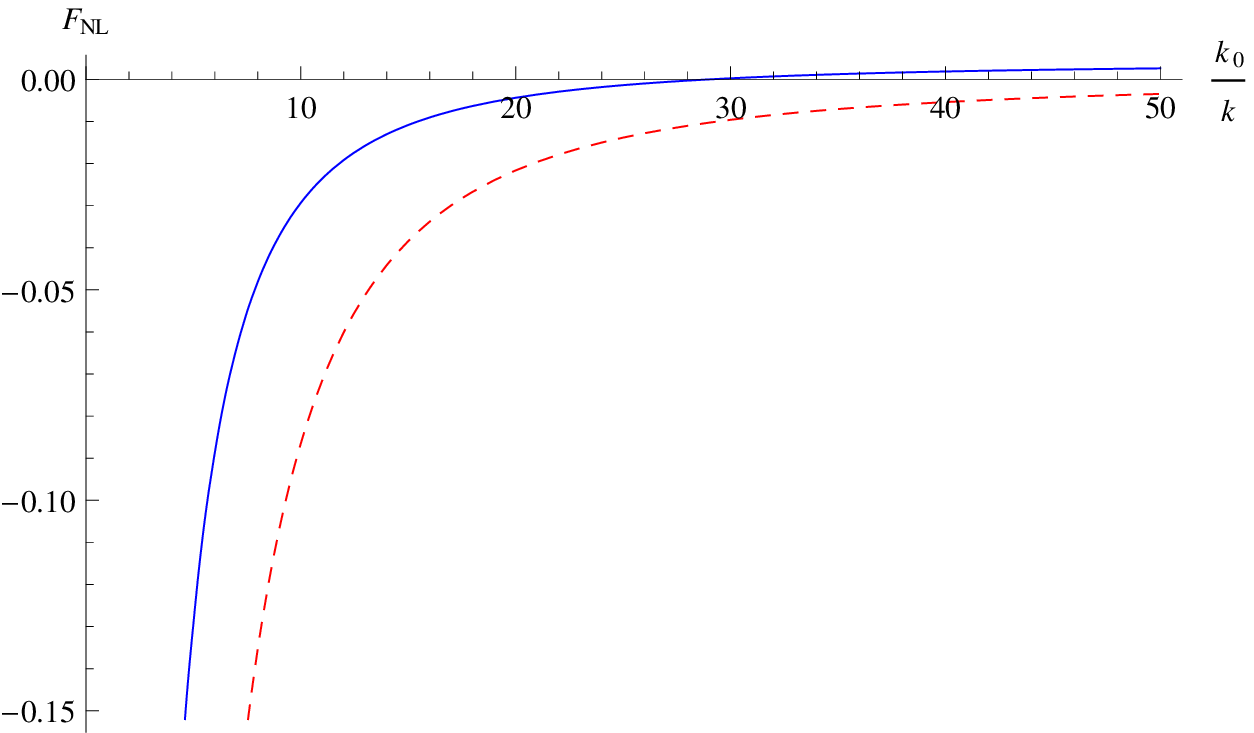}
 \end{minipage}
 \begin{minipage}{.5\textwidth}
  \includegraphics[scale=0.55]{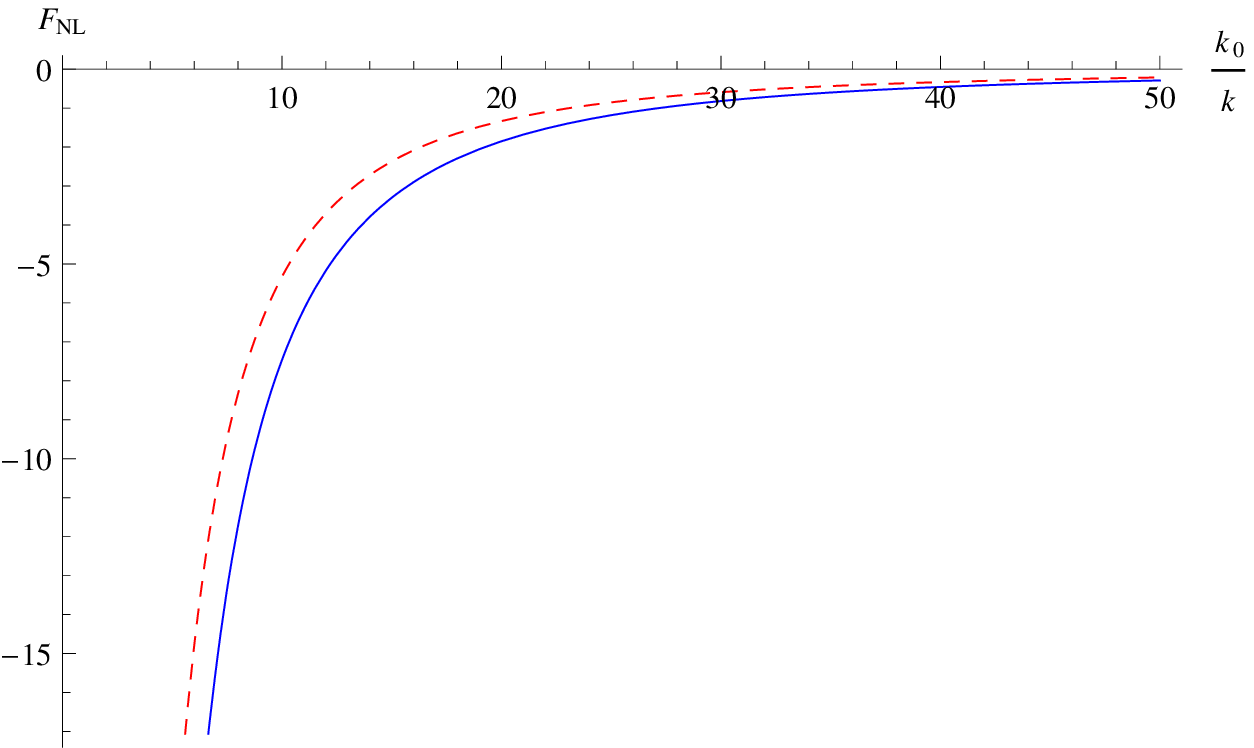}
 \end{minipage}
 \caption{The squeezed limit of the numerically computed (blue lines) and analytic (red lines) bispectrum $F_{NL}(k_0/500,k,k)$ is plotted for a large scale $k_0/500$. From left to right we keep $\lambda$ constant, $\lambda=3.8\times10^{-19}$, while $n=3$, and $n=4$, respectively.}
\label{FNLfnlslb1}
\end{figure}

\begin{figure}
 \begin{minipage}{.3\textwidth}
  \includegraphics[scale=0.6]{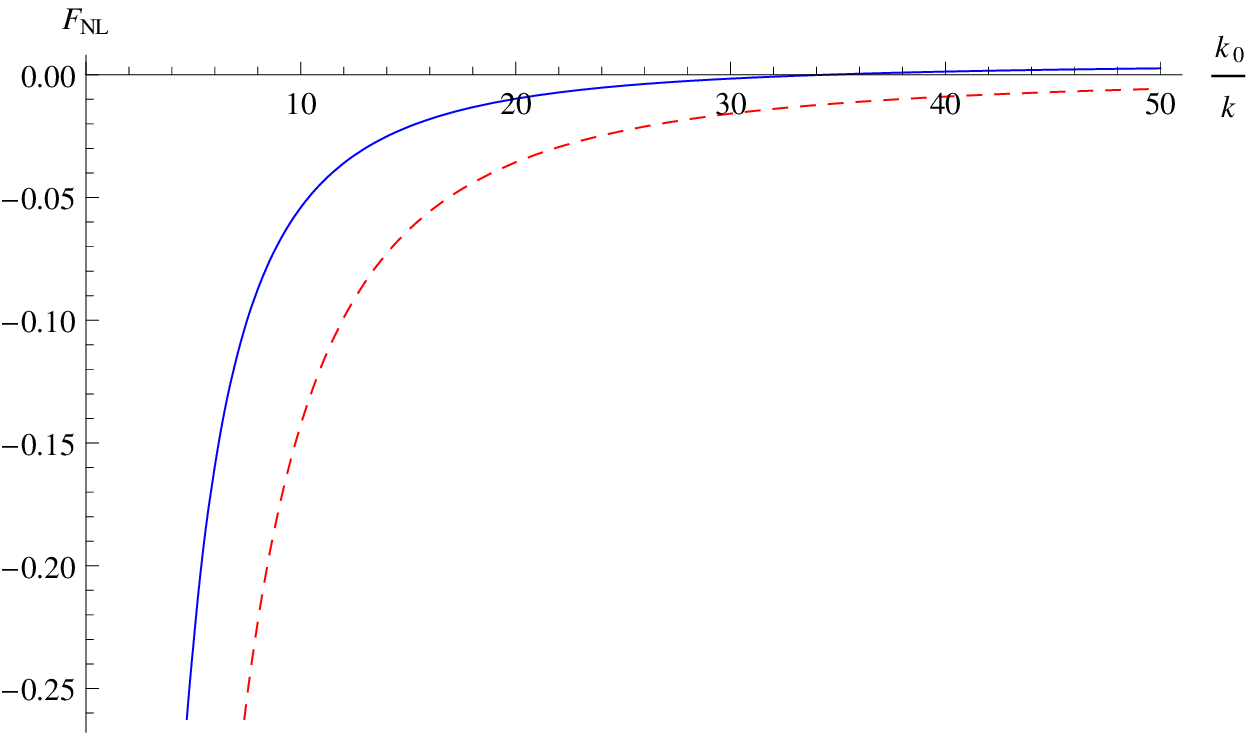}
  \end{minipage}
 \begin{minipage}{.45\textwidth}
  \includegraphics[scale=0.6]{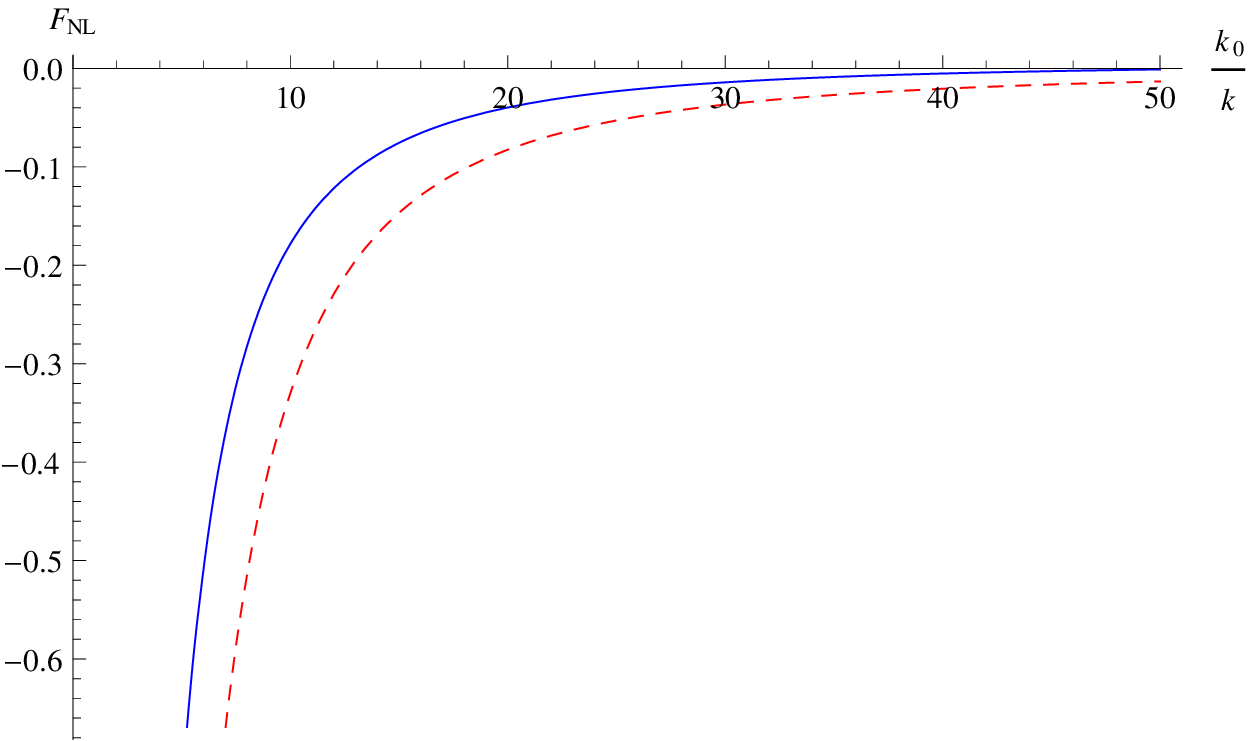}
 \end{minipage}
 \begin{minipage}{.3\textwidth}
  \includegraphics[scale=0.6]{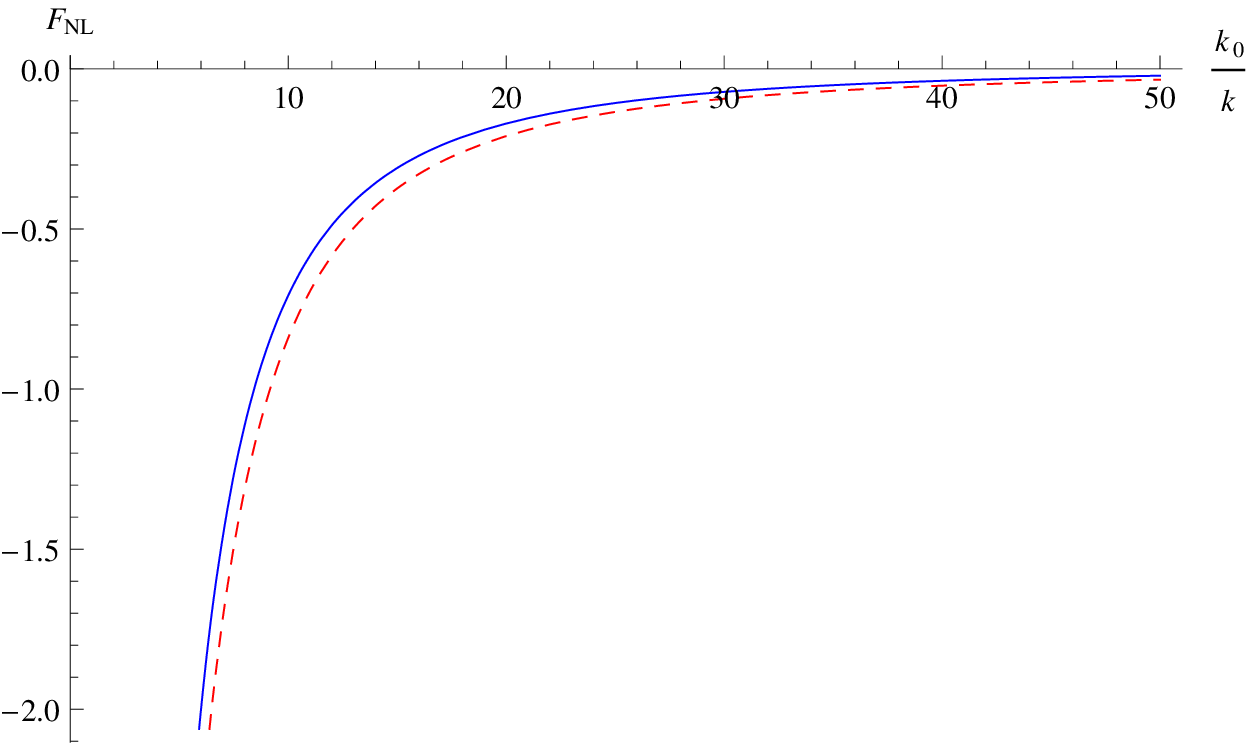}
 \end{minipage}
 \caption{The squeezed limit of the numerically computed (blue lines) and analytic (red lines) bispectrum $F_{NL}(k_0/500,k,k)$ is plotted for a large scale $k_0/500$. From left to right we keep $n$ constant, $n=3$, while $\lambda=6.0\times10^{-19}$, $\lambda=1.2\times10^{-18}$, and $\lambda=2.4\times10^{-18}$, respectively.}
\label{FNLfnlslb2}
\end{figure}
\begin{figure}
  \begin{minipage}{.5\textwidth}
  \includegraphics[scale=0.55]{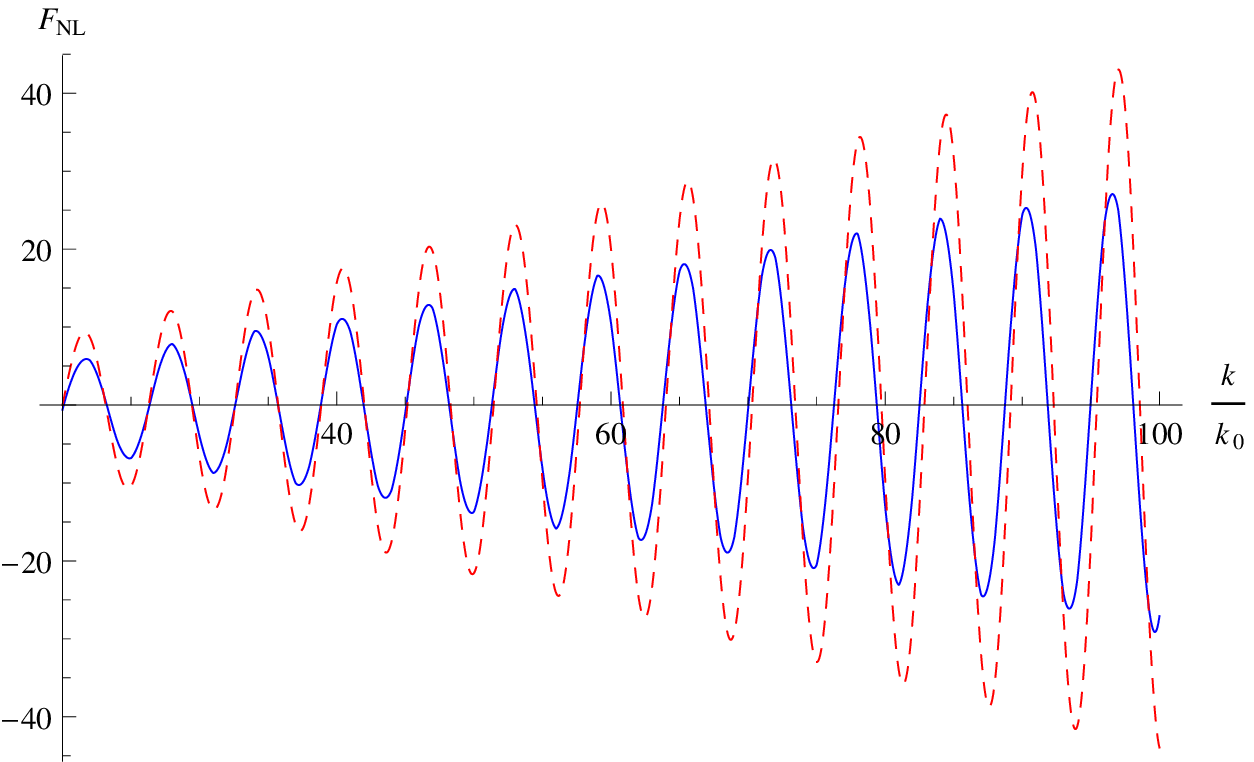}
 \end{minipage}
 \begin{minipage}{.5\textwidth}
  \includegraphics[scale=0.55]{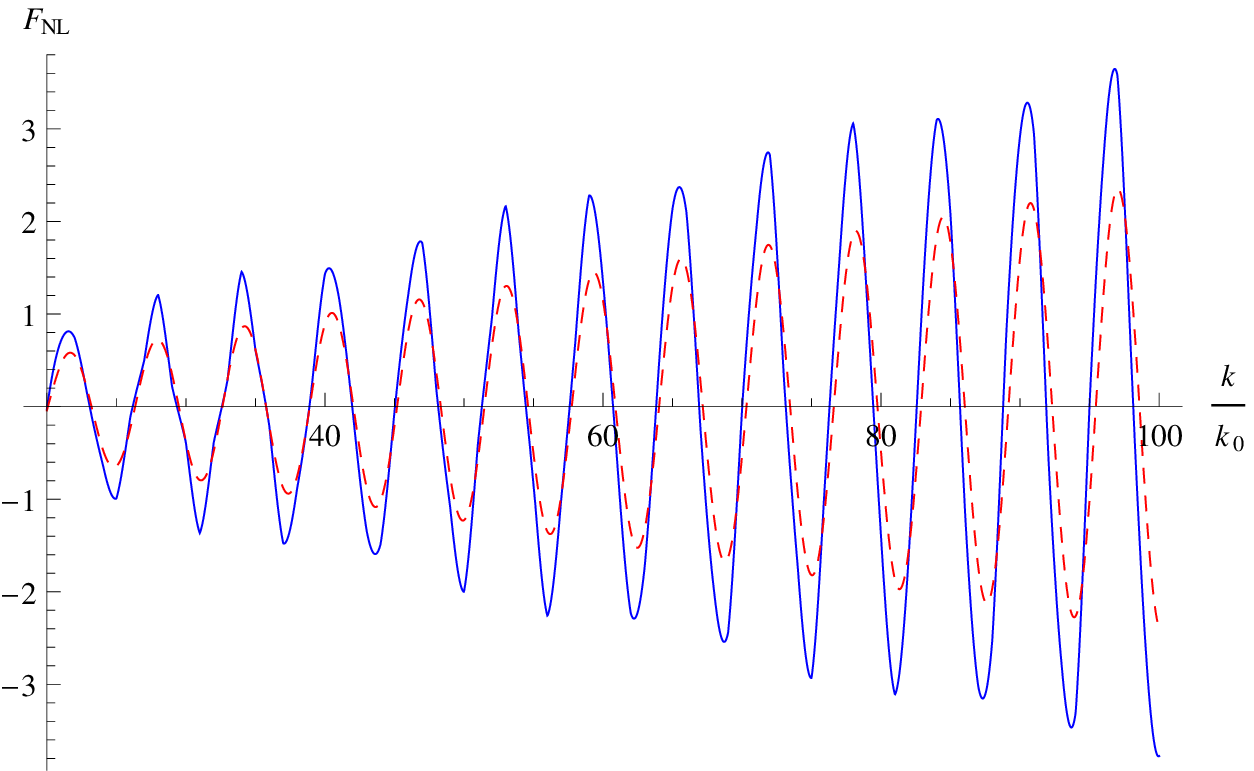}
 \end{minipage}
 \caption{The squeezed limit of the numerically computed (blue lines) and analytic (red lines) bispectrum $F_{NL}(k,1000k_0,1000k_0)$ is plotted for a small scale $1000 k_0$. From left to right we keep $\lambda$ constant, $\lambda=3.8\times10^{-19}$, while $n=3$, and $n=4$, respectively.}
\label{FNLfnlsla1}
\end{figure}

\begin{figure}
 \begin{minipage}{.3\textwidth}
  \includegraphics[scale=0.6]{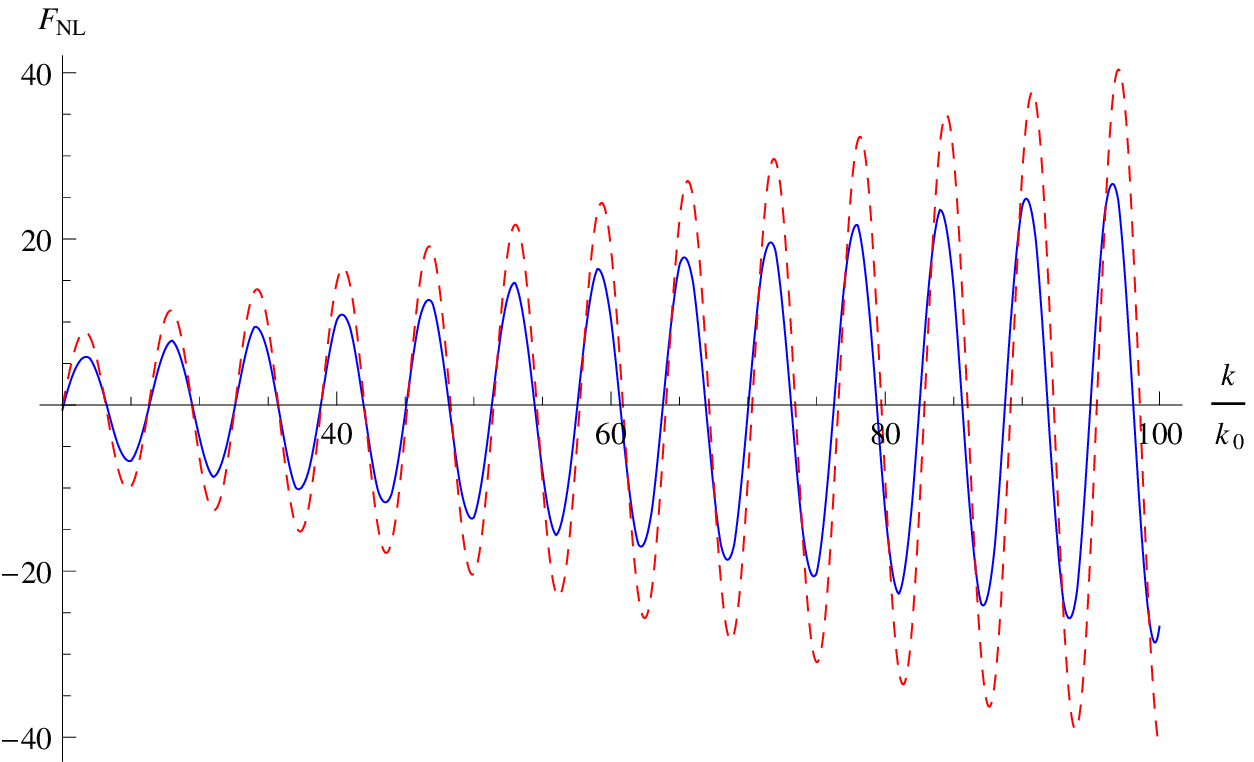}
  \end{minipage}
 \begin{minipage}{.45\textwidth}
  \includegraphics[scale=0.6]{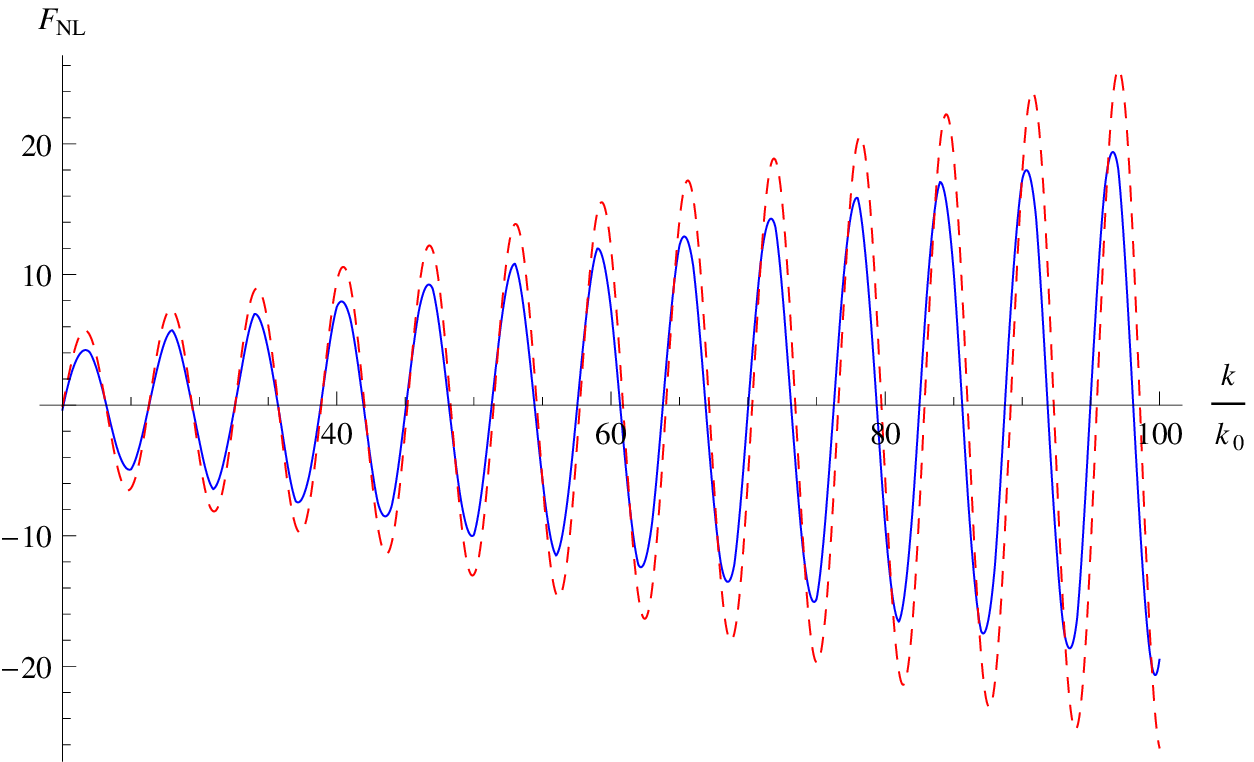}
 \end{minipage}
 \begin{minipage}{.3\textwidth}
  \includegraphics[scale=0.6]{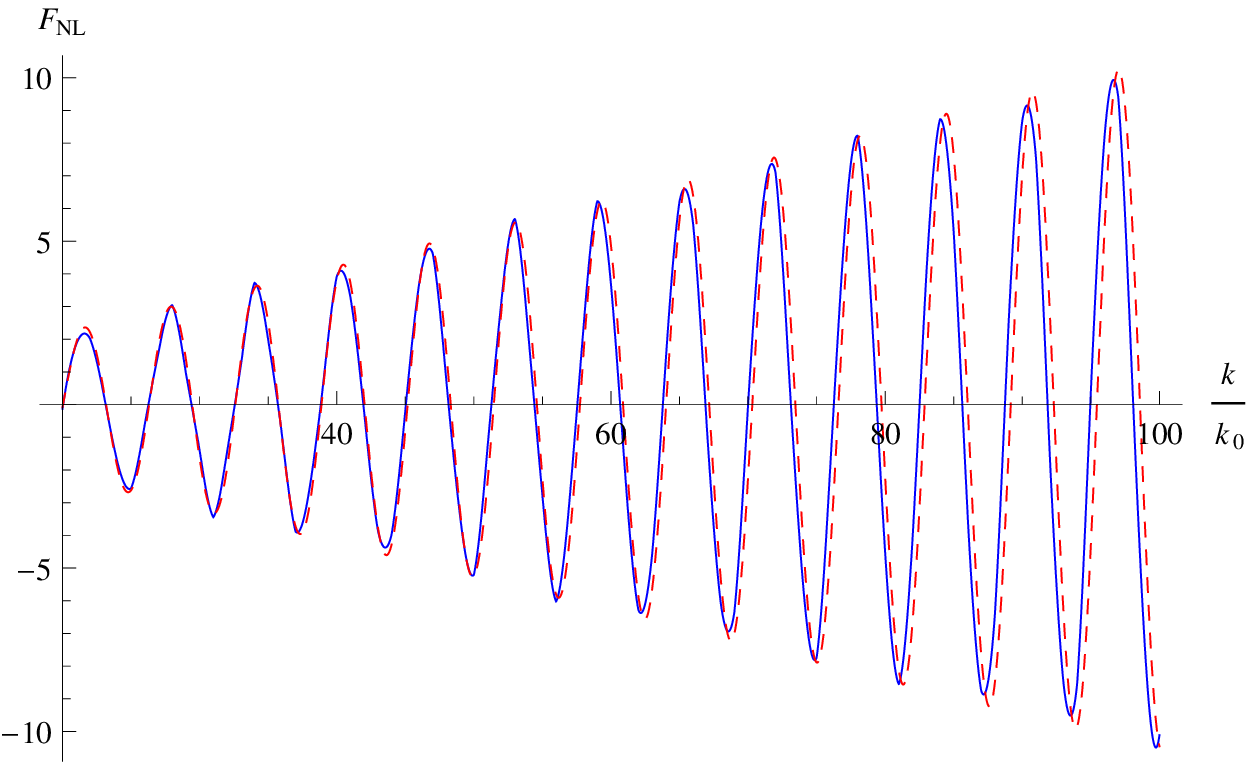}
 \end{minipage}
 \caption{The squeezed limit of the numerically computed (blue lines) and analytic (red lines) bispectrum $F_{NL}(k,1000k_0,1000k_0)$ is plotted for a small scale $1000 k_0$. From left to right we keep $n$ constant, $n=3$, while $\lambda=6.0\times10^{-19}$, $\lambda=1.2\times10^{-18}$, and $\lambda=2.4\times10^{-18}$, respectively.}
\label{FNLfnlsla2}
\end{figure}
\begin{figure}
  \begin{minipage}{.5\textwidth}
  \includegraphics[scale=0.6]{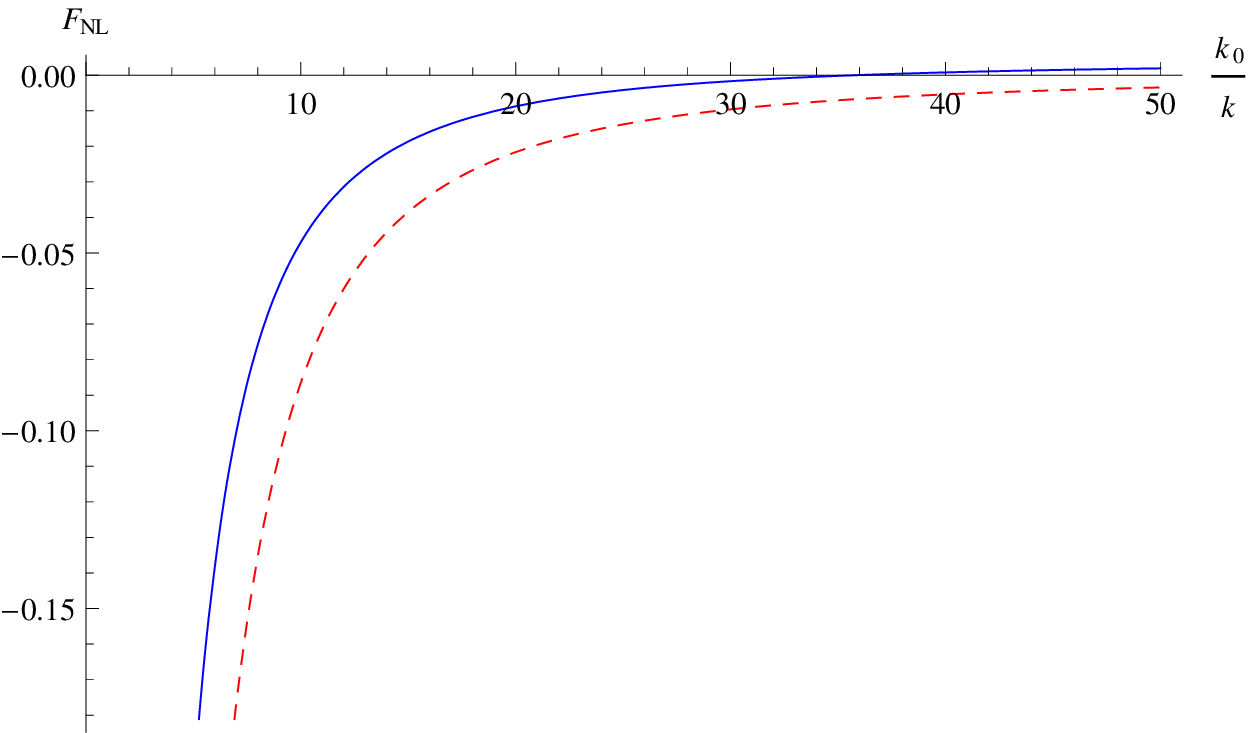}
 \end{minipage}
 \begin{minipage}{.5\textwidth}
  \includegraphics[scale=0.6]{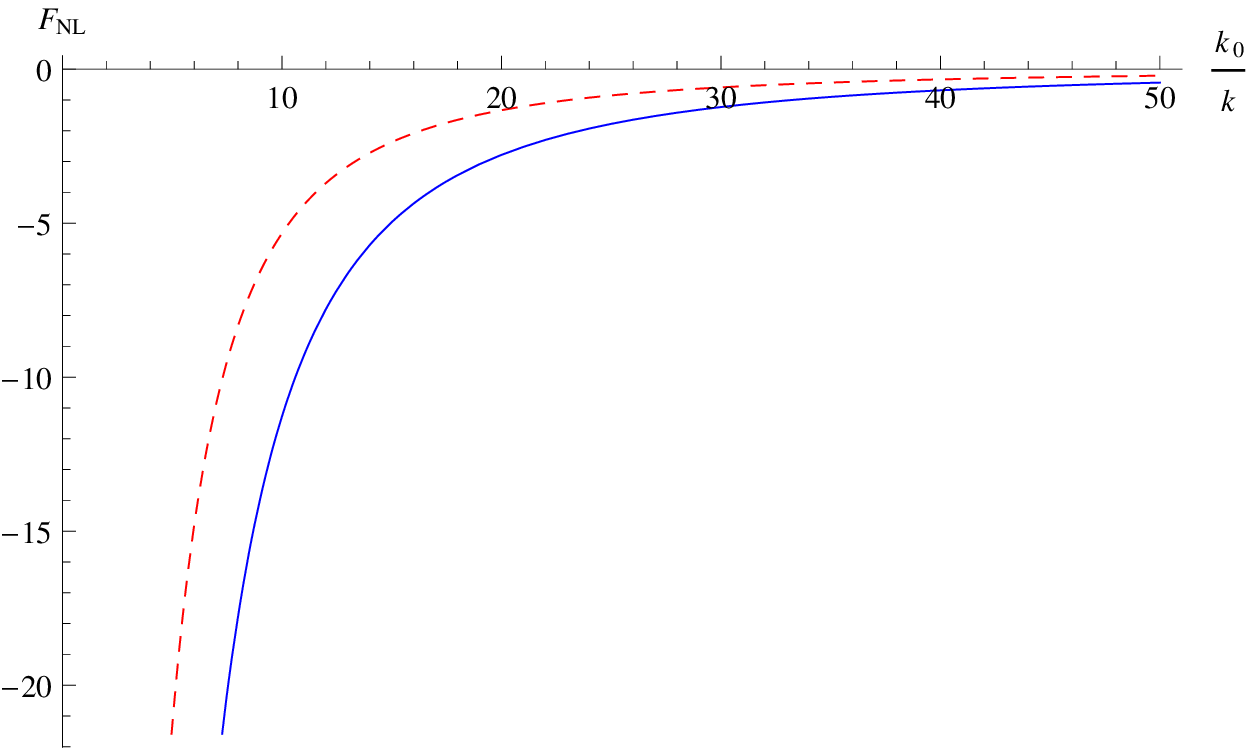}
 \end{minipage}
 \caption{The equilateral limit of the numerically computed (blue lines) and analytic (red lines) bispectrum $F_{NL}(k,k,k)$ is plotted for large scales. From left to right we keep $\lambda$ constant, $\lambda=3.8\times10^{-19}$, while $n=3$, and $n=4$, respectively.}
\label{FNLfnlelb1}
\end{figure}

\begin{figure}
 \begin{minipage}{.3\textwidth}
  \includegraphics[scale=0.6]{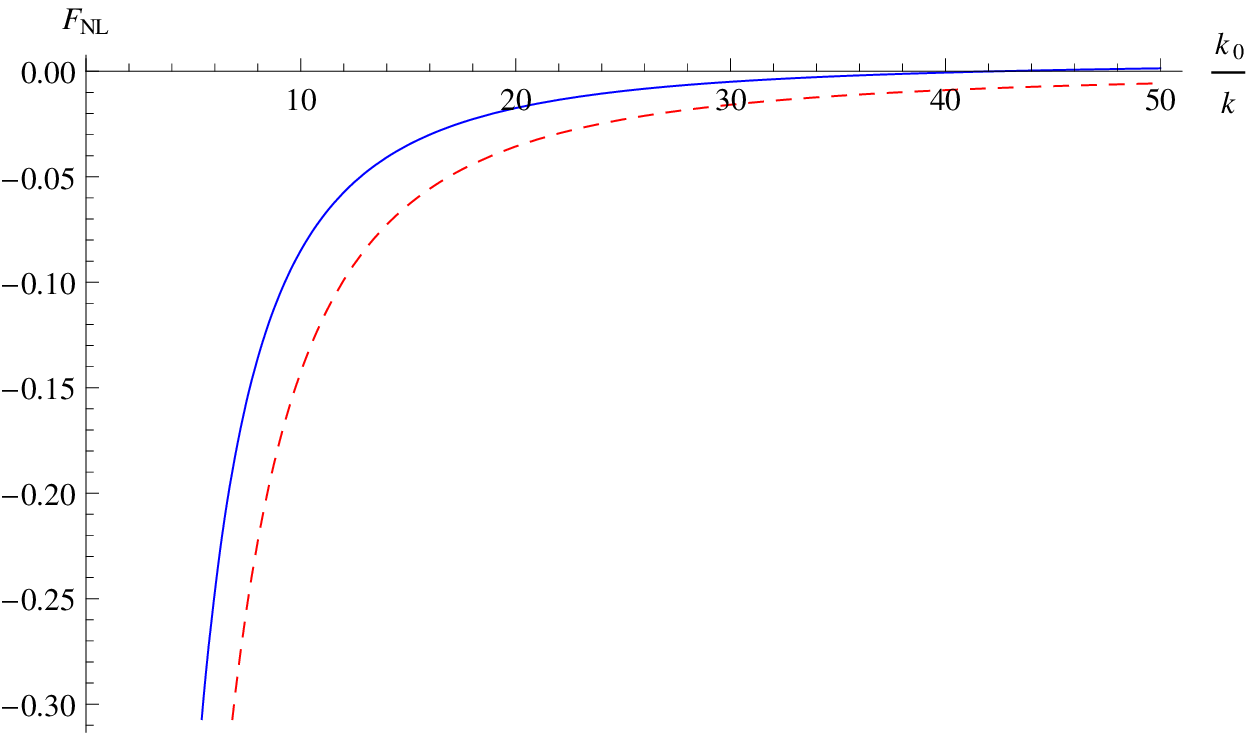}
  \end{minipage}
 \begin{minipage}{.45\textwidth}
  \includegraphics[scale=0.6]{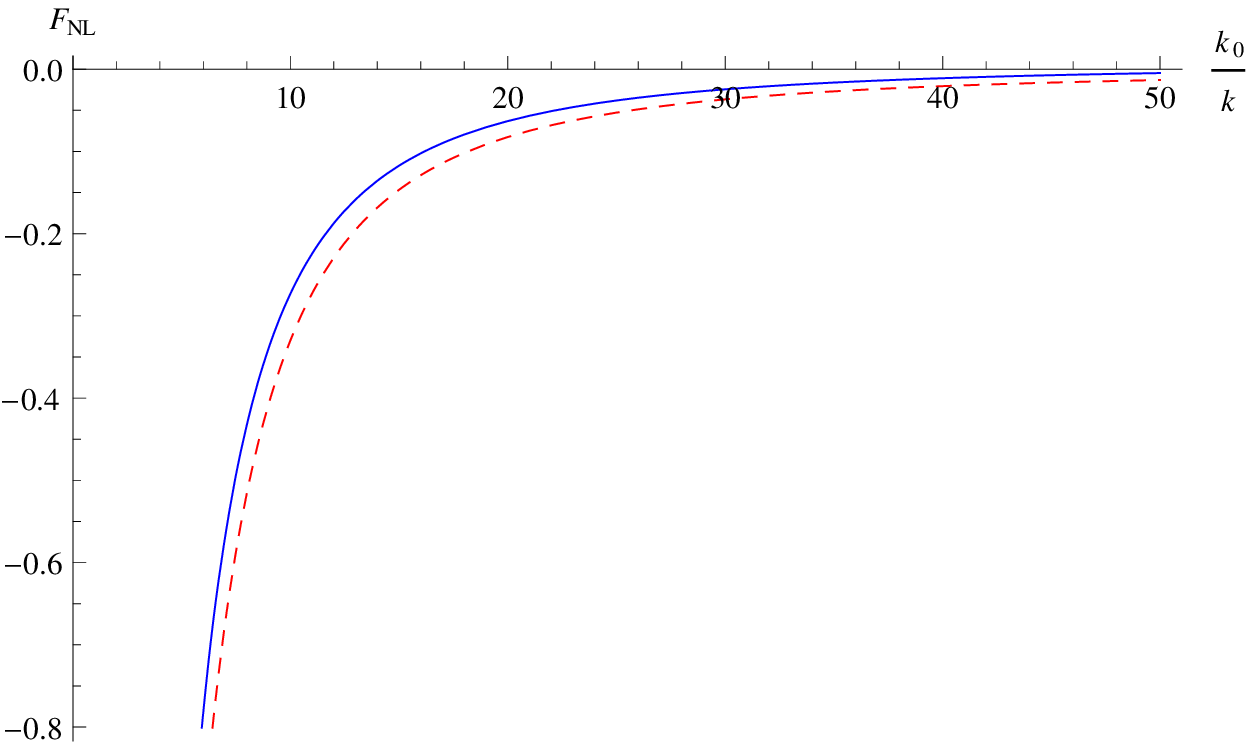}
 \end{minipage}
 \begin{minipage}{.3\textwidth}
  \includegraphics[scale=0.6]{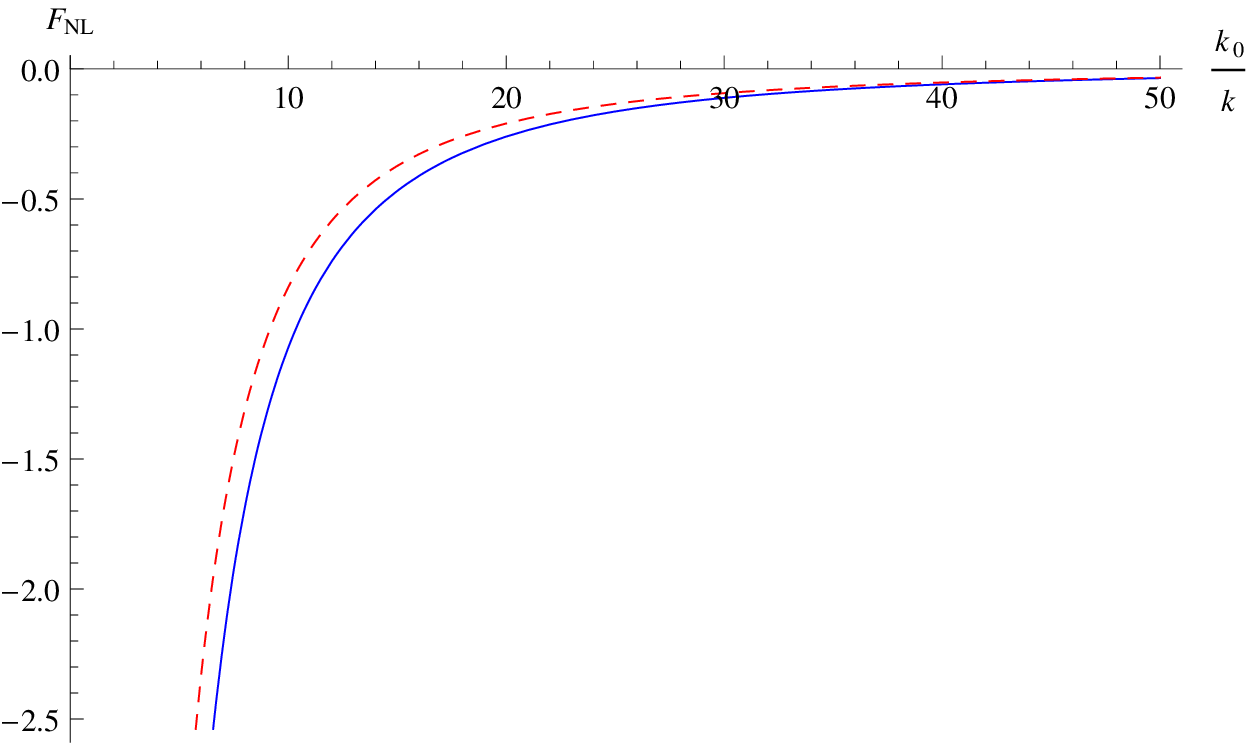}
 \end{minipage}
 \caption{The equilateral limit of the numerically computed (blue lines) and analytic (red lines) bispectrum $F_{NL}(k,k,k)$ is plotted for large scales. From left to right we keep $n$ constant, $n=3$, while $\lambda=6.0\times10^{-19}$, $\lambda=1.2\times10^{-18}$, and $\lambda=2.4\times10^{-18}$, respectively.}
\label{FNLfnlelb2}
\end{figure}
\begin{figure}
  \begin{minipage}{.5\textwidth}
  \includegraphics[scale=0.6]{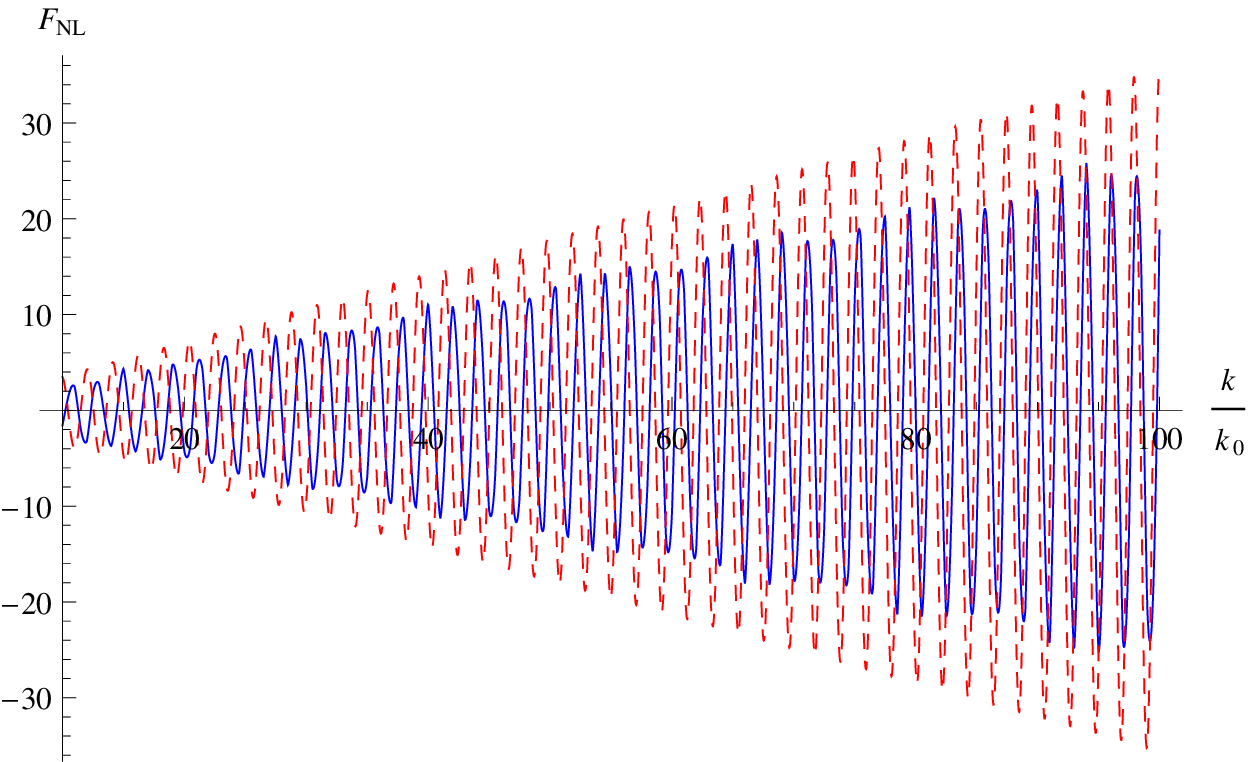}
 \end{minipage}
 \begin{minipage}{.5\textwidth}
  \includegraphics[scale=0.6]{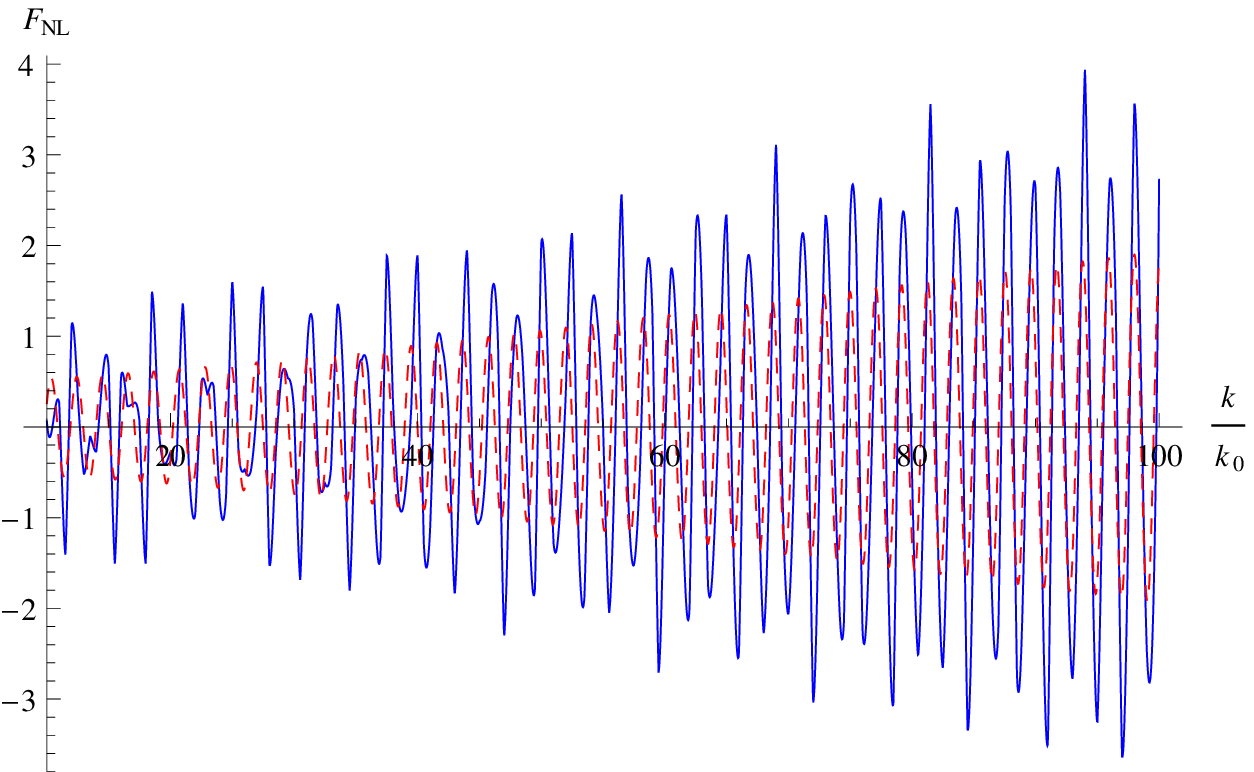}
 \end{minipage}
 \caption{The equilateral limit of the numerically computed (blue lines) and analytic (red lines) bispectrum $F_{NL}(k,k,k)$ is plotted for small scales. From left to right we keep $\lambda$ constant, $\lambda=3.8\times10^{-19}$, while $n=3$, and $n=4$, respectively.}
\label{FNLfnlela1}
\end{figure}

\begin{figure}
 \begin{minipage}{.3\textwidth}
  \includegraphics[scale=0.6]{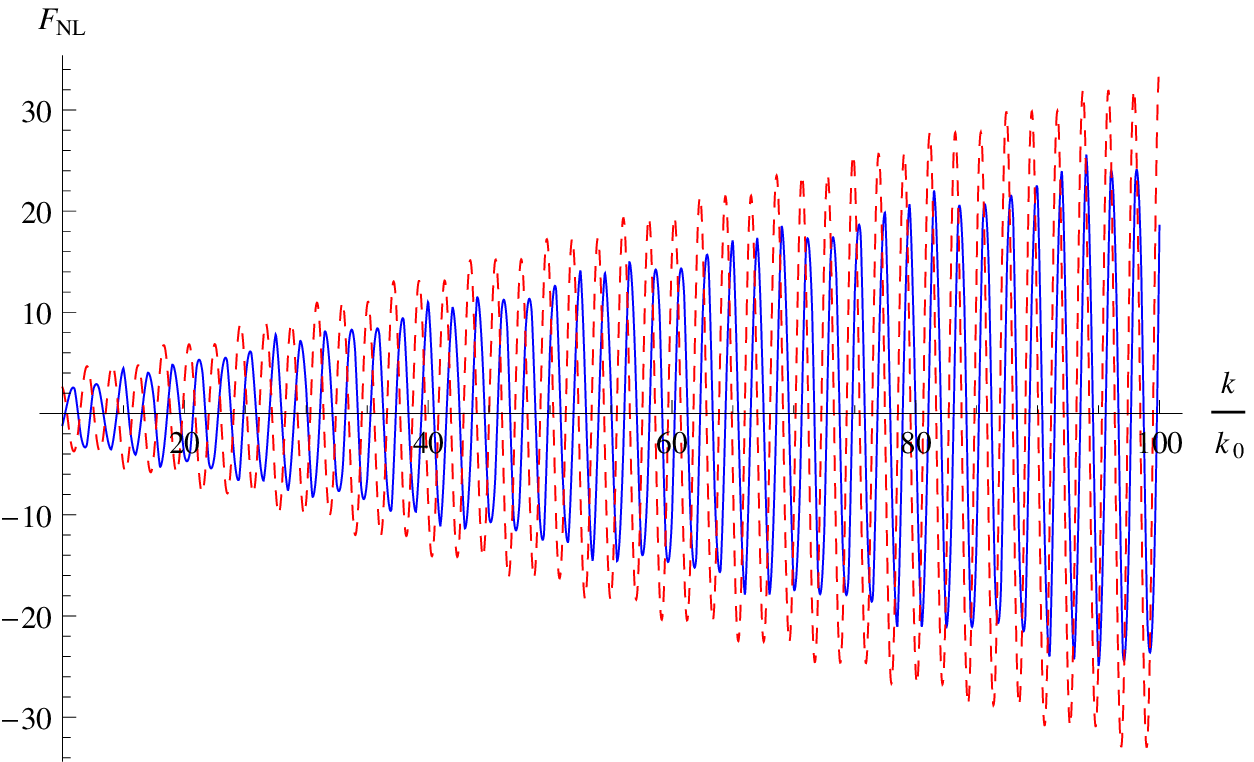}
  \end{minipage}
 \begin{minipage}{.45\textwidth}
  \includegraphics[scale=0.6]{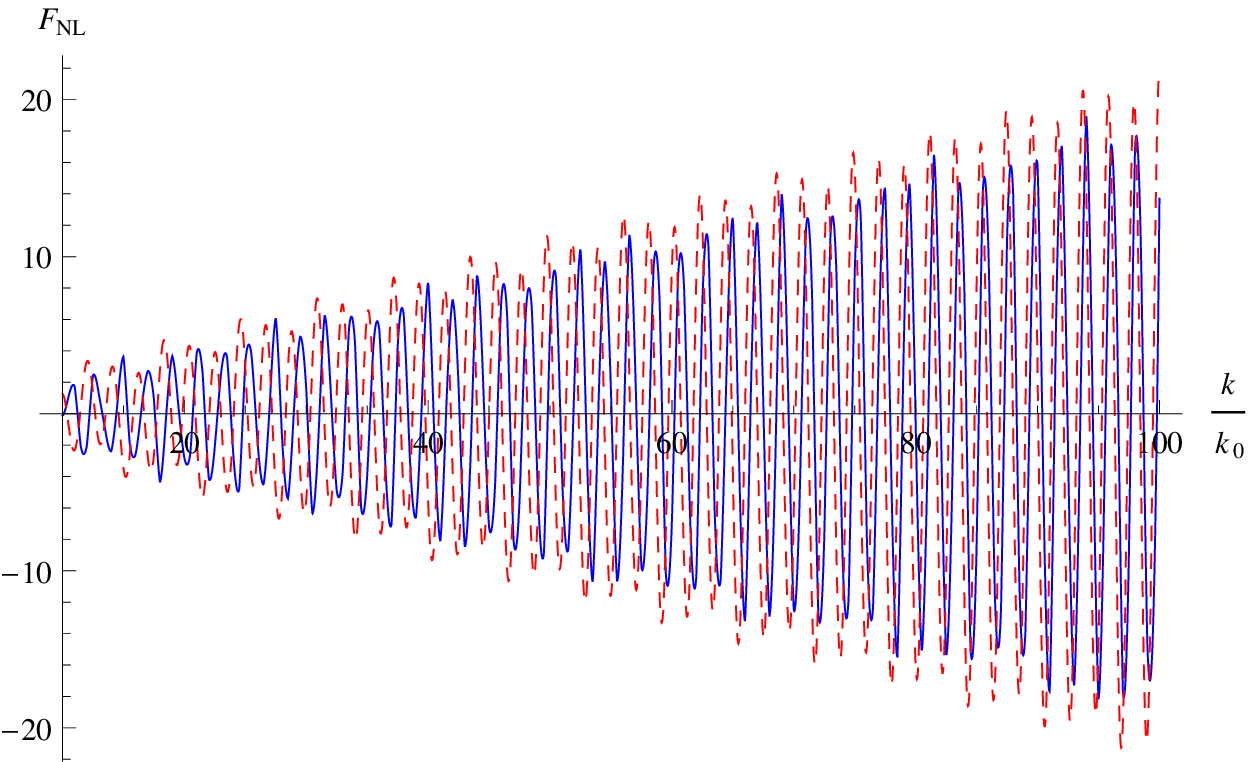}
 \end{minipage}
 \begin{minipage}{.3\textwidth}
  \includegraphics[scale=0.6]{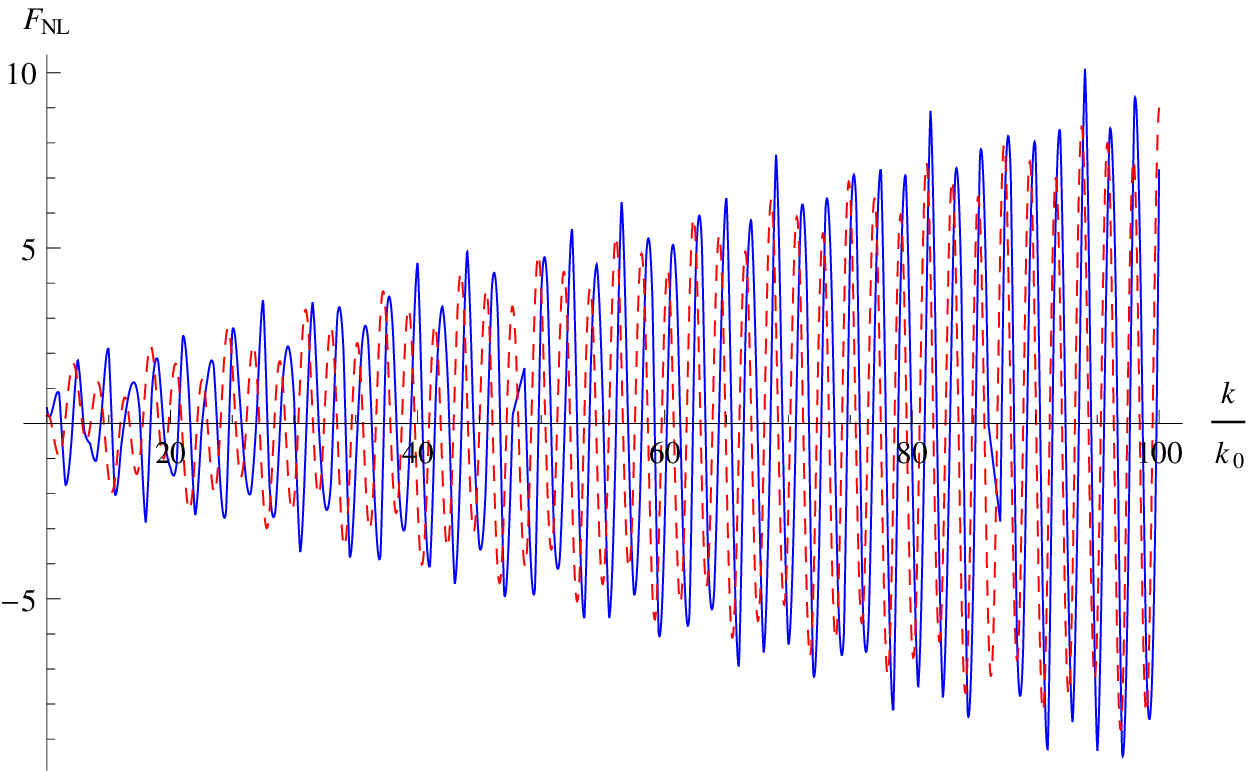}
 \end{minipage}
 \caption{The equilateral limit of the numerically computed (blue lines) and analytic (red lines) bispectrum $F_{NL}(k,k,k)$ is plotted for small scales. From left to right we keep $n$ constant, $n=3$, while $\lambda=6.0\times10^{-19}$, $\lambda=1.2\times10^{-18}$, and $\lambda=2.4\times10^{-18}$, respectively.}
\label{FNLfnlela2}
\end{figure}
\subsubsection{Models with the same bispectrum}
We have also found some interesting effects in the $F_{NL}$ functions Eq. \eqref{FNL} when the appropriate choice of parameters give the same spectrum as shown in Fig. \ref{P123plotall3}. In this case in the squeezed and equilateral limits at large scales the $F_{NL}$ functions are positive for $n<2$ and negative for $n>2$. While in both limits for small scales the $F_{NL}$ functions could be the same when $n$ and $\lambda$ fulfill Eq. \ref{condition}. This corresponds to a break in the degeneracy since, as we saw above, the power spectrum was indistinguishable for different values of $n$ and $\lambda$. While in case of the squeezed and equilateral limits for small scales and for different values of $n$ and $\lambda$ we found that the $F_{NL}$ functions can be the same.
\begin{figure}
  \centering
  \includegraphics[scale=0.8]{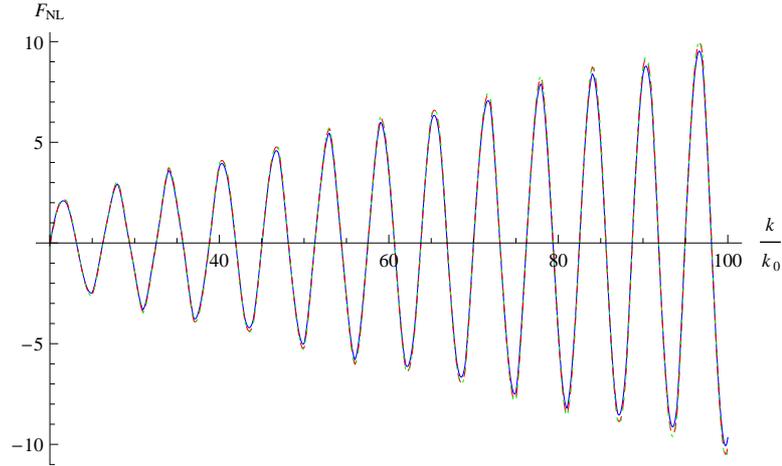}
  \caption{Plots of $F_{NL}(k,1000k_0,1000k_0)$ in the squeezed limit of small scales and for three different values of $n$ and $\lambda$. The parameters used are (blue) $n=2/3 \mbox{ and } \lambda=2.3\times10^{-15}$, (dashed-red) $n=3 \mbox{ and } \lambda=2.4\times10^{-18}$, and  (dashed-green) $n=4 \mbox{ and } \lambda=1.8\times10^{-19}$. In this case the $F_{NL}$ functions are the same for the appropriate values of $n$ and $\lambda$ for which the bispectrum is the same.}
  \label{FNL123plotsla}
\end{figure}

\begin{figure}
 \centering
 \includegraphics[scale=0.7]{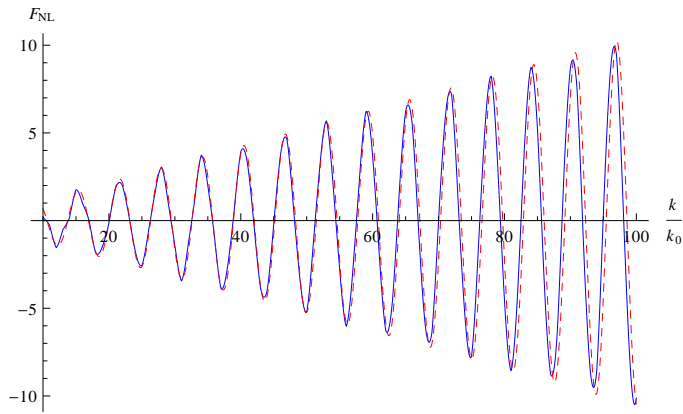}
 \caption{Numerical (solid-line) and analytic (dashed-line) plots of $F_{NL}(k,1000k_0,1000k_0)$ in the squeezed limit of small scales and for $n=3 \mbox{ and } \lambda=2.4\times10^{-18}$. This figure shows the agreement between the numerical and analytic results.}
 \label{FNLfnlsla}
\end{figure}

In the squeezed limit and for small scales we found that $F_{NL}$ is the same for different values of $n$ and $\lambda$ as can be seen in Fig. \ref{FNL123plotsla}. Fig. \ref{FNLfnlsla} shows the comparison between the numerical result and the analytic expression. For $n>2$ we have that $F_{NL}<0$ as can be seen in Fig. \ref{FNL123plotslb} for which the numerical result is plotted for different values of $n$ and $\lambda$. The comparison between the numerical and analytic results in cases of $n<2$ and $n>2$ are shown in Figs. \ref{FNLfnlslb1} and \ref{FNLfnlslb2}, respectively. It can be seen that the analytic expression Eq. \eqref{fnlslb} determines the change of sign in $F_{NL}$ correctly. 
\begin{figure}
  \centering
  \includegraphics[scale=0.7]{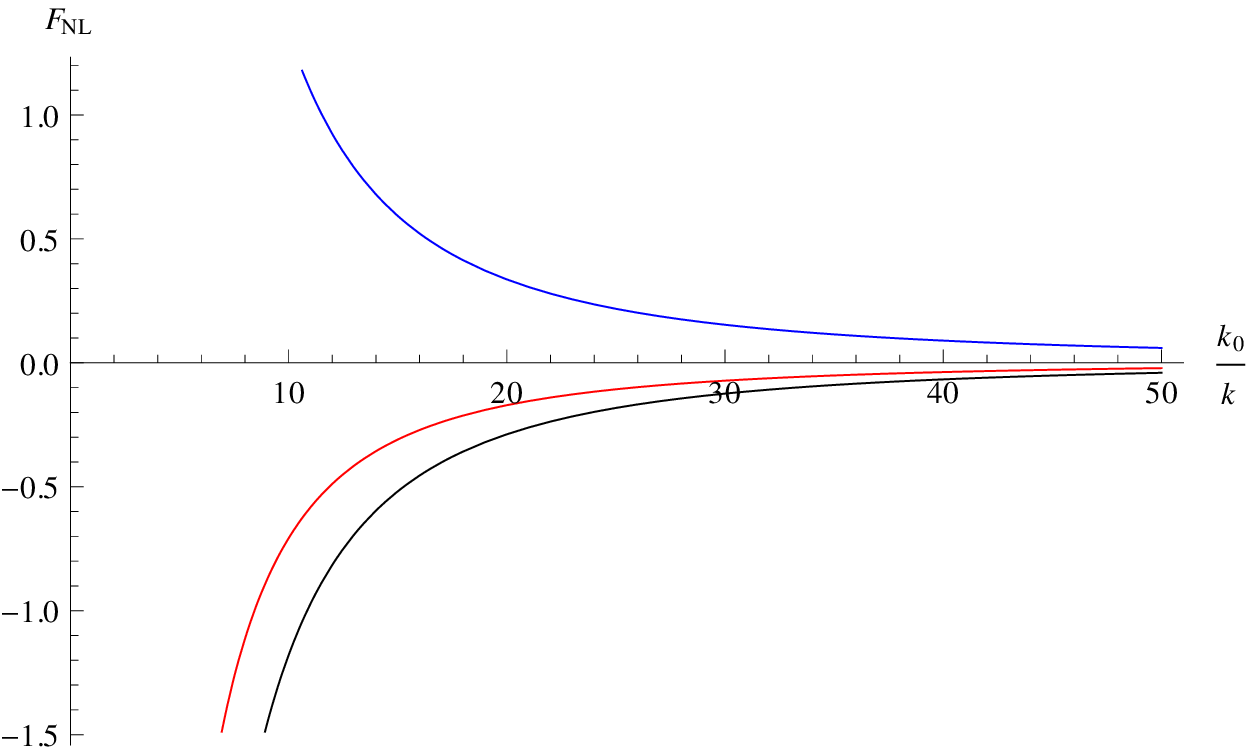}
  \caption{Numerical plots of $F_{NL}(k_0/500,k,k)$ in the squeezed limit of large scales and for three different values of $n$ and $\lambda$. The parameters used are (blue) $n=2/3 \mbox{ and } \lambda=2.3\times10^{-15}$, (red) $n=3 \mbox{ and } \lambda=2.4\times10^{-18}$, and (black) $n=4 \mbox{ and } \lambda=1.8\times10^{-19}$. In this case the $F_{NL}$ functions are different for the particular values of $n$ and $\lambda$, although the power spectrum for this values is the same.}
  \label{FNL123plotslb}
\end{figure}

\begin{figure}
 \centering
 \includegraphics[scale=0.7]{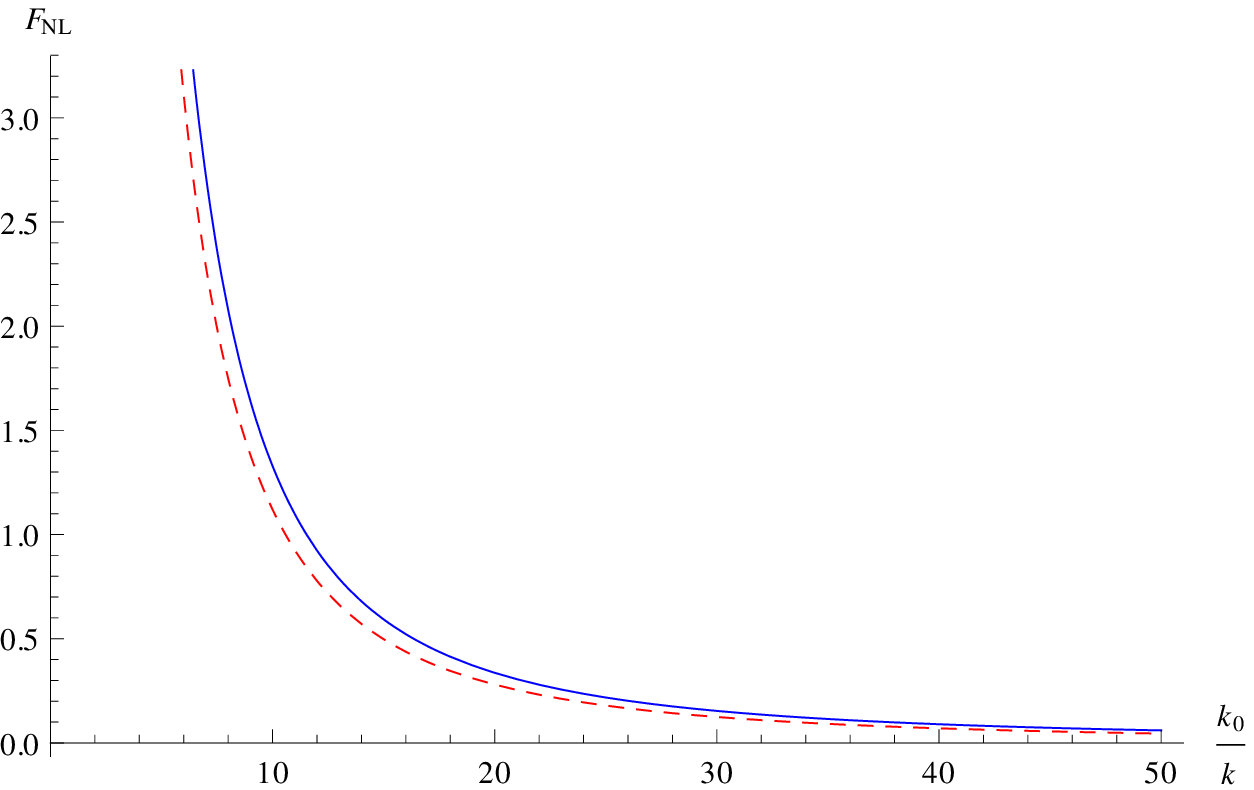}
 \caption{Numerical (solid-line) and analytic (dashed-line) plots of $F_{NL}(k_0/500,k,k)$ in the squeezed limit of large scales and for $n=2/3 \mbox{ and } \lambda=2.3\times10^{-15}$. This figure shows the agreement between the numerical and analytic results.}
 \label{FNLfnlslb1}
\end{figure}

\begin{figure}
 \centering
 \includegraphics[scale=0.7]{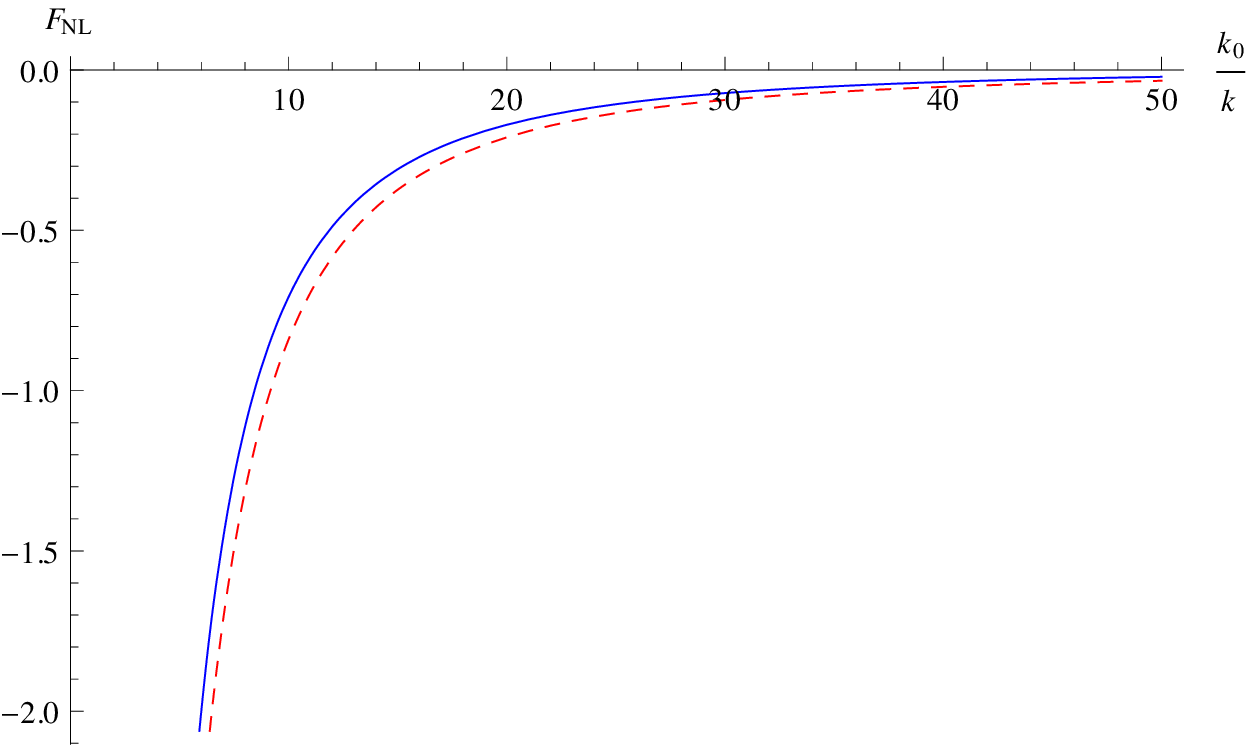}
 \caption{Numerical (solid-line) and analytic (dashed-line) plots of $F_{NL}(k_0/500,k,k)$ in the squeezed limit of large scales and for $n=3 \mbox{ and } \lambda=2.4\times10^{-18}$. This figure shows the agreement between the numerical and analytic results.}
 \label{FNLfnlslb2}
\end{figure}

In case of the equilateral limit and small scales we again found, Fig. \ref{FNL123plotela}, that the $F_{NL}$ functions are the same for different values of $n$ and $\lambda$. 
\begin{figure}
 \centering
 \includegraphics[scale=0.8]{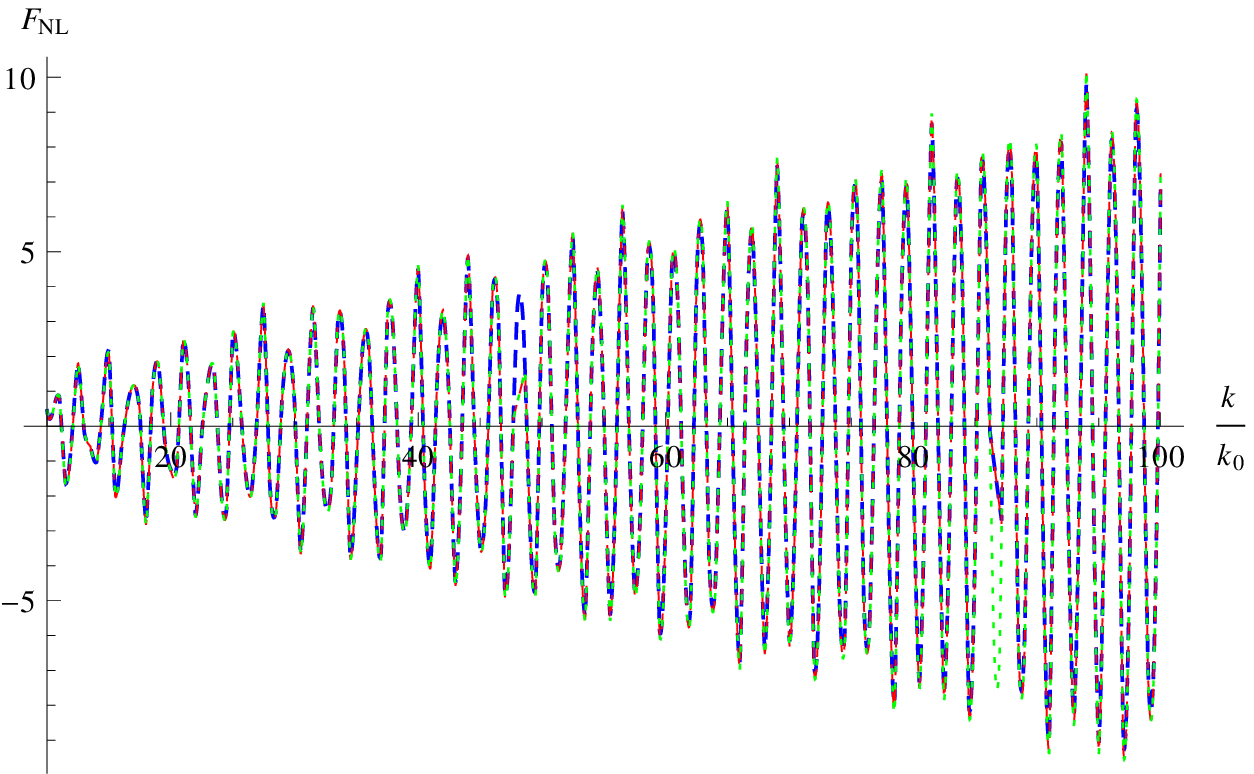}
 \caption{Plots of $F_{NL}(k,k,k)$ in the equilateral limit of small scales and for three different values of $n$ and $\lambda$ for which the bispectrum is exactly the same. The parameters used are (blue) $n=2/3 \mbox{ and } \lambda=2.3\times10^{-15}$, (dashed-red) $n=3 \mbox{ and } \lambda=2.4\times10^{-18}$, and  (dashed-green) $n=4 \mbox{ and } \lambda=1.8\times10^{-19}$. The errors marks in the plots correspond to a small path in the interpolation function to be plotted. In this case the $F_{NL}$ functions are the same for the particular values of $n$ and $\lambda$.}
 \label{FNL123plotela}
\end{figure}
In the equilateral limit and for large scales the sign of the $F_{NL}$ function is determined again by $n$. Figs. \ref{FNLfnlelb1} and \ref{FNLfnlelb2} show this change. The analytic expression is compared with the numerical results in Fig. \ref{FNL123plotelb}.
\begin{figure}
 \centering
 \includegraphics[scale=0.7]{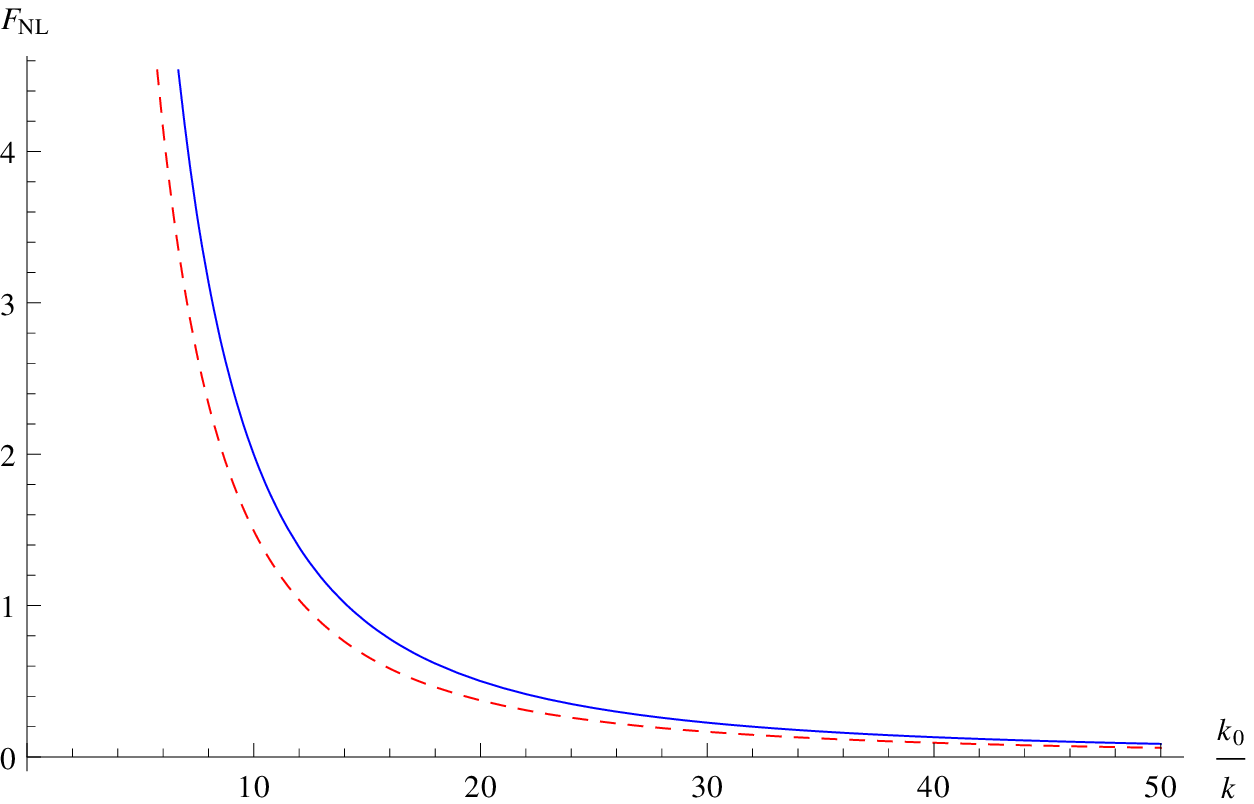}
 \caption{Numerical (solid-line) and analytic (dashed-line) plots of $F_{NL}(k,k,k)$ in the equilateral limit of large scales and for $n=2/3 \mbox{ and } \lambda=2.3\times10^{-15}$. This figure shows the agreement between the numerical and analytic results.}
 \label{FNLfnlelb1}
\end{figure}

\begin{figure}
 \centering
 \includegraphics[scale=0.7]{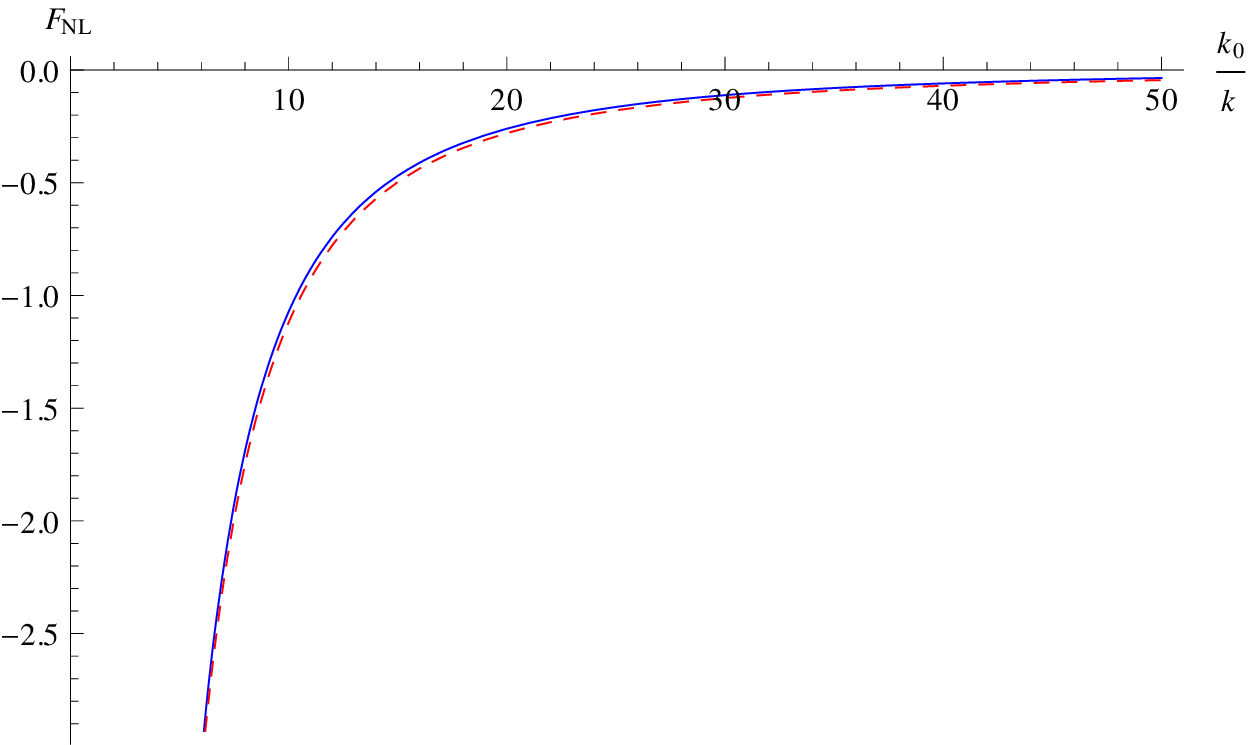}
 \caption{Numerical (solid-line) and analytic (dashed-line) plots of $F_{NL}(k,k,k)$ in the equilateral limit of large scales and for $n=3 \mbox{ and } \lambda=2.4\times10^{-18}$. This figure shows the agreement between the numerical and analytic results.}
 \label{FNLfnlelb2}
\end{figure}

\begin{figure}
 \centering
 \includegraphics[scale=0.7]{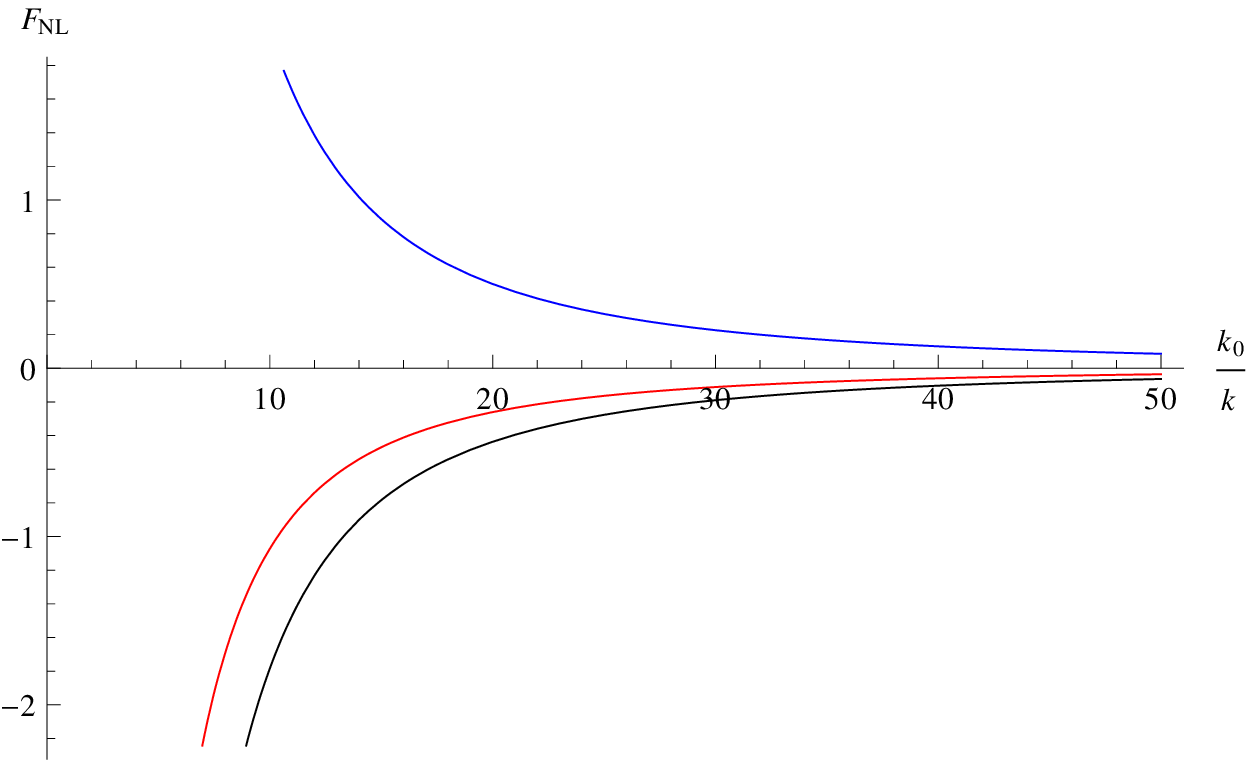}
 \caption{Numerical plots of $F_{NL}(k,k,k)$ in the equilateral limit of large scales and for three different values of $n$ and $\lambda$. The parameters used are (blue) $n=2/3 \mbox{ and } \lambda=2.3\times10^{-15}$, (red) $n=3 \mbox{ and } \lambda=2.4\times10^{-18}$, and (black) $n=4 \mbox{ and } \lambda=1.8\times10^{-19}$. In this case the $F_{NL}$ functions are different an appropriate choice of $n$ and $\lambda$, although the power spectrum for this values is the same}
 \label{FNL123plotelb}
\end{figure}
The feature in the potential studied in this thesis can generate the same power spectrum for an appropriate choice of values on $n$ and $\lambda$. Although at the bispectrum level this degeneracy is removed. In this case we can see that at large scales and for values of $n<2$ the $F_{NL}$ functions are positive in the squeezed and equilateral limits. While in the squeezed and equilateral limits for small scales and for different values of $n$ and $\lambda$ the $F_{NL}$ functions are the same. 

We summarized our results in Tables \ref{table1} and \ref{table2}. For large scales we have $k<k_0$ while for small scales $k > k_0$. The ``Equal'' indicates that the $F_{NL}$ functions can be the same for an appropriate choice of the values of the parameters for which the power spectrum is also the same.
\begin{table}
  \begin{center}
   \begin{tabular}{| c | c | c |}
    \hline
		& $k\le k_0$ & $k>k_0$ \\ \hline
    $P_{\zeta}$ &  Equal & Equal \\
    \hline
  \end{tabular}
  \caption{Equality/Inequality of the power spectrum for an appropriate choice of $n$ and $\lambda$ for which the spectrum can be the same.}\label{table1}
  \end{center}
  \end{table}

  \begin{table}
  \begin{center}
   \begin{tabular}{| c | c | c | c | c |}
    \hline
		& SL $k<k_0$ & SL $k \ge k_0$ & EL $k<k_0$ & EL $k \ge k_0$ \\ \hline
    $F_{NL}$ & Not equal & Equal & Not equal & Equal \\
    \hline
  \end{tabular}
  \caption{Equality/Inequality of the $F_{NL}$ functions for an appropriate choice of $n$ and $\lambda$ for which the spectrum can be the same. SL stands for squeezed limit and EQ for equilateral limit. For large scales we have $k<k_0$ while for small scales $k>k_0$.}\label{table2}
  \end{center}
  \end{table} 
\chapter{Conclusions} 
\label{Conclusions} 
\lhead{Chapter 5. \emph{Conclusions}} 
In this thesis we studied the generation of primordial perturbations in the context of inflationary models with features in the inflaton potential. In Chapter \ref{Chapter1} we introduced a justification for the study of non-Gaussianities of cosmological perturbations. In Chapter \ref{Chapter2}, we reviewed the theoretical framework, assumptions and observational evidence which contributed to the establishment of the Big Bang model and highlighted its successes and problems. We then moved on to an extension of the Big Bang model in order to account for solutions to the flatness and horizon problems as well as the formation of primordial perturbations. The resulting model is presented in Chapter \ref{Chapter3} as the Inflationary model, in which a scalar field minimally coupled to gravity drives an accelerated expansion of the universe solving the Big Bang problems and giving rise to the seeds of primordial perturbations. In this chapter we also introduced the calculation of the $n$-point correlation functions which relate the  fluctuations of the spacetime metric and the inflaton at different places of the early universe. As explained in Section \ref{NG} many single-field slow-roll models of inflation predict small amounts of non-Gaussianity (correlation functions of order larger than 2), even though there is still a mechanism for producing a large and detectable amount of non-Gaussianity, which corresponds among others, to the violation the slow-roll conditions \citep{et}. In Section \ref{curvatureperturbations} we reviewed the Maldacena's calculation \citep{m} of the second and third order terms in the inflationary action. The power spectrum and bispectrum of the primordial curvature perturbations are defined in Section \ref{psb}. 

In Chapter \ref{Chapter4} we studied the effects on the power spectrum and bispectrum of primordial curvature perturbations produced by a discontinuity in the $n$-th order derivative in the potential of the inflaton. In Section \ref{sdf} we reviewed the importance and justification of these features on the potential with respect to the observational data. As we mentioned there these models can provide better fits to the apparent low multipole glitch at $l \sim 20-40$ in the study of the angular spectrum of the CMB radiation \citep{constraints1,constraints2}. One important characteristic of our model is that it can generate large non-Gaussianities through the violation of the slow-roll conditions. We also found that each different type of feature has distinctive effects on the spectrum and bispectrum of curvature perturbations which depend both on the order $n$ of the discontinuous derivative and on the amplitude $\lambda$ of discontinuity. We have found that the spectra of primordial curvature perturbations have an oscillatory behavior for small scales. Also in the squeezed and equilateral limits for small scales the bispectrum has an oscillatory behavior whose phase depends on the parameters determining the discontinuity, and whose amplitude is inversely proportional to the scale. 

One of the most important results of  our work is that the features in the potential can also generate the same power spectrum for an appropriate choice of values on $n$ and $\lambda$ Eq. \ref{condition}, although at the bispectrum level this degeneracy is removed. In this case in the squeezed and equilateral limits at large scales the $F_{NL}$ functions are positive for $n<2$ and negative for $n>2$. While in both limits for small scales the $F_{NL}$ functions could be the same when $n$ and $\lambda$ fulfill Eq. \ref{condition}. The importance of this result is that if we would only considered the power spectrum of curvature perturbations different models of inflation could be indistinguishable. It is then important to study the bispectrum of primordial perturbation in order to distinguish between several models of inflation \cite{pxxii}. We show again our results in Tables \ref{table1} and \ref{table2}. For large scales we have $k<k_0$ while for small scales $k > k_0$. The ``Equal'' indicates that the $F_{NL}$ functions can be the same.
\begin{table}
  \begin{center}
   \begin{tabular}{| c | c | c |}
    \hline
		& $k\le k_0$ & $k>k_0$ \\ \hline
    $P_{\zeta}$ &  Equal & Equal \\
    \hline
  \end{tabular}
  \caption{Equality of the power spectrum for an appropriate choice of $n$ and $\lambda$ for which the spectrum can be the same.}\label{table1}
  \end{center}
  \end{table}

  \begin{table}
  \begin{center}
   \begin{tabular}{| c | c | c | c | c |}
    \hline
		& SL $k<k_0$ & SL $k \ge k_0$ & EL $k<k_0$ & EL $k \ge k_0$ \\ \hline
    $F_{NL}$ & Not equal & Equal & Not equal & Equal \\
    \hline
  \end{tabular}
  \caption{Equality/Inequality of the $F_{NL}$ functions for an appropriate choice of $n$ and $\lambda$ for which the spectrum can be the same. SL stands for squeezed limit and EQ for equilateral limit. For large scales we have $k<k_0$ while for small scales $k>k_0$.}\label{table2}
  \end{center}
  \end{table}
Our numerical results of the power spectra are in agreement with the observational data from WMAP  and Planck \citep{wmap,pxxii}. The next step in the study of our model is to contrast the numerical results of the bispectra with the observational data. The interpretation of the CMB polarization from Planck data \citep{pxxii} or other experiments such as Bicep2 have not been released yet. Even though our results of the bispectra agreed with the previous results of WMAP \citep{wmap} and the preliminary results from Planck \citep{pxxii,pxxiv}. After comparing the data we could discriminate between different inflationary cosmological models in order to yield models that better describe the evolution of our universe.

Also up to this point our analysis has been focused on the study of the two- and three-point correlation functions of primordial perturbations. We have been able to characterize different inflationary model according to their predictions and comparison with the observational data present up-to-date. But there is still a long way to go. Since different models of inflation can produce the same spectra and bispectra (Figs. \ref{FNL123plotsla} and \ref{FNL123plotela}) we should move on to the study of higher order correlation functions such as the four-point correlation function known as the \textit{trispectrum}.  Then we could distinguish between different models of inflation. To achieve this task properly our numerical code would have to be improved. Moreover the resolution of experiments which detect tiny correlations of temperature at different points in the sky would also have to be improved if we really want to compare the observational data with the predictions of inflationary models which produce non-Gaussianities.

It is an exciting time for cosmologists! 
\addtocontents{toc}{\vspace{2em}} 
\appendix 
\addtocontents{toc}{\vspace{2em}} 
\backmatter
\label{Bibliography}
\lhead{\emph{Bibliography}} 
\bibliographystyle{unsrtnat}
\bibliography{../../Books/Bibliography} 

\begin{thebibliography}{57}
\providecommand{\natexlab}[1]{#1}
\providecommand{\url}[1]{\texttt{#1}}
\expandafter\ifx\csname urlstyle\endcsname\relax
  \providecommand{\doi}[1]{doi: #1}\else
  \providecommand{\doi}{doi: \begingroup \urlstyle{rm}\Url}\fi

\bibitem[Komatsu et~al.(2009)Komatsu, Afshordi, Bartolo, Baumann, Bond,
  et~al.]{et}
E.~Komatsu, N.~Afshordi, N.~Bartolo, D.~Baumann, J.R. Bond, et~al.
\newblock {Non-Gaussianity as a Probe of the Physics of the Primordial Universe
  and the Astrophysics of the Low Redshift Universe}.
\newblock 2009.

\bibitem[Hinshaw et~al.(2013)]{wmapcpr}
G.~Hinshaw et~al.
\newblock {Nine-Year Wilkinson Microwave Anisotropy Probe (WMAP) Observations:
  Cosmological Parameter Results}.
\newblock \emph{Astrophys.J.Suppl.}, 208:\penalty0 19, 2013.
\newblock \doi{10.1088/0067-0049/208/2/19}.

\bibitem[Ade et~al.(2013{\natexlab{a}})]{pxvi}
P.A.R. Ade et~al.
\newblock {Planck 2013 results. XVI. Cosmological parameters}.
\newblock 2013{\natexlab{a}}.

\bibitem[Reichardt et~al.(2012)Reichardt, Shaw, Zahn, Aird, Benson,
  et~al.]{gbe1}
C.L. Reichardt, L.~Shaw, O.~Zahn, K.A. Aird, B.A. Benson, et~al.
\newblock {A measurement of secondary cosmic microwave background anisotropies
  with two years of South Pole Telescope observations}.
\newblock \emph{Astrophys.J.}, 755:\penalty0 70, 2012.
\newblock \doi{10.1088/0004-637X/755/1/70}.

\bibitem[Sievers et~al.(2013)Sievers, Hlozek, Nolta, Acquaviva, Addison,
  et~al.]{gbe2}
Jonathan~L. Sievers, Renee~A. Hlozek, Michael~R. Nolta, Viviana Acquaviva,
  Graeme~E. Addison, et~al.
\newblock {The Atacama Cosmology Telescope: Cosmological parameters from three
  seasons of data}.
\newblock 2013.

\bibitem[Yadav and Wandelt(2010)]{pncmb}
Amit~P.S. Yadav and Benjamin~D. Wandelt.
\newblock {Primordial Non-Gaussianity in the Cosmic Microwave Background}.
\newblock \emph{Adv.Astron.}, 2010:\penalty0 565248, 2010.
\newblock \doi{10.1155/2010/565248}.

\bibitem[Dodelson(2003)]{mc}
Scott Dodelson.
\newblock {Modern cosmology}.
\newblock 2003.

\bibitem[Kolb and Turner(1990)]{eu}
Edward~W. Kolb and Michael~S. Turner.
\newblock {The Early universe}.
\newblock \emph{Front.Phys.}, 69:\penalty0 1--547, 1990.

\bibitem[Bennett et~al.(2013)]{wmapfmr}
C.L. Bennett et~al.
\newblock {Nine-Year Wilkinson Microwave Anisotropy Probe (WMAP) Observations:
  Final Maps and Results}.
\newblock \emph{Astrophys.J.Suppl.}, 208:\penalty0 20, 2013.
\newblock \doi{10.1088/0067-0049/208/2/20}.

\bibitem[Ade et~al.(2013{\natexlab{b}})]{pxxii}
P.A.R. Ade et~al.
\newblock {Planck 2013 results. XXII. Constraints on inflation}.
\newblock 2013{\natexlab{b}}.

\bibitem[Chen(2010)]{xc}
Xingang Chen.
\newblock {Primordial Non-Gaussianities from Inflation Models}.
\newblock \emph{Adv.Astron.}, 2010:\penalty0 638979, 2010.
\newblock \doi{10.1155/2010/638979}.

\bibitem[Liddle and Lyth(2000)]{inf}
Andrew~R. Liddle and D.H. Lyth.
\newblock {Cosmological inflation and large scale structure}.
\newblock 2000.

\bibitem[Collins(2011)]{hael}
Hael Collins.
\newblock {Primordial non-Gaussianities from inflation}.
\newblock 2011.

\bibitem[Chung and Enea~Romano(2006)]{aersi}
Daniel~J.H. Chung and Antonio Enea~Romano.
\newblock {Approximate consistency condition from running spectral index in
  slow-roll inflationary models}.
\newblock \emph{Phys.Rev.}, D73:\penalty0 103510, 2006.
\newblock \doi{10.1103/PhysRevD.73.103510}.

\bibitem[Guth(1981)]{ag}
Alan~H. Guth.
\newblock {The Inflationary Universe: A Possible Solution to the Horizon and
  Flatness Problems}.
\newblock \emph{Phys.Rev.}, D23:\penalty0 347--356, 1981.
\newblock \doi{10.1103/PhysRevD.23.347}.

\bibitem[Linde(1990)]{linde}
Andrei~D. Linde.
\newblock {Particle physics and inflationary cosmology}.
\newblock \emph{Contemp.Concepts Phys.}, 5:\penalty0 1--362, 1990.

\bibitem[Maldacena(2003)]{m}
Juan~Martin Maldacena.
\newblock {Non-Gaussian features of primordial fluctuations in single field
  inflationary models}.
\newblock \emph{JHEP}, 0305:\penalty0 013, 2003.

\bibitem[Brandenberger(2010)]{rb}
Robert~H. Brandenberger.
\newblock {Introduction to Early Universe Cosmology}.
\newblock \emph{PoS}, ICFI2010:\penalty0 001, 2010.

\bibitem[Jain et~al.(2013)Jain, Joyce, Thompson, Upadhye, Battat,
  et~al.]{modifiedg}
Bhuvnesh Jain, Austin Joyce, Rodger Thompson, Amol Upadhye, James Battat,
  et~al.
\newblock {Novel Probes of Gravity and Dark Energy}.
\newblock 2013.

\bibitem[Hogg et~al.(2005)Hogg, Eisenstein, Blanton, Bahcall, Brinkmann,
  et~al.]{cosmichomogeneity}
David~W. Hogg, Daniel~J. Eisenstein, Michael~R. Blanton, Neta~A. Bahcall,
  J.~Brinkmann, et~al.
\newblock {Cosmic homogeneity demonstrated with luminous red galaxies}.
\newblock \emph{Astrophys.J.}, 624:\penalty0 54--58, 2005.
\newblock \doi{10.1086/429084}.

\bibitem[Adelman-McCarthy et~al.(2008)]{sdss}
Jennifer~K. Adelman-McCarthy et~al.
\newblock {The Sixth Data Release of the Sloan Digital Sky Survey}.
\newblock \emph{Astrophys.J.Suppl.}, 175:\penalty0 297--313, 2008.
\newblock \doi{10.1086/524984}.

\bibitem[Peebles(1998)]{peeblessm}
P.J.E. Peebles.
\newblock {The Standard cosmological model}.
\newblock 1998.

\bibitem[Primack(2005)]{primack}
Joel~R. Primack.
\newblock {Precision cosmology}.
\newblock \emph{New Astron.Rev.}, 49:\penalty0 25--34, 2005.
\newblock \doi{10.1016/j.newar.2005.01.039}.

\bibitem[Ryden(2003)]{itoc}
Barbara Ryden.
\newblock {Introduction to Cosmology}.
\newblock 2003.

\bibitem[Coc et~al.(2013)Coc, Uzan, and Vangioni]{bbn}
Alain Coc, Jean-Philippe Uzan, and Elisabeth Vangioni.
\newblock {Standard Big-Bang Nucleosynthesis after Planck}.
\newblock 2013.

\bibitem[Wald(1984)]{w}
Robert~M. Wald.
\newblock {General Relativity}.
\newblock 1984.

\bibitem[Oyvind~G(2007)]{gron}
.~Sigbjorn~H. Oyvind~G.
\newblock Einstein's general theory of relativity.
\newblock 2007.

\bibitem[Freedman et~al.(2001)]{hubblediagram}
W.L. Freedman et~al.
\newblock {Final results from the Hubble Space Telescope key project to measure
  the Hubble constant}.
\newblock \emph{Astrophys.J.}, 553:\penalty0 47--72, 2001.
\newblock \doi{10.1086/320638}.

\bibitem[Riess et~al.(1998)]{acceleration1}
Adam~G. Riess et~al.
\newblock {Observational evidence from supernovae for an accelerating universe
  and a cosmological constant}.
\newblock \emph{Astron.J.}, 116:\penalty0 1009--1038, 1998.
\newblock \doi{10.1086/300499}.

\bibitem[Perlmutter et~al.(1999)]{acceleration2}
S.~Perlmutter et~al.
\newblock {Measurements of Omega and Lambda from 42 high redshift supernovae}.
\newblock \emph{Astrophys.J.}, 517:\penalty0 565--586, 1999.
\newblock \doi{10.1086/307221}.

\bibitem[Burles et~al.(1999)Burles, Nollett, and Turner]{nucleosynthesis}
S.~Burles, K.M. Nollett, and Michael~S. Turner.
\newblock {Big bang nucleosynthesis: Linking inner space and outer space}.
\newblock 1999.

\bibitem[Peebles(1994)]{peebles}
P.J.E. Peebles.
\newblock {Principles of physical cosmology}.
\newblock 1994.

\bibitem[Fixsen(2009)]{temperature}
D.J. Fixsen.
\newblock {The Temperature of the Cosmic Microwave Background}.
\newblock \emph{Astrophys.J.}, 707:\penalty0 916--920, 2009.
\newblock \doi{10.1088/0004-637X/707/2/916}.

\bibitem[Smoot(1997)]{cmbspectrum}
George~F. Smoot.
\newblock {The CMB spectrum}.
\newblock pages 407--440, 1997.

\bibitem[Langlois(2010)]{linf}
D.~Langlois.
\newblock {Lectures on inflation and cosmological perturbations}.
\newblock \emph{Lect.Notes Phys.}, 800:\penalty0 1--57, 2010.
\newblock \doi{10.1007/978-3-642-10598-2_1}.

\bibitem[Ade et~al.(2013{\natexlab{c}})]{pi}
P.A.R. Ade et~al.
\newblock {Planck 2013 results. I. Overview of products and scientific
  results}.
\newblock 2013{\natexlab{c}}.

\bibitem[{Guth} and {Weinberg}(1983)]{agw}
A.~H. {Guth} and E.~J. {Weinberg}.
\newblock {Could the universe have recovered from a slow first-order phase
  transition?}
\newblock \emph{Nuclear Physics B}, 212:\penalty0 321--364, February 1983.
\newblock \doi{10.1016/0550-3213(83)90307-3}.

\bibitem[Arnowitt et~al.(2008)Arnowitt, Deser, and Misner]{adm}
Richard~L. Arnowitt, Stanley Deser, and Charles~W. Misner.
\newblock {The Dynamics of general relativity}.
\newblock \emph{Gen.Rel.Grav.}, 40:\penalty0 1997--2027, 2008.
\newblock \doi{10.1007/s10714-008-0661-1}.

\bibitem[Peskin and Schroeder(1995)]{pe}
M.~Peskin and D.~Schroeder.
\newblock \emph{An introduction to quantum field theory}.
\newblock Addison-Wesley Publishing Company, 1995.

\bibitem[Chen et~al.(2007)Chen, Easther, and Lim]{a2}
Xingang Chen, Richard Easther, and Eugene~A. Lim.
\newblock {Large Non-Gaussianities in Single Field Inflation}.
\newblock \emph{JCAP}, 0706:\penalty0 023, 2007.
\newblock \doi{10.1088/1475-7516/2007/06/023}.

\bibitem[Arroja et~al.(2011)Arroja, Romano, and Sasaki]{aer}
Frederico Arroja, Antonio~Enea Romano, and Misao Sasaki.
\newblock {Large and strong scale dependent bispectrum in single field
  inflation from a sharp feature in the mass}.
\newblock \emph{Phys.Rev.}, D84:\penalty0 123503, 2011.
\newblock \doi{10.1103/PhysRevD.84.123503}.

\bibitem[Adshead et~al.(2012)Adshead, Dvorkin, Hu, and Lim]{a1}
Peter Adshead, Cora Dvorkin, Wayne Hu, and Eugene~A. Lim.
\newblock {Non-Gaussianity from Step Features in the Inflationary Potential}.
\newblock \emph{Phys.Rev.}, D85:\penalty0 023531, 2012.
\newblock \doi{10.1103/PhysRevD.85.023531}.

\bibitem[Ade et~al.(2013{\natexlab{d}})]{pxxiv}
P.A.R. Ade et~al.
\newblock {Planck 2013 Results. XXIV. Constraints on primordial
  non-Gaussianity}.
\newblock 2013{\natexlab{d}}.

\bibitem[Salopek and Bond(1990)]{salopek}
D.S. Salopek and J.R. Bond.
\newblock {Nonlinear evolution of long wavelength metric fluctuations in
  inflationary models}.
\newblock \emph{Phys.Rev.}, D42:\penalty0 3936--3962, 1990.
\newblock \doi{10.1103/PhysRevD.42.3936}.

\bibitem[Figueroa et~al.(2012)Figueroa, Sefusatti, Riotto, and
  Vernizzi]{figueroa}
D.G. Figueroa, E.~Sefusatti, A.~Riotto, and F.~Vernizzi.
\newblock {The Effect of Local non-Gaussianity on the Matter Bispectrum at
  Small Scales}.
\newblock \emph{JCAP}, 1208:\penalty0 036, 2012.
\newblock \doi{10.1088/1475-7516/2012/08/036}.

\bibitem[Komatsu and Spergel(2001)]{komatsufnl}
Eiichiro Komatsu and David~N. Spergel.
\newblock {Acoustic signatures in the primary microwave background bispectrum}.
\newblock \emph{Phys.Rev.}, D63:\penalty0 063002, 2001.
\newblock \doi{10.1103/PhysRevD.63.063002}.

\bibitem[Weinberg(2005)]{inin}
Steven Weinberg.
\newblock {Quantum contributions to cosmological correlations}.
\newblock \emph{Phys.Rev.}, D72:\penalty0 043514, 2005.
\newblock \doi{10.1103/PhysRevD.72.043514}.

\bibitem[Chen et~al.(2008)Chen, Easther, and Lim]{a3}
Xingang Chen, Richard Easther, and Eugene~A. Lim.
\newblock {Generation and Characterization of Large Non-Gaussianities in Single
  Field Inflation}.
\newblock \emph{JCAP}, 0804:\penalty0 010, 2008.
\newblock \doi{10.1088/1475-7516/2008/04/010}.

\bibitem[Hannestad et~al.(2010)Hannestad, Haugbolle, Jarnhus, and Sloth]{hann}
Steen Hannestad, Troels Haugbolle, Philip~R. Jarnhus, and Martin~S. Sloth.
\newblock {Non-Gaussianity from Axion Monodromy Inflation}.
\newblock \emph{JCAP}, 1006:\penalty0 001, 2010.
\newblock \doi{10.1088/1475-7516/2010/06/001}.

\bibitem[Starobinsky(1992)]{starobinsky}
Alexei~A. Starobinsky.
\newblock {Spectrum of adiabatic perturbations in the universe when there are
  singularities in the inflation potential}.
\newblock \emph{JETP Lett.}, 55:\penalty0 489--494, 1992.

\bibitem[Hamann et~al.(2007)Hamann, Covi, Melchiorri, and Slosar]{constraints1}
Jan Hamann, Laura Covi, Alessandro Melchiorri, and Anze Slosar.
\newblock {New Constraints on Oscillations in the Primordial Spectrum of
  Inflationary Perturbations}.
\newblock \emph{Phys.Rev.}, D76:\penalty0 023503, 2007.
\newblock \doi{10.1103/PhysRevD.76.023503}.

\bibitem[Hazra et~al.(2010)Hazra, Aich, Jain, Sriramkumar, and
  Souradeep]{constraints2}
Dhiraj~Kumar Hazra, Moumita Aich, Rajeev~Kumar Jain, L.~Sriramkumar, and Tarun
  Souradeep.
\newblock {Primordial features due to a step in the inflaton potential}.
\newblock \emph{JCAP}, 1010:\penalty0 008, 2010.
\newblock \doi{10.1088/1475-7516/2010/10/008}.

\bibitem[Martin et~al.(2013)Martin, Ringeval, and Vennin]{encyclopaedia}
Jerome Martin, Christophe Ringeval, and Vincent Vennin.
\newblock {Encyclopaedia Inflationaris}.
\newblock 2013.

\bibitem[Adams et~al.(1997)Adams, Ross, and Sarkar]{Adams2}
Jennifer~A. Adams, Graham~G. Ross, and Subir Sarkar.
\newblock {Multiple inflation}.
\newblock \emph{Nucl.Phys.}, B503:\penalty0 405--425, 1997.
\newblock \doi{10.1016/S0550-3213(97)00431-8}.

\bibitem[Adams et~al.(2001)Adams, Cresswell, and Easther]{Adams}
Jennifer~A. Adams, Bevan Cresswell, and Richard Easther.
\newblock {Inflationary perturbations from a potential with a step}.
\newblock \emph{Phys.Rev.}, D64:\penalty0 123514, 2001.
\newblock \doi{10.1103/PhysRevD.64.123514}.

\bibitem[Romano and Sasaki(2008)]{pp}
Antonio~Enea Romano and Misao Sasaki.
\newblock {Effects of particle production during inflation}.
\newblock \emph{Phys.Rev.}, D78:\penalty0 103522, 2008.
\newblock \doi{10.1103/PhysRevD.78.103522}.

\bibitem[Larson et~al.(2011)Larson, Dunkley, Hinshaw, Komatsu, Nolta,
  et~al.]{wmap}
D.~Larson, J.~Dunkley, G.~Hinshaw, E.~Komatsu, M.R. Nolta, et~al.
\newblock {Seven-Year Wilkinson Microwave Anisotropy Probe (WMAP) Observations:
  Power Spectra and WMAP-Derived Parameters}.
\newblock \emph{Astrophys.J.Suppl.}, 192:\penalty0 16, 2011.
\newblock \doi{10.1088/0067-0049/192/2/16}.

\end{thebibliography}

\end{document}